\documentclass[12pt,preprint]{aastex}
\usepackage{xcolor}
\usepackage{ulem}
\usepackage{graphicx}
\usepackage{threeparttable}
\usepackage[colorlinks=true,citecolor=blue]{hyperref}

\title{The Extreme Red Excess in Blazar Ultraviolet Broad Emission Lines}
\author{Brian Punsly\altaffilmark{1}, Paola Marziani\altaffilmark{2}, Marco Berton\altaffilmark{3,4} and Preeti Kharb\altaffilmark{5}}\altaffiltext{1}{1415 Granvia Altamira, Palos Verdes Estates CA, USA 90274: ICRANet, Piazza della Repubblica 10 Pescara 65100, Italy and ICRA, Physics Department, University La Sapienza, Roma, Italy,
brian.punsly@cox.net}\altaffiltext{2}{INAF, Osservatorio Astronomico di Padova, Italy}\altaffiltext{3}{Finnish Centre for Astronomy with ESO (FINCA), University of Turku, Vesilinnantie 5, FI-20014 University of Turku, Finland}\altaffiltext{4}{Aalto University Mets{\"a}hovi Radio Observatory, Mets{\"a}hovintie 114, FI-02540 Kylm{\"a}l{\"a}, Finland}\altaffiltext{5}{National Centre for Radio Astrophysics - Tata Institute of Fundamental Research, S. P. Pune University Campus, Ganeshkhind, Pune 411007, India}
\begin{document}
\begin{abstract}
We present a study of quasars with very redward asymmetric (RA) ultraviolet (UV) broad emission lines (BELs). An excess of redshifted emission has been previously shown to occur in the BELs of radio loud quasars and is most extreme in certain blazars. Paradoxically, blazars are objects that are characterized by a highly relativistic blue-shifted outflow towards Earth. We show that the red emitting gas resides in a very broad component (VBC) that is typical of Population B quasars that are defined by a wide H$\beta$ BEL profile. Empirically, we find that RA BEL blazars have both low Eddington rates ($\lesssim1\%$) and an inordinately large (order unity) ratio of long term time averaged jet power to accretion luminosity. The latter circumstance has been previously shown to be associated with a depressed extreme UV ionizing continuum. Both properties conspire to produce a low flux of ionizing photons, two orders of magnitude less than typical Population B quasars. We use CLOUDY models to demonstrate that a weak ionizing flux is required for gas near the central black hole to be optimally ionized to radiate BELs with high efficiency (most quasars over-ionize nearby gas, resulting in low radiative efficiency). The large gravitational redshift and transverse Doppler shift results in a VBC that is redshifted by $\sim 2000 -5000$~km~s$^{-1}$ with a correspondingly large line width. The RA BELs result from an enhanced efficiency (relative to typical Population B quasars) to produce a luminous, redshifted VBC near the central black hole.
\end{abstract}

\section{Introduction}
Quasars are characterized by bright broadband emission from a galactic nucleus with a spectral energy distribution (SED) that peaks in the ultraviolet (UV). This bright quasi-stellar emission is considered to arise from viscous dissipation from an accretion flow onto a supermassive black hole \citep{mal83}. Furthermore, quasar optical/UV spectra typically contain luminous broad emission lines (BELs). The quasar broadband continuum and the BELs offer clues to the state of the accreting gas. Another prominent feature occurs in $\sim 10\%$ of quasars. These quasars possess powerful relativistic radio jets and are known as radio loud quasars (RLQs). The preponderance of quasars have weak or absent radio jets and are known as radio quiet quasars (RQQs). Surprisingly, despite the on or off condition for the extremely powerful jet, there are no conspicuous differences between the observed signatures of the accretion states in RLQs and RQQs. To first order, the continua and BELs in RLQS and RQQs are remarkably similar \citep{cor94,zhe97,tel02,ric02}. Systematic differences in the two families of spectra are only revealed by the study of subtle lower order spectral features \citep{cor98,tel02,pun14}. Consequently, the difference between the black hole accretion system (the putative central engine) is still poorly constrained by observation and therefore so is our understanding of the dynamics of relativistic jet launching more than 50 years after their discovery in quasars.

This similarity of accretion signatures is perplexing from a theoretical point of view. Numerical models of the accretion have offered little, if any, insight into the subtle spectral difference. From a continuum point of view, the first difference that has been found involves the extreme ultraviolet (EUV) tail of the spectral energy distribution \citep{zhe97,tel02}. RLQs have a deficit of EUV emission (the EUV deficit) relative to RQQs and the degree of this deficit increases with the relative strength of the long term time averaged jet power, $\overline{Q}$, compared to the accretion flow bolometric luminosity, $L_{\rm{bol}}$ \citep{pun15}. Namely, the magnitude of the EUV luminosity below $1100 \AA$ relative to the SED peak luminosity at $\sim 1100 \AA$ is anti-correlated with $\overline{Q}/L_{\rm{bol}}$.

In this paper, we concentrate on the most conspicuous difference seen in the BELs between RLQs and RQQs in order to see if there are more clues to the difference in the accretion state or nuclear environment. The differences in line shapes between RQQ and RLQ BELs was summarized in \citet{wbr96}: ``The CIV BEL generally has stronger red than blue wings in RLQs, but the blue wing is often stronger than the red in RQQs. The H$\beta$ BEL also often has stronger red wings in RLQs, with the RQQs showing similar frequency of red and blue asymmetries." The difference in the high ionization CIV BEL was later quantified. There is a highly significant correlation between the spectral index from 10 GHz to 1350 $\AA$ and the amount of excess luminosity in the red wing of quasar CIV BELs, at $>99.9999\%$ statistical significance \citep{pun10}. The prominence of the redward excess is apparently associated with the radio jet emission mechanism and is most pronounced for lines of sight close to the jet axis.

In order to explore this strange occurrence of an excess of red emission in sources in which the plasma is beamed towards Earth, we look at the most extreme blazar red-winged BELs in detail. We also consider these sources in the context of complete samples of blazars. In order to construct a large sample, we need a UV line that is easily accessible and has been observed for entire samples of quasars that qualify as extreme blazars (objects that display extreme blazar properties as formally defined in Section 4). The only suitable line is MgII$\lambda 2798$ in spite of the difficulty of extracting the FeII emission. There are three very redward asymmetric (RA) examples of a MgII BEL, 3C 279 (z = 0.536), 0954+556 (z = 0.899, 4C +55.17) and S5 1803+784 (z = 0.684). They are very high polarization quasars (HPQs) that have been detected with $\sim 10\%$ optical polarization \citep{imp90,lis16}. The MgII profiles are shown in Figure 1 in order to demonstrate the extreme circumstance under consideration. This is not a subtle effect nor a consequence of FeII multiplets. The preponderance of the line shape is redshifted emission from, ostensibly, receding gas even though these objects have powerful jets relativistic of plasma beamed toward Earth. The motivation of this paper is that this is not coincidental. We surmise that even though the BEL gas is far from the central engine, there is valuable evidence on the nature of the accretion flow in RLQs that can be gleaned from this paradoxical circumstance.

\begin{figure}
\begin{center}
\includegraphics[width=100 mm, angle= 0]{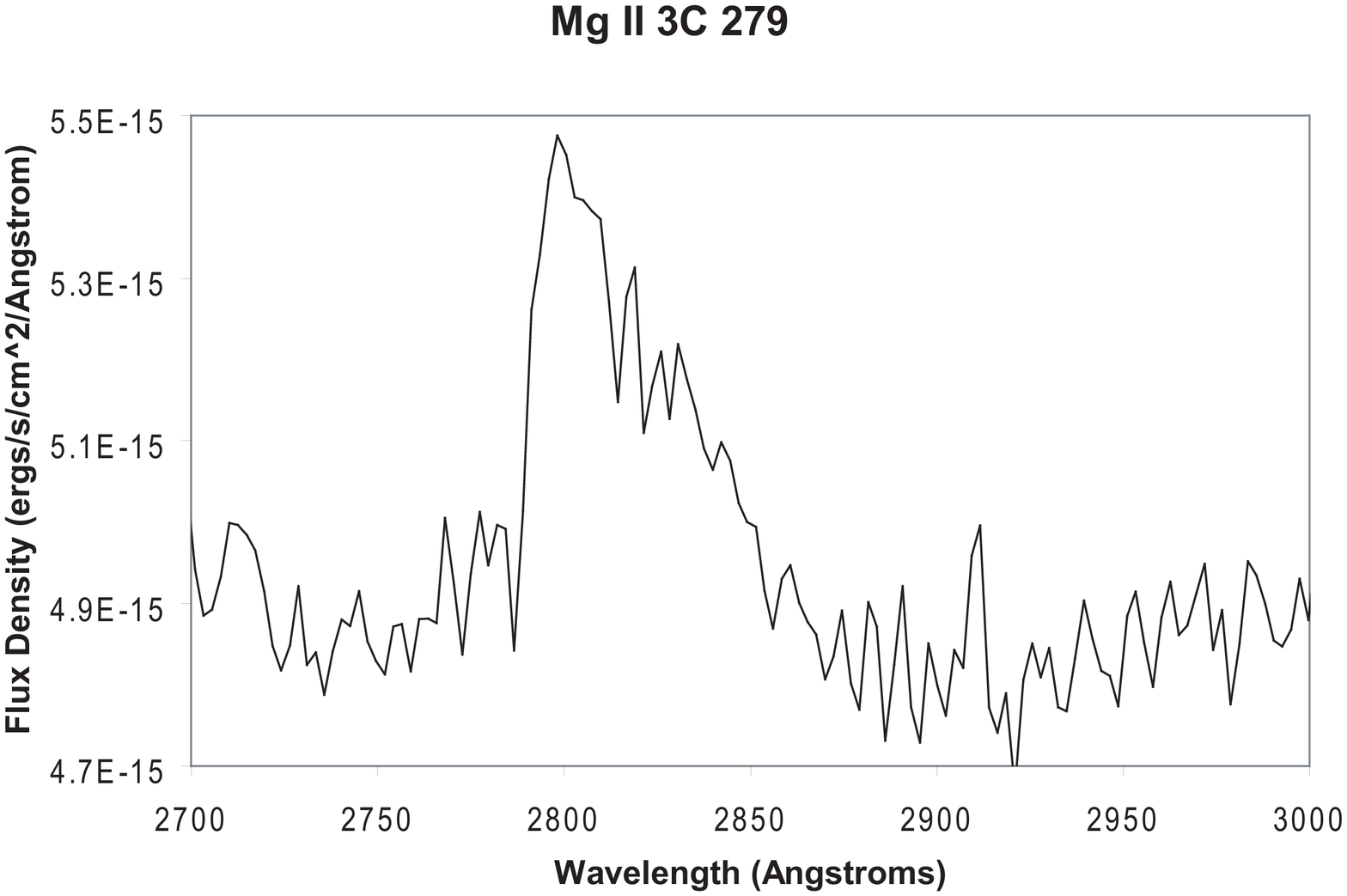}
\includegraphics[width=100 mm, angle= 0]{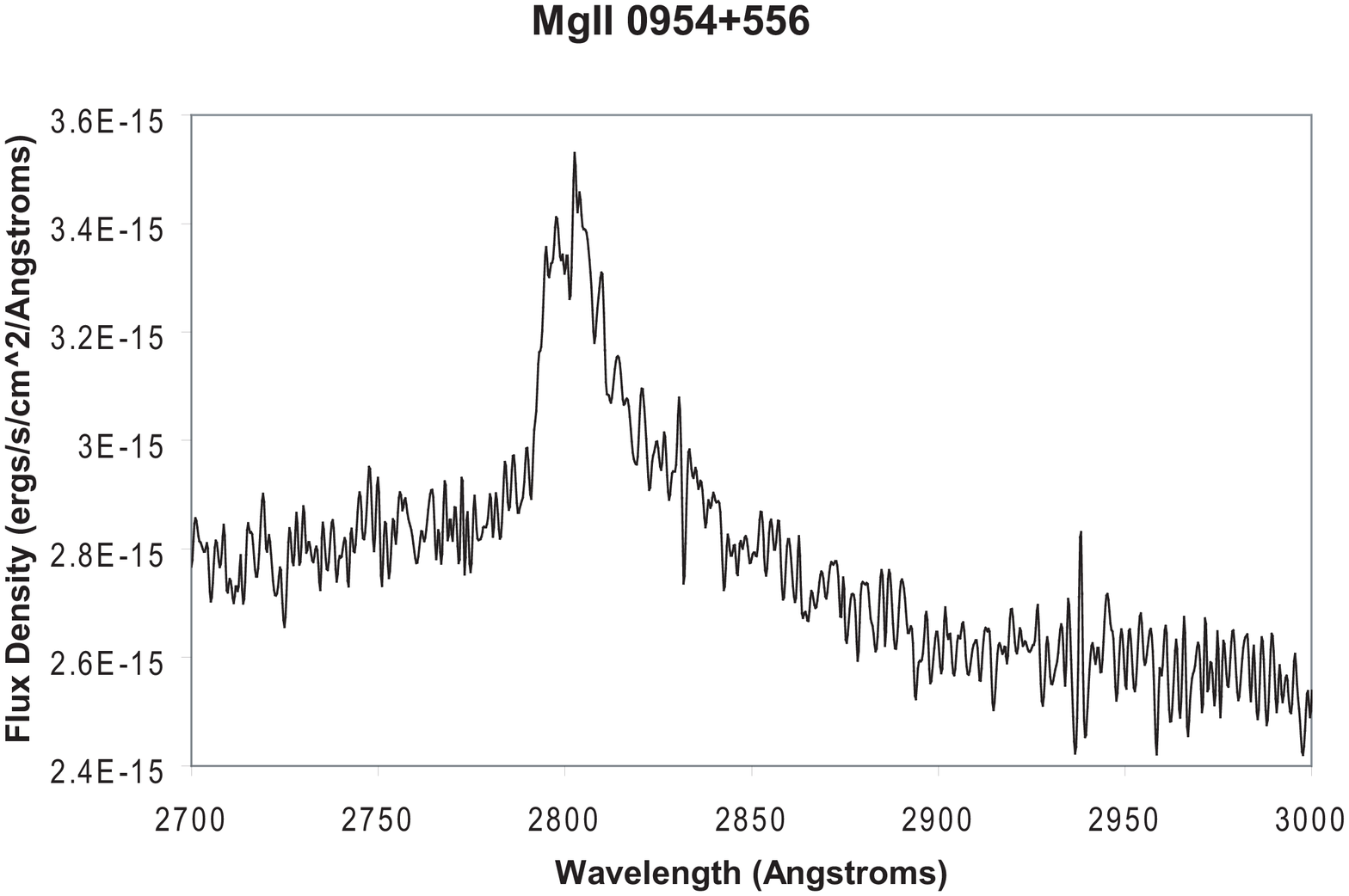}
\includegraphics[width=100 mm, angle= 0]{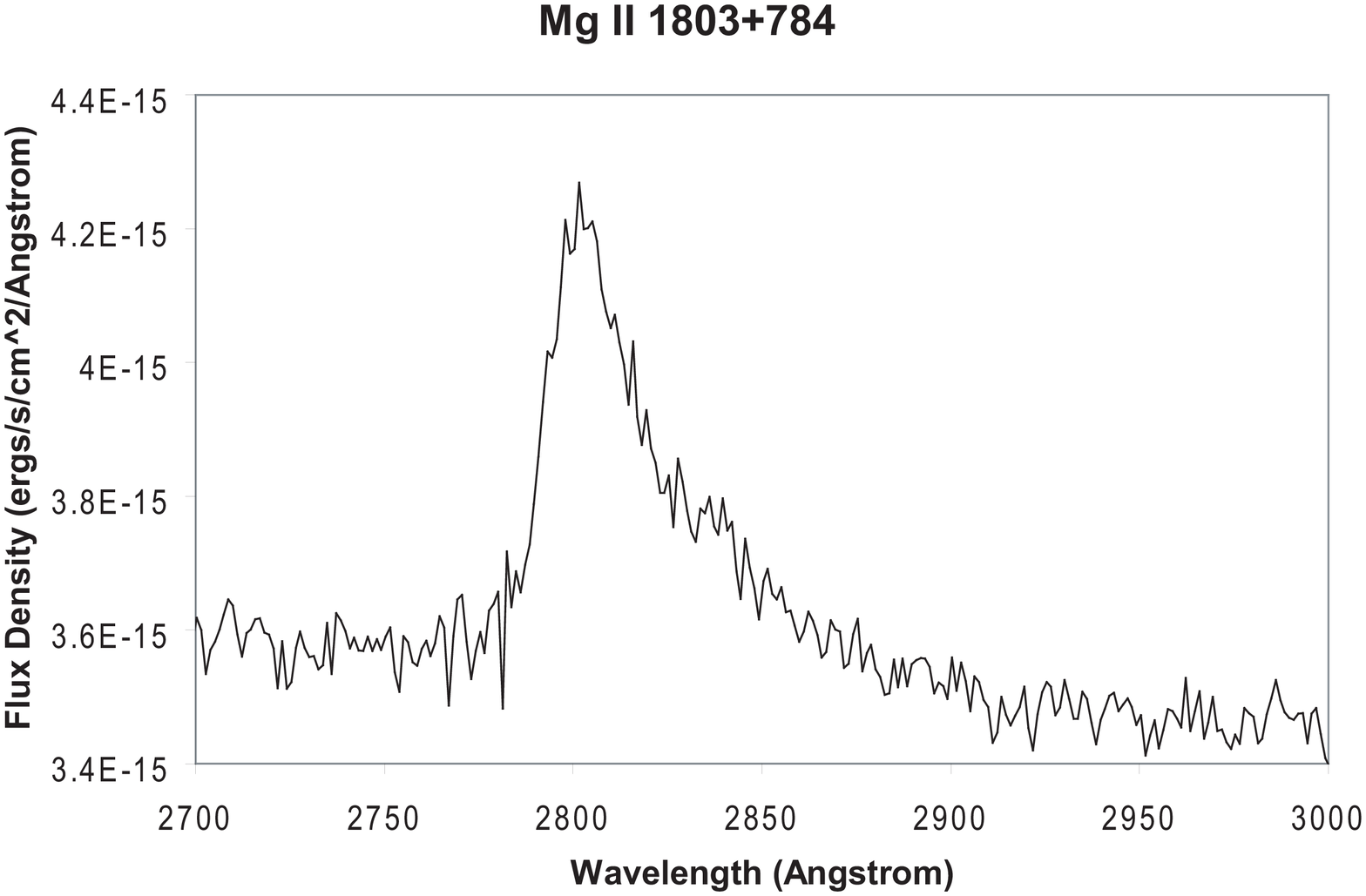}
\caption{The extremely RA MgII BELs of the three
HPQs, 3C 279 (4/9/1992), 0954+556 (3/24/2012) and 1803+784
(07/04/1986). The flux densities are evaluated in the quasar rest
frame.}
\end{center}
\end{figure}

In Section 2, we explore the BELs of four HPQs in detail, the
three extreme RA blazars, 3C 279, 0954+556, S5
1803+784 and 3C 345 which has recently developed a significant
redward asymmetry. The MgII line is used to compare these sources to
complete samples of extreme blazars. However, we also have CIV and
H$\beta$ lines that we consider as well for these quasars. In
Section 3 and the Appendix, we introduce new radio images in order to define
estimates of the long term time averaged jet power, $\overline{Q}$.
Section 4 is a comparison of these sources to larger complete
samples in order to see what makes them different than other extreme
blazars. In the next section, we interpret the difference as a very low Eddington rate for a quasar and a very strong jet. In Section 6, we explore the effect of the polar line of sight on central black hole mass estimates and the inferred Eddington rate. Section 7 is a computation of the gravitational and transverse Doppler shift of a Keplerian disk of gas. In Section 8, we describe the ionization state resulting from a low Eddington rate combined with the EUV deficit associated with a strong jet. This motivates a CLOUDY analysis that compares the blazars with the RA BELs to conventional quasars and low Eddington rate Seyfert galaxies. The weak ionizing continuum seems to explain why these particular blazars have such large red asymmetries in their BEL profiles. We compare our explanation of highly RA UV BELs to other explanations that have been posited in the literature or might be possible in Section 10.

\section{Blazars with Extreme RA Broad Emission Line Profiles}
3C 279, 0954+556 and 1803+784 have very extreme blazar properties as are
indicated in Table 1. Other extreme blazars are tabulated
as well for future reference in Section 4. All three have a very
high optical polarization. Both 3C 279 and 1803+784 are extremely
superluminal. No component motion between epochs has been analyzed
for 0954+556. All three are part of the \citet{ghi10} bright gamma
ray blazar sample. All of which is indicative of an outflow with
very large Lorentz factors with a velocity vector aligned to
within a few degrees of the line of sight resulting in large Doppler
enhancement \citep{lin85}.
\par The quasars 3C 279 and 0954+556 have the largest
redward asymmetry, $A$ (defined in Equation (1)), of the C IV BEL in
the sample of HST spectra from the Evans and Koratkar Atlas of
Hubble Space Telescope Faint Object Spectrograph Spectra of Active
Galactic Nuclei and Quasars \citep{eva04,pun10}. The quantity, $A$,
is defined in terms of the full width half maximum, FWHM, in $\AA$,
the midpoint of an imaginary line connecting a point defined at 1/4
of the peak flux density of the BEL on the red side of the BEL to
1/4 of the peak flux density on the blue side of the BEL,
$\lambda_{25}$, and a similar midpoint defined at 8/10 of the flux
density maximum, $\lambda_{80}$, as
\begin{equation}
A = \frac{\lambda_{25} - \lambda_{80}}{FWHM} \;.
\end{equation}
A positive value of $A$ means that there is excess flux in the red
broad wing of the BEL. For 3C 279 and 0954 + 556, the values were
0.64 and 1.25, respectively \citep{wil95}. This is extreme compared
to the average of the sample of 95 quasar C IV BELs in \citet{pun10}, $A =0.039 \pm
0.219 $.
\par In this section we
begin the analysis by providing formal fits to the BELs of the most
RA blazars. Our fits are decompositions based on the
analysis of \citet{bro94}
\begin{itemize}
\item The narrow line (NL, hereafter) region is a distant distribution of gas with emission lines characterized by a FWHM (full width at half maximum)
$\leq2000$~km~s$^{-1}$. Radio galaxies are known to often possess very strong narrow line regions with a FWHM between 1500 - 2000~km~s$^{-1}$, even if there is no evidence of a photo-ionizing source \citep{bal14,bes99}. Theoretically, relativistic jets are expected to energize gas thereby inducing narrow line emission \citep{bic98}. This is verified empirically. The correlation of $\overline{Q}$ with narrow line luminosity is stronger than the correlation of $L_{\rm{bol}}$ with the narrow line luminosity in radio loud quasars \citep{pun11}. Thus, strong narrow lines with $\rm{FWHM} \sim 1500 -2000$~km~s$^{-1}$ are expected in our fits to blazar spectra.
\item The broad component (BC, hereafter) is produced by gas that is virialized within gravitational potential of the central supermasssive black hole, The definition of this region is sometimes conflated with the notion of the intermediate BLR \citep{bro94}. The BC component of the emission lines are characterized by a FWHM $\gtrsim 2500$~km~s$^{-1}$. The interpretation of lines with a FWHM $\sim 2000$~km~s$^{-1}$ as a BC or a NL is not obvious in general. Line width is strongly affected by viewing angle, in ways that are likely different from line to line (e.g. H$\beta$\ and CIV). In the case of blazars, the low value of the viewing angle makes narrow widths of the low ionization lines more likely than in a general sample of quasars (see also Section 2.1).
\item The very broad component (VBC, hereafter) is produced by gas that is virialized within gravitational potential of
the central supermasssive black hole. The larger FWHM than the BC,
by definition, indicates that its source lies deeper within the
gravitational potential than the BC, i.e., closer to the central
supermassive black hole. This second broad line component is often
required in fits to BELs. The combination of the BC and VBC comprise
the ``full BC" component.
\end{itemize}
The background FeII emission was subtracted using a template computed by \citet{bru08}.   Line luminosity will be computed
using the following cosmological parameters: $H_{0}$=69.6 km s$^{-1}$ Mpc$^{-1}$, $\Omega_{\Lambda}=0.714$ and $\Omega_{m}=0.286$. The MgII BC component
fit is sensitive to the fact that MgII is a doublet at vacuum
wavelengths of $\lambda=2976.35\AA$ and $\lambda=2803.53\AA$. The
intensity ratio depends on the properties of the gas in the BLR. In \citet{mar13}, a ratio of the intensity of
the short wavelength line to the intensity of the long wavelength
line of 1.25 was estimated for quasars, in general. This
prescription is used in our MgII BC fits. The VBC is too broad for
the doublet nature of the MgII BEL to be significant and is ignored
in our fits.

\begin{table}
\caption{Broadline Blazar Properties}
{\footnotesize\begin{tabular}{ccc} \tableline\tableline \rule{0mm}{3mm}
 Quasar & Median Optical Polarization & Jet Component Speed    \\
 & percent  &   \\
& \citet{wil92}  & MOJAVE   \\
\tableline \rule{0mm}{1mm}\\
\hline
\multicolumn{3}{c}{Redward   Asymmetric   Blazars } \\
\tableline \rule{0mm}{1mm}
3C 279\tablenotemark{a} & 8.6 &  $ 20.58 \pm 0.79$c \tablenotemark{d}  \\
0954+556\tablenotemark{a}  & 8.68 &  No Measurement   \\
1803+784\tablenotemark{a} & 35.2 &  $ 9.36 \pm 0.87 $c \tablenotemark{e}   \\
3C 345 & 6.65 &  $19.28 \pm 0.51$c \tablenotemark{d} \\
\hline
\multicolumn{3}{c}{\citet{pea81} Sample   Extreme   Blazars } \\
\hline
0016+731 & 1.1 &  $8.23 \pm 0.34$c \tablenotemark{e}  \\
0133+476 & 20.8\tablenotemark{b} &  $16.54 \pm 0.56$c \tablenotemark{e}   \\
0212+735 & 7.8 &  $ 6.58 \pm 0.18 $c \tablenotemark{e}  \\
0804+499 & 6.48 &  $ 1.02 \pm 0.26  $c \tablenotemark{e}   \\
0836+710 & 1.1\tablenotemark{b} &  $  21.10 \pm 0.77  $c \tablenotemark{d}  \\
0850+581 & 0.4\tablenotemark{c} &  $  7.56 \pm 0.38  $c \tablenotemark{e}   \\
0859+470 & 1.0\tablenotemark{b} &  $  16.1 \pm 1.3  $c \tablenotemark{d}   \\
0945+408 & 1.12 &  $  20.23 \pm 0.95  $c \tablenotemark{d}  \\
1458+718  & 1.0\tablenotemark{c} &  $  6.72 \pm 0.27    $c \tablenotemark{e}   \\
1633+382 & 1.1 &  $  29.3 \pm 1.3    $c \tablenotemark{d}  \\
1637+574 & 1.2 &  $  13.61 \pm 0.89    $c \tablenotemark{d}   \\
1739+522 & 3.7\tablenotemark{b} & No Measurement   \\
1828+487  & 0.89 &  $  13.06 \pm 0.14 $c \tablenotemark{d}   \\
\hline
\multicolumn{3}{c}{3CR  Extreme  Blazars } \\
\hline
3C 273  & 0.28 &  $ 14.85 \pm 0.17    $c \tablenotemark{d}  \\
3C 309.1  & 1.0\tablenotemark{c} &  $  6.72 \pm 0.27    $c \tablenotemark{e}  \\
3C 380  & 0.89 &  $  13.06 \pm 0.14 $c \tablenotemark{d}  \\
3C 418  & ... &  $   6.7 \pm 1.4  $c \tablenotemark{e} \\
3C 454.3  & 3.1 &  $   13.79 \pm 0.49  $c \tablenotemark{e}  \\
\hline
\multicolumn{3}{c}{Additional 3C   Extreme   Blazars } \\
\hline
3C 120  & 2.3\tablenotemark{f} &  $   8.70 \pm 0.13  $c \tablenotemark{g}  \\
3C 395  & ... &  $   8.81 \pm 0.24  $c \tablenotemark{e}  \\
3C 446  & 8.8 &  $   17.7 \pm 3.1  $c \tablenotemark{e}  \\
\tableline \rule{0mm}{1mm}
\end{tabular}}
\tablenotetext{a}{Bright gamma ray blazar \citep{ghi10}}\tablenotetext{b}{\citet{imp90}}\tablenotetext{c}{\citet{imp91}}
\tablenotetext{d}{\citet{lis13}}\tablenotetext{e}{\citet{lis16}}\tablenotetext{f}{http://www.bu.edu/blazars}\tablenotetext{g}{\citet{jor17}}
\end{table}
For the sake of scientific rigor all the resultant fits need to be
analyzed within the same formalism. This means that previously
analyzed results will be refit. We choose the formalism of
\citet{mar96} in order to decompose the lines as above. The first
step is to redefine the asymmetry parameter in Equation (1) as the
asymmetry index, AI. First express all wavelengths in terms of
velocity relative to the laboratory of wavelength of the emission
line, $\lambda_{o}$, i.e., $v=c(\lambda-\lambda_{o})/\lambda_{o}$.
The first velocity is the peak of the emission, $v_{\rm{peak}}$.
Then we require the red and blue velocities at 1/4 maximum and 3/4
maximum, $v_{\rm{R}}(1/4)$, $v_{\rm{B}}(1/4)$, $v_{\rm{R}}(3/4)$ and
$v_{\rm{B}}(3/4)$, respectively. The line asymmetry is described by
\begin{eqnarray}
&& \rm{AI} \equiv
\frac{v_{\rm{R}}(1/4)+v_{\rm{B}}(1/4)-2v_{\rm{peak}}}{v_{\rm{R}}(1/4)-v_{\rm{B}}(1/4)}\;,\\
&& \rm{C(1/4)}\equiv \frac{v_{\rm{R}}(1/4)+v_{\rm{B}}(1/4)}{2}\;.
\end{eqnarray}
We rely primarily on these empirical measures as they are more
robust than parameters that are dependent on the line fitting
method.
\subsection{The Narrow Line/Broad Component Distinction} \label{ncbc}
An important issue is to properly identify components that ostensibly appear to be narrow lines, FWHM$\sim 1500-2000$~km~s$^{-1}$, for these blazars. There are two competing circumstances that need to be differentiated. The ambiguity is most apparent for the low ionization species, MgII and H$\beta$, that are generally believed to originate in a planar distribution of gas in near Keplerian motion \citep{bro86}. Thus, the polar line of sight associated with a blazar will significantly reduce the FWHM that is associated with the virialized motion. Consequently, a planar virialized component can appear to have a FWHM$<2000$~km~s$^{-1}$ for a blazar when it would appear as an unambiguous BC for a more oblique quasar line of sight. The other complicating factor is that ``narrow" lines can actually often be quite wide in radio sources (even narrow line radio galaxies) with a FWHM$\sim 1500-2000$~km~s$^{-1}$ being common \citep{mcc93,bes99}. So we have we need to be careful with our identifications of the BC. We use the following guide lines
\begin{itemize}
\item H$\beta$: We use an inflection point (an abrupt change in line shape or a narrow kink) between the NL and the BC to define our decomposition.
\item MgII: The NL is very weak in general and there is no evidence of strong NLs in radio sources \citep{bes99}. Thus, we identify a component with FWHM$\sim 1500-2000$~km~s$^{-1}$ as a BC.
\item Ly$\alpha$: The Ly$\alpha$ NL can be strong in radio sources \citep{mcc93}. There is one incidence of a fitted component with FWHM$\sim 1500-2000$~km~s$^{-1}$ (see Table 2 for 0954+556). However, there is a clear inflection point that allows us to separate a much narrower NL (FWHM$ < 1000$~km~s$^{-1}$) from the BC.
\item CIV: The CIV NL can also be strong in radio sources \citep{mcc93}. There is one incidence of a fitted component with FWHM$\sim 1500-2000$~km~s$^{-1}$ (see Table 2 for 0954+556). We separate a NL component by identifying a central spike (see Figure 3) with the FHWM of the Ly$\alpha$ NL mentioned above.
\end{itemize}

\subsection{The Broad Emission Lines of 3C 279}
The MgII and CIV BELs were previously analyzed, but are re-analyzed
here in order to get a uniform data reduction of all the the lines
in all of our blazars \citep{pun12,wil95,net95}.
\begin{table}
\caption{RA Blazar Broad Emission Line Fits}
{\tiny\begin{tabular}{ccccccccccc} \tableline\rule{0mm}{3mm}
1 &  2  &  3 &  4  & 5  & 6 & 7 & 8 & 9 & 10&11 \\
 Date &  Line & VBC & VBC &  VBC & BC &BC &Full BC  & Full BC & AI & C(1/4) \\
 &   &  Peak & FWHM & Luminosity & FWHM & Luminosity &FWHM & Luminosity & & \\
 &    &  km/s &  km/s  & ergs/s  & km/s & ergs/s &km/s & ergs/s & & km/s \\
\tableline \rule{0mm}{3mm}
3C279 &    &   &    &   &  &  & & & & \\
\hline
4/8/1994\tablenotemark{j,i} & H$\beta$ & $2141\pm 72$ & $13980\pm 8058 $ & $7.64 \times 10^{42}$  & 4810\tablenotemark{i} & $2.22 \times 10^{42}$ & 9477 & $9.86 \times 10^{42}$ &0.18 & 2099 \\
4/9/1992\tablenotemark{a} & MgII & $2018\pm 95$ & $6263\pm 235 $ & $1.20 \times 10^{43}$  & $1036\pm 106$ & $2.40 \times 10^{42}$ & 4652 & $1.44 \times 10^{43}$ &0.49 & 1915 \\
4/29/2009\tablenotemark{b} & MgII & $1351\pm 154$ & $6404\pm 2297$ & $5.13 \times 10^{42}$  & $2742\pm 241$& $4.29 \times 10^{42}$ & 3739 & $9.42 \times 10^{42}$ &0.24 & 567 \\
4/11/2010\tablenotemark{b} & MgII & $1555\pm 170$ & $6599\pm 764$ & $5.03 \times 10^{42}$  & $2500\pm 320$&$2.96 \times 10^{42}$  & 3819 & $7.99 \times 10^{42}$ &0.29 & 1025 \\
4/8/1992\tablenotemark{c} & CIV & $5252\pm 139$ & $10517\pm 196$ & $3.22 \times 10^{43}$  & $3860\pm 90$& $1.41 \times 10^{43}$ & 5419 & $4.62 \times 10^{43}$ &0.49 & 3762 \\
4/8/1992\tablenotemark{c} & Ly$\alpha$ & $2391\pm 161$ & $10373\pm 275$ & $8.62 \times 10^{43}$  & $2548\pm 85 $ & $7.76 \times 10^{43}$ & 3452 & $1.44 \times 10^{44}$ &0.13 & 616 \\
\hline
0954+556 &    &   &    &   &  &  & & & &\\
\hline
12/8/1983\tablenotemark{d,i} & H$\beta$ & $2306\pm 238$ & $9000\pm 843$ & $3.35 \times 10^{43}$  & $3364\pm 356$ & $4.75 \times 10^{42}$ & 9787 & $3.82 \times 10^{43}$ &0.14 & 4173 \\
3/24/2012\tablenotemark{e,i} & H$\beta$ & $2281\pm 396$ & $8040\pm 419$ & $2.53 \times 10^{43}$  & $2061\pm 589$ & $6.64 \times 10^{42}$  & 4515 & $3.19 \times 10^{43}$ &0.09 & 1454 \\
12/8/1983\tablenotemark{d} & MgII & $2032\pm 148$ & $5507\pm 314$ & $1.61 \times 10^{43}$  & 1664 & $7.55 \times 10^{42}$ & 2856 & $2.37 \times 10^{43}$ &0.28 & 1401 \\
1/12/2003\tablenotemark{e} & MgII & $1662\pm 236$ & $8020\pm 360$ & $1.83 \times 10^{43}$  & 1532 & $7.63 \times 10^{42}$  & 3250 & $2.59 \times 10^{43}$ &0.30 & 1610 \\
3/24/2012\tablenotemark{e} & MgII & $1812\pm 130$ & $7501\pm 908$ & $2.22 \times 10^{43}$  & 1467 & $6.64 \times 10^{42}$  & 3028 & $2.89 \times 10^{43}$ &0.41 & 1466 \\
1/20/1993\tablenotemark{c} & CIV & $3790\pm 2447$ & $8776\pm 501$ & $7.82 \times 10^{43}$  & $1971\pm 59$ & $3.42 \times 10^{43}$ & 2610 & $1.12 \times 10^{44}$ &0.60 & 2964 \\
1/20/1993\tablenotemark{c} & Ly$\alpha$ & $2640\pm 369$ & $6292\pm 656$ & $1.02 \times 10^{44}$  & $1780\pm 93$ & $7.99 \times 10^{43}$ & $6292$ & $1.87 \times 10^{44}$ &0.42 & 1338\\
1/20/1993\tablenotemark{c} & OVI & $3675\pm 314$ & $10031\pm 220$ & $5.89 \times 10^{43}$  & $1100\pm 204$ & $9.18 \times 10^{42}$ & $10031$ & $5.89 \times 10^{43}$ &0.45 & 3850 \\
\hline
1803+784 &    &   &    &   &  &  & & & &\\ \hline
7/04/1986\tablenotemark{d} & H$\beta$ & $4236$ & 5415 & $4.18 \times 10^{42}$  & $3765$ & $9.19 \times 10^{42}$ & 4462 & $1.34 \times 10^{43}$ & 0.23 & 2263 \\
7/04/1986\tablenotemark{d} & MgII & $2465$ & $6380$ & $1.36 \times 10^{43}$  & 1785 & $6.89 \times 10^{42}$ & 2950 & $2.05 \times 10^{43}$ &0.47 & 2016 \\
\hline
3C 345 &    &   &    &   &  &  & & & &\\
\hline
6/11/2012\tablenotemark{e} & H$\beta$ & $2382\pm 205$ & $11123\pm 412$ & $6.83\times 10^{42}$  & $3185\pm 125$ & $4.79 \times 10^{42}$ & 4010 & $1.16 \times 10^{43}$ &0.22 & 1342 \\
5/27/2016\tablenotemark{f} & H$\beta$ & $453\pm 108$ & $15429\pm 1044$ & $1.44 \times 10^{43}$  & $2620\pm 106$ & $2.92 \times 10^{42}$  & 4508 & $1.73 \times 10^{43}$ &-0.01 & 482 \\
6/7/1992\tablenotemark{c,j} & MgII & $1454\pm 577$ & $7130\pm 524$ & $5.43 \times 10^{43}$  & $3355\pm 1541$ & $4.76 \times 10^{43}$ & 4382 & $1.02 \times 10^{44}$ &0.13 & 800 \\
8/20/1995\tablenotemark{c} & MgII & $3038\pm 23$ & $9405\pm 2422$ & $2.64 \times 10^{43}$  & $3641\pm 134$ & $4.24 \times 10^{43}$ & 4181 & $7.47 \times 10^{43}$ &0.10 & 391 \\
3/24/2009\tablenotemark{b} & MgII & $1474\pm 105$ & $4538\pm 72$ & $6.71 \times 10^{43}$  & $2737\pm 30$ & $2.96 \times 10^{43}$ & 4229 & $9.67 \times 10^{43}$ & 0.16 & 1017 \\
6/11/2012\tablenotemark{e} & MgII & $1368\pm 71$ & $5963\pm 152$ & $2.54 \times 10^{43}$  & $3330\pm 62$ & $2.17 \times 10^{43}$ & 4225 & $4.71 \times 10^{43}$ &0.11 & 610 \\
5/30/2016\tablenotemark{g} & MgII & $2290\pm 165$ & $4510\pm 206$ & $1.62 \times 10^{43}$  & $3503\pm 81$ & $2.45 \times 10^{43}$ & 4578 & $4.07 \times 10^{43}$ &0.16 & 833 \\
8/29/2016\tablenotemark{h} & MgII & $2194\pm 207$ & $6471\pm 368$ & $2.01 \times 10^{43}$  & $3045\pm 499$ & $9.48 \times 10^{42}$ & 5385 & $2.96 \times 10^{43}$ &0.30 & 1586 \\
9/17/2017\tablenotemark{h} & MgII & $2265\pm 397$ & $7026\pm 858$ & $1.57 \times 10^{43}$  & $2993\pm 290$ & $1.00 \times 10^{43}$ & 4543 & $2.57 \times 10^{43}$ &0.32 & 1480 \\
6/7/1992\tablenotemark{c,i} & CIV & $2055\pm 2713$ & $8336\pm 1401$ & $2.50 \times 10^{44}$  & $3136\pm 86 $ & $1.40 \times 10^{44}$ & 4471 & $3.90 \times 10^{44}$ &0.28 & 1414 \\
8/20/1995\tablenotemark{c} & CIV & $3029\pm 1702$ & $10079\pm 734$ & $7.82 \times 10^{43}$  & $3365\pm 92 $ & $ 1.39 \times 10^{44}$ & 3727 & $2.18 \times 10^{44}$ &0.07 & 503 \\
\end{tabular}}
\tablenotetext{a}{Cerro Tololo Inter-American Observatory 4 Meter
\citep{net95,pun12}} \tablenotetext{b}{Steward Fermi Monitoring
Program \citep{pun12}} \tablenotetext{c}{HST} \tablenotetext{d}{Hale
Telescope
\citep{law96}}\tablenotetext{e}{SDSS}
\tablenotetext{f}{Nordic Optical Telescope}
\tablenotetext{g}{Telescopio Nazionale Galileo}
\tablenotetext{h}{Copernico Telescope, 1.8 Meter}
\tablenotetext{i}{Noisy Spectrum. FWHM of BC has a large
uncertainty due to line shape and noise.}
\tablenotetext{j}{San Pedro Martir 2.2 Meter \citep{mar96}}
\end{table}
Table 2 summarizes the results of the line fits. The first two
columns are the date and line that was observed. the third column
gives the redward shift in velocity space of the VBC in the fit,
followed by the FWHM and luminosity of the VBC. The next two columns
are the FWHM and luminosity of the narrower BC. Columns (8) and (9) denote the FWHM
and luminosity of the composite broad line. These are used in the
estimators of the central black hole mass and the bolometric
luminosity. The last two columns are the measures of line asymmetry
that were introduced in Equations (2) and (3).
\subsubsection{Redward Asymmetry and the Ionization State} The MgII line is a low ionization ($\approx$15~eV)
emission line and CIV is a high ionization ($\approx$54~eV) emission line and in principle the lines might originate in different regions
of the broad line region (BLR). Fortunately, both the high and low
ionization lines of 3C279 were observed simultaneously in April 1992. The
lines and their fits are compared in Figure 2. The high ionization
VBC is shifted more towards the red than the low ionization VBC as shown in columns (3) of Table
2. In general, all of the 3C279 low ionization VBCs in Table 2, MgII
H$\beta$ and Ly$\alpha$, are shifted less than half as much as the
high ionization VBC of CIV.

\begin{figure}
\begin{center}
\includegraphics[width=75 mm, angle= 0]{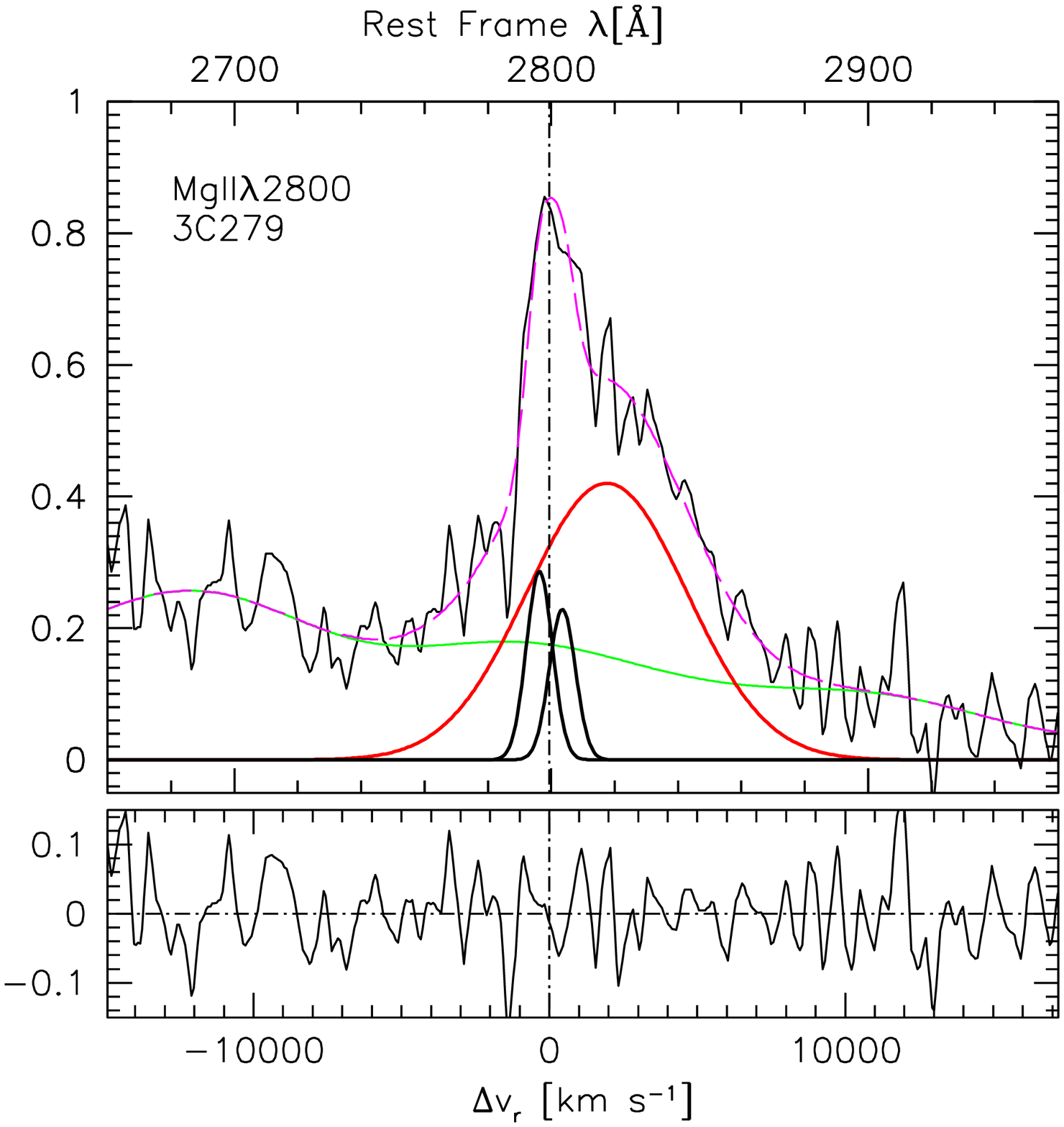}
\includegraphics[width=75 mm, angle= 0]{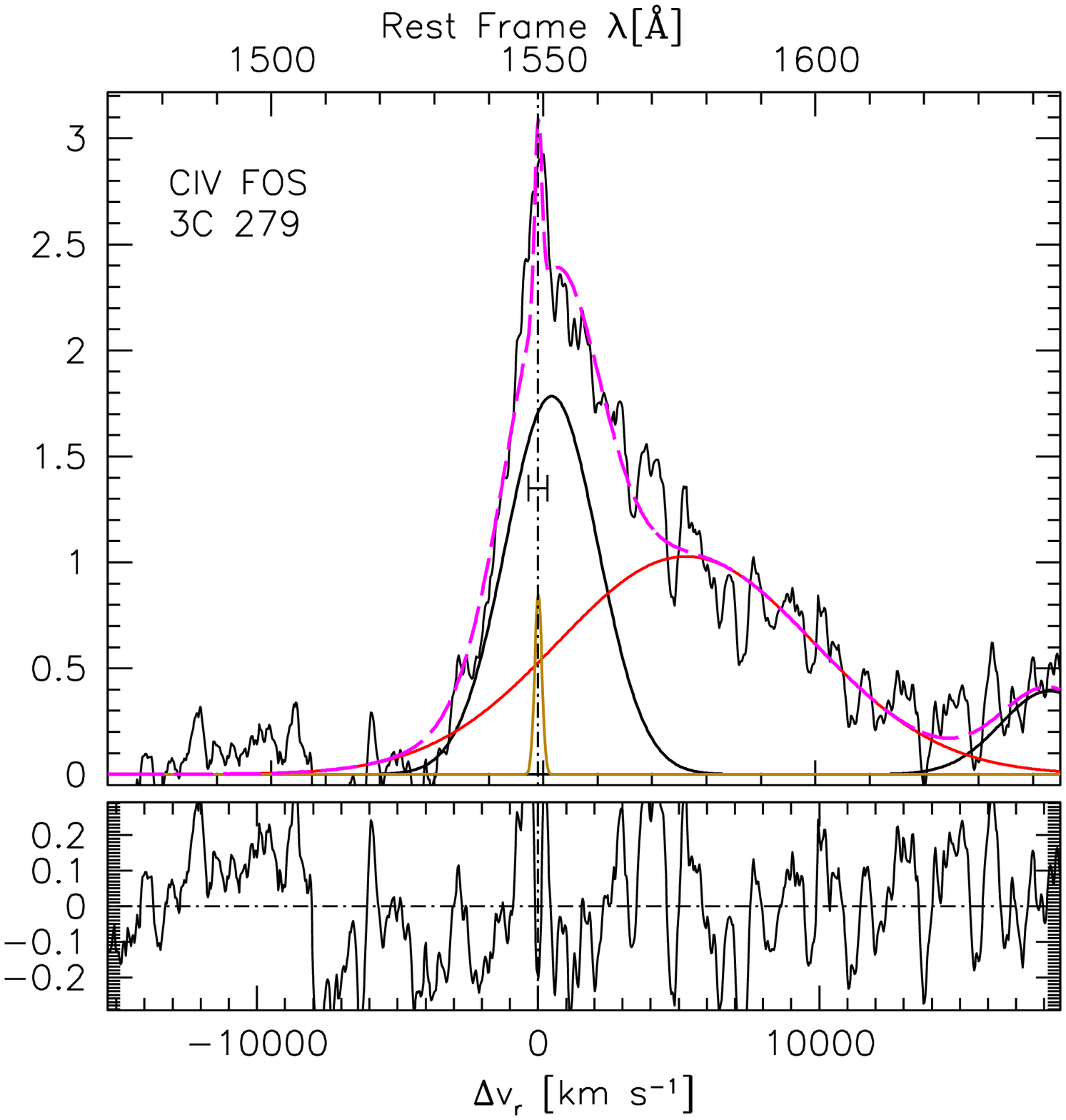}
\caption{Simultaneous observations of 3C 279 MgII (left) and CIV (right).
Notice that the VBC is the source of the large redward asymmetry.
Its center frequency is shifted towards the red. This shift is
larger for the high ionization line, CIV (see Table 2).}
\end{center}
\end{figure}

\subsubsection{Estimating the Bolometric Luminosity of the Accretion Flow}
For the blazars that are under consideration in Table 1, the
spectrum in the optical and UV bands is often dominated by the
synchrotron emission from the radio jet. Thus, the continuum cannot
be used to estimate the bolometric luminosity of the accretion flow,
$L_{\rm{bol}}$. In this paper, the notion of bolometric luminosity
does not include reprocessed radiation in the infrared from distant
molecular clouds. This would be double counting the thermal
accretion emission that is reprocessed at mid-latitudes
\citep{dav11}. Another method of estimating $L_{\rm{bol}}$ is to use
the luminosity of the BELs \citep{wan04}. This method is more
indirect. However, it is not affected by line of sight effects due
to Doppler enhancement of jet radiation that contaminates the
accretion flow continuum. Thus, we establish bolometric corrections
for the BELs based on the composite quasar spectra in
\citet{lao97,zhe97} and the line strength ratios from these HST
composites \citep{tel02}. From the luminosity of the UV lines, we
obtain bolometric corrections based on $L(MgII)$ and $L(CIV)$,
respectively
\begin{equation}
L_{\mathrm{bol}} \approx 251L(MgII),\quad
L_{\mathrm{bol}}\approx107L(CIV)\;,
\end{equation}
as in \citet{pun16} \footnote{where we have corrected the mistake in \citet{pun16} on the multiplier for MgII (it was 151 not 251 in that paper)}.
We use the composite ratio from \citet{wan04}, $L(H\beta)/L(MgII)=
16.35/25.62$ in combination with Equation (4) in order to obtain a
third $L_{\rm{bol}}$ estimator
\begin{equation}
L_{\mathrm{bol}}\approx 393 L(H\beta)\;.
\end{equation}
Since RLQs can have very strong narrow lines that are enhanced by
the jet excitation of the NL region, it is understood that in
Equations (4) and (5) that the full BC luminosity, less the NL luminosity, is to be used in
the estimate of the photo-ionizing source, $L_{\rm{bol}}$. Equations
(4) and (5) are used in conjunction with the full broad line
luminosity from Table 2 to estimate $L_{\rm{bol}}$ using various
BELs in column (2) of Table 3.
\subsubsection{Estimating the Mass of the Central Supermassive Black Hole}
This section is a preliminary discussion of virial mass estimates of the supermassive central black hole. It does not include the very important line of sight corrections that are required to reliably apply these standard formulae to blazars. This will be addressed in Section 6. The traditional estimates presented here will be large underestimates of the black hole mass.

As discussed above, for the blazars that are under consideration the optical
and UV synchrotron emission tends to mask the thermal emission from
the accretion flow. Thus, the continuum cannot be used in a virial
estimate of the central black hole mass, $M_{bh}$. Fortunately there
are estimators that are based solely on BELs, The low ionization
lines are much more reliable than the high ionization CIV line for
estimating $M_{bh}$ \citep{tra12,mar13}. There is considerable
scatter in the estimations in single epoch estimates. Thus, to
moderate dispersion in our results, we choose two estimators for
both MgII and H$\beta$ from the literature and multiple epochs. The
formula from \citet{she12} using the MgII BEL is:

\begin{eqnarray}
&&\log\left(\frac{M_{bh}}{M_{\odot}}\right) = 3.979
+0.698\log\left(\frac{L(MgII)}{10^{44} \,\rm{erg/s}}\right) +
1.382\log\left(\frac{\rm{FWHM}}{\rm{km/s}}\right)\;.
\end{eqnarray}
Alternatively, the formula of \citet{tra12} yields a different
estimate
\begin{equation}
\frac{M_{bh}}{M_{\odot}} = 6.79\times 10^{6}
\left(\frac{L(MgII)}{10^{42}
\,\rm{erg/s}}\right)^{0.5}\left(\frac{\rm{FWHM}}{1000
\,\rm{km/s}}\right)^{2} \;.
\end{equation}
The formula from \citet{she12} using the H$\beta$ BEL is
\begin{equation}
\log\left(\frac{M_{bh}}{M_{\odot}}\right) = 1.963
+0.401\log\left(\frac{L(H\beta)}{10^{44} \,\rm{erg/s}}\right) +
1.959\log\left(\frac{\rm{FWHM}}{\rm{km/s}}\right)\;.
\end{equation}
The formula of \citet{gre05} gives us a different estimate
\begin{equation}
\log\left(\frac{M_{bh}}{M_{\odot}}\right) = 1.67
+0.56\log\left(\frac{L(H\beta)}{10^{44} \,\rm{erg/s}}\right) +
2\log\left(\frac{\rm{FWHM}}{\rm{km/s}}\right)\;.
\end{equation}
Table 3 displays our estimated masses from these formulae. Table 3 justifies our motivation for using multiple lines and
multiple estimators for each line, the single epoch virial BEL
estimates have a large scatter as is evident in the entries for the
same object. By averaging all the MgII line estimates together we
should obtain a more robust MgII estimate. Similarly, we do the same
to improve the H$\beta$ line estimate of $M_{bh}$. Then we average
these together in order to get our most reliable estimate.
\par  Note that there is a
large difference in the MgII estimates and the H$\beta$ estimate
which is disconcerting. There is a factor of $\sim 5$ difference in the estimates for 3C 279. Furthermore, the average is driven by a noisy H$\beta$ measurement that needs to be reconfirmed with a deep spectrum during a state of low synchrotron emission. More importantly, in Section 6, we consider the affects on the
$M_{bh}$ estimate for the near polar line of sight in these extreme blazars.
\subsection{The Broad Emission Lines of 0954+556} Table 1 indicates
that this \citet{pea81} complete sample source is a powerful gamma
ray source that has a very high median optical polarization. There
are no VLBI measurements of component motion, so it is unknown if
this is a superluminal source. There are no MgII measurements that
are contemporaneous with a high ionization line observations. The
redshift of 0.899 renders the MgII line too red to have been a natural
target for HST and, unfortunately, was not observed. Furthermore,
there was no simultaneous ground based observation. However, MgII
has been very stable in the three observations spread out over 30
years as indicated in Table 2. We chose to refit the MgII line from
1983 in \citet{law96} since it is such an unusual profile and
requires special attention in this study. Our broad component fit in
Table 2 was narrower and slightly stronger than the values in
\citet{law96}, 6677 km/s and $1.53 \times 10^{42}$ ergs/s,
respectively. Note that the three fits to the MgII line indicates
that the ``broad component" has a FWHM $\approx$ 1600 km/s. It is
formally a narrow line, but as we discussed in Section 2.1 this is most likely a BC that appears narrow due to the polar, blazar line of sight. This is an important distinction since this interpretation drastically decreases the FWHM of the total broad component that drives the virial mass estimates in Table 3.
\par The best signal to noise of all the observations of MgII was the 2012
SDSS observation. Thus, it is meaningful to compare it to the CIV
observations. Even though the observations are not contemporaneous,
as in 3C 279, the high ionization CIV is shifted much farther
towards the red than all three of the MgII fits in Table 2 and
Figure 3. The estimated $M_{bh} \gtrsim 10^{9}M_{\odot}$ in Table 3,
similar to 3C 279. However, as for 3C 279, there is a noisy H$\beta$ observation from 1983 that drives this number high. If we ignore this measurement, the average mass estimate is $M_{bh} \lesssim 5\times 10^{8}M_{\odot}$.
\begin{figure}
\begin{center}
\includegraphics[width=75 mm, angle= 0]{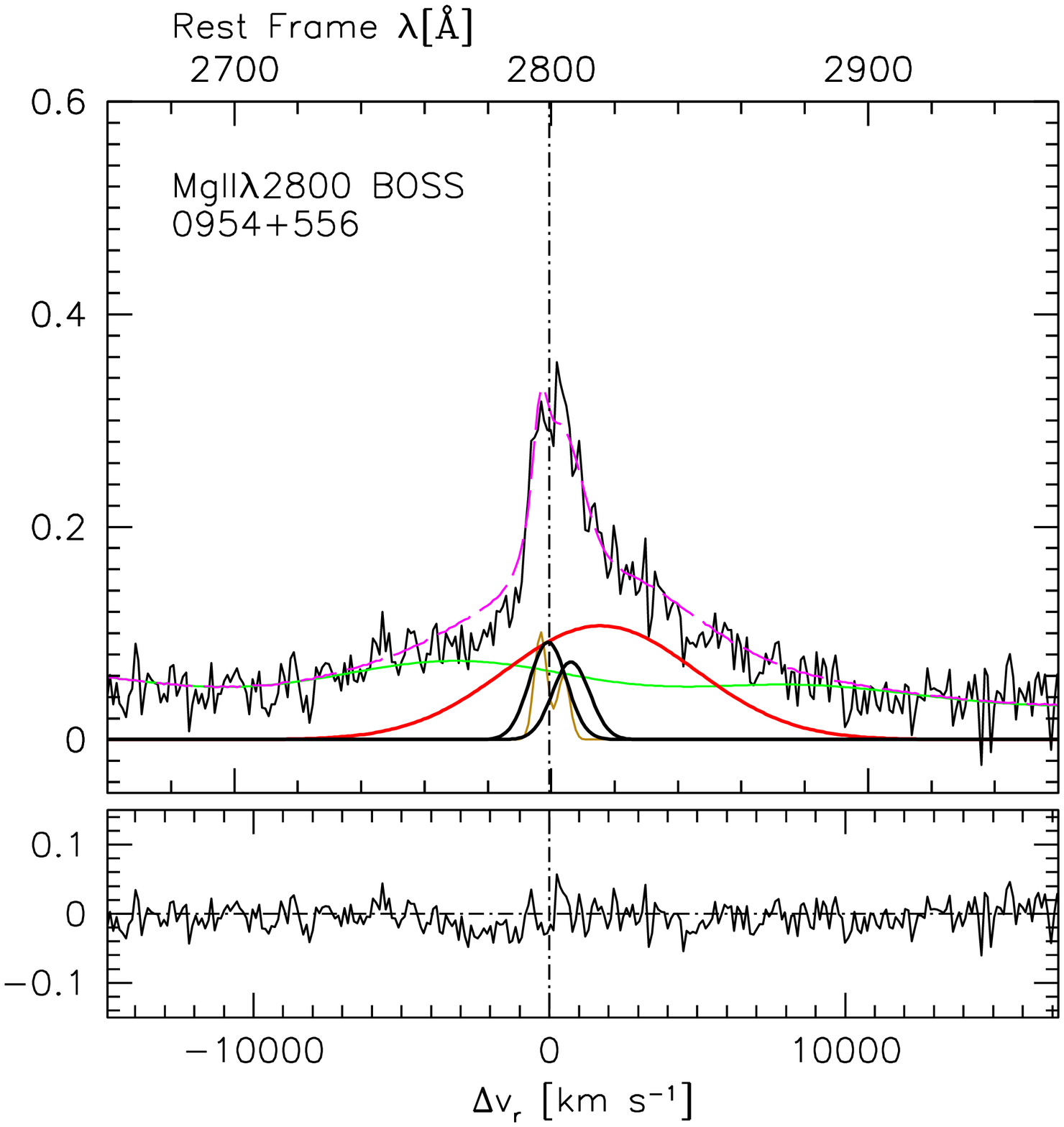}
\includegraphics[width=75 mm, angle= 0]{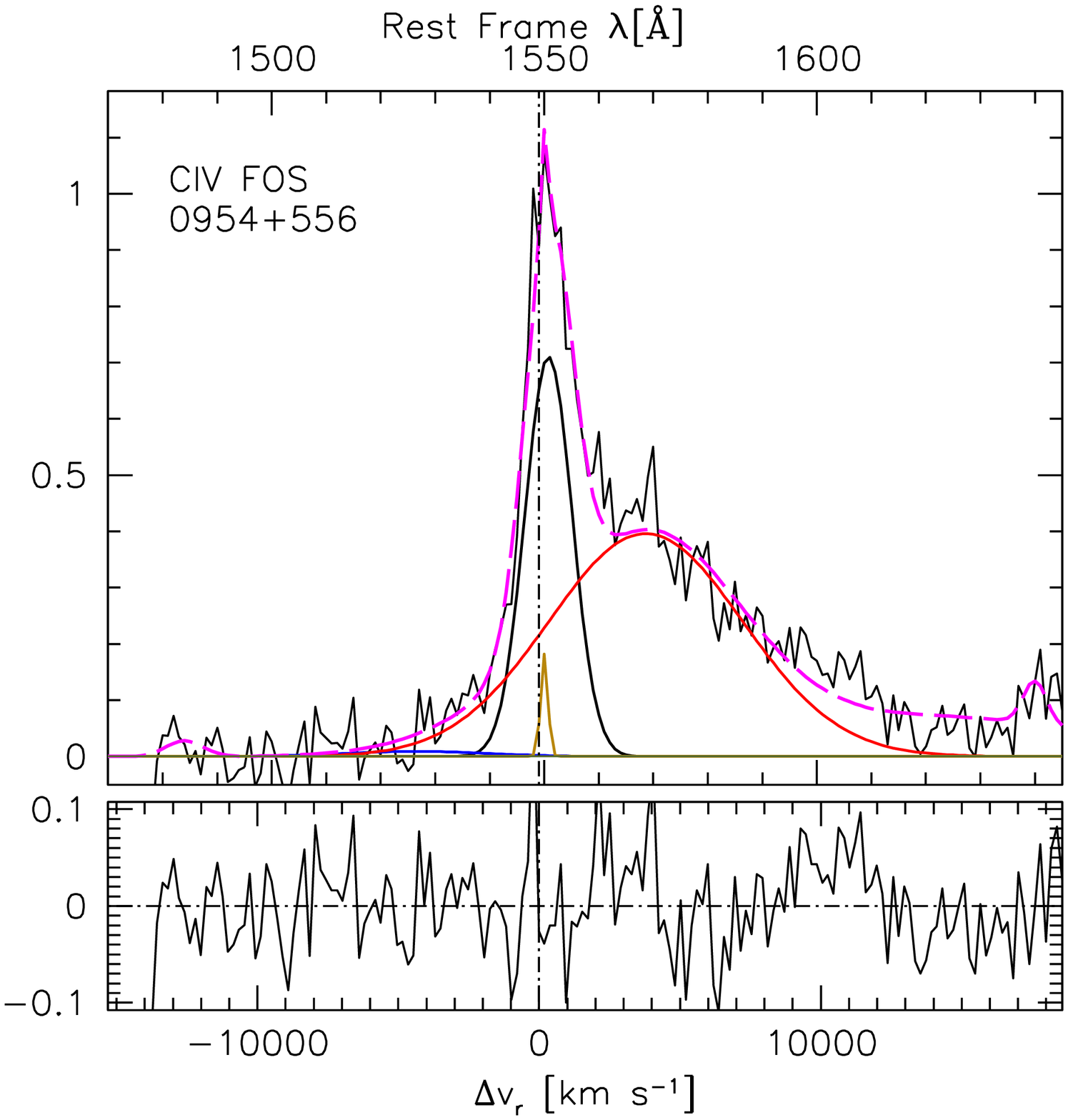}
\caption{Non-simultaneous observations of MgII (left) on 2/24/2012
and CIV (right) on 1/20/1993 for the blazar 0954+556. Notice that
the VBC is the source of the large redward asymmetry. Its center
frequency is shifted towards the red. This shift is larger for the
high ionization line, CIV (see Table 2).}
\end{center}
\end{figure}
\subsection{The Broad Emission Lines of 1803+784}
There is only one high signal to noise observation of 1803+784, a
very deep Hale telescope observation when the blazar was in a low
optical state. Unfortunately there was not a CIV line observation
with HST. However, this is the most striking example of a redward
asymmetric line MgII line in the \citet{pea81} complete sample.
Thus, it is worthy of a detailed investigation.
\begin{figure}
\begin{center}
\includegraphics[width=75 mm, angle= 0]{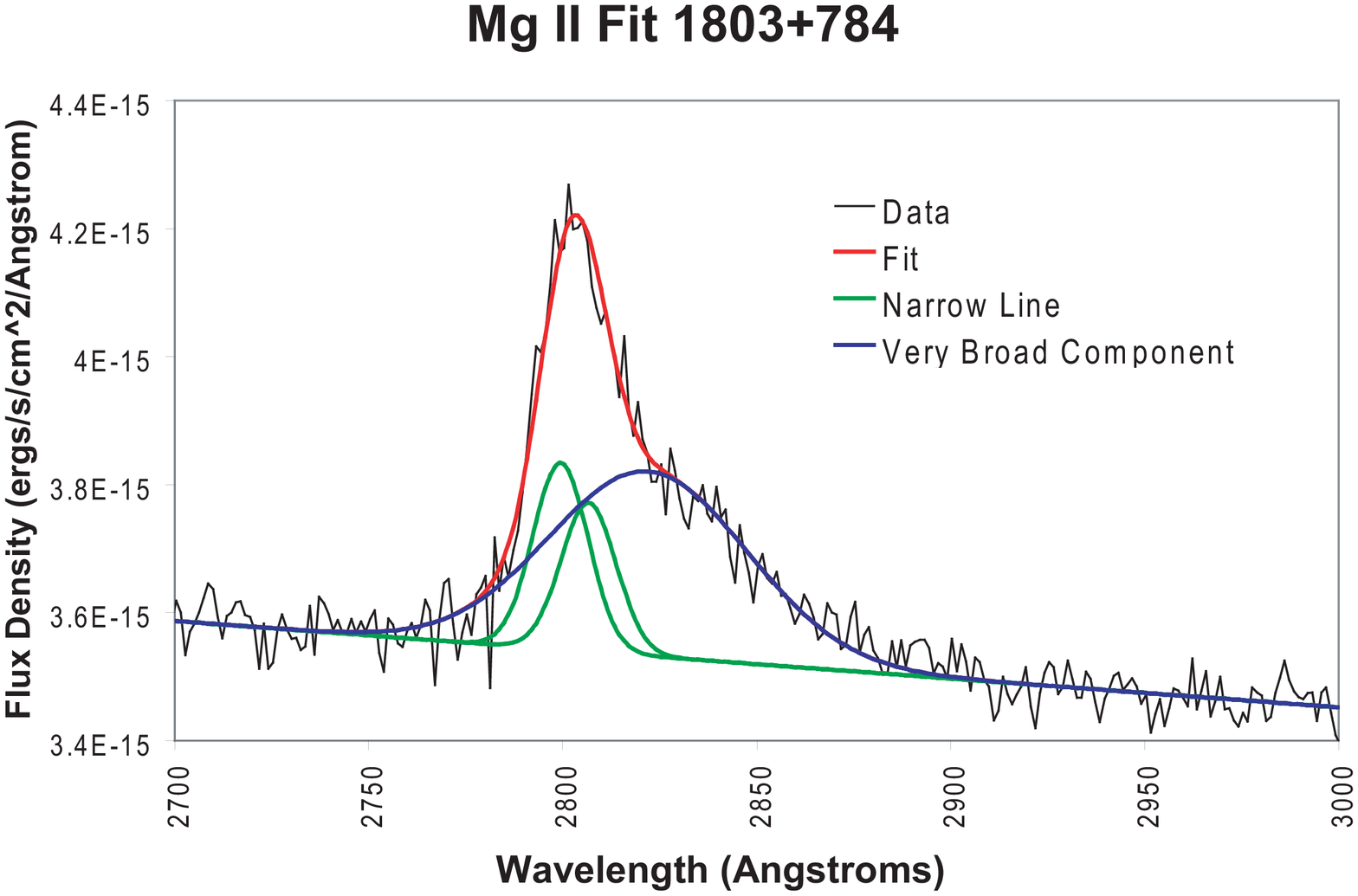}
\includegraphics[width=75 mm, angle= 0]{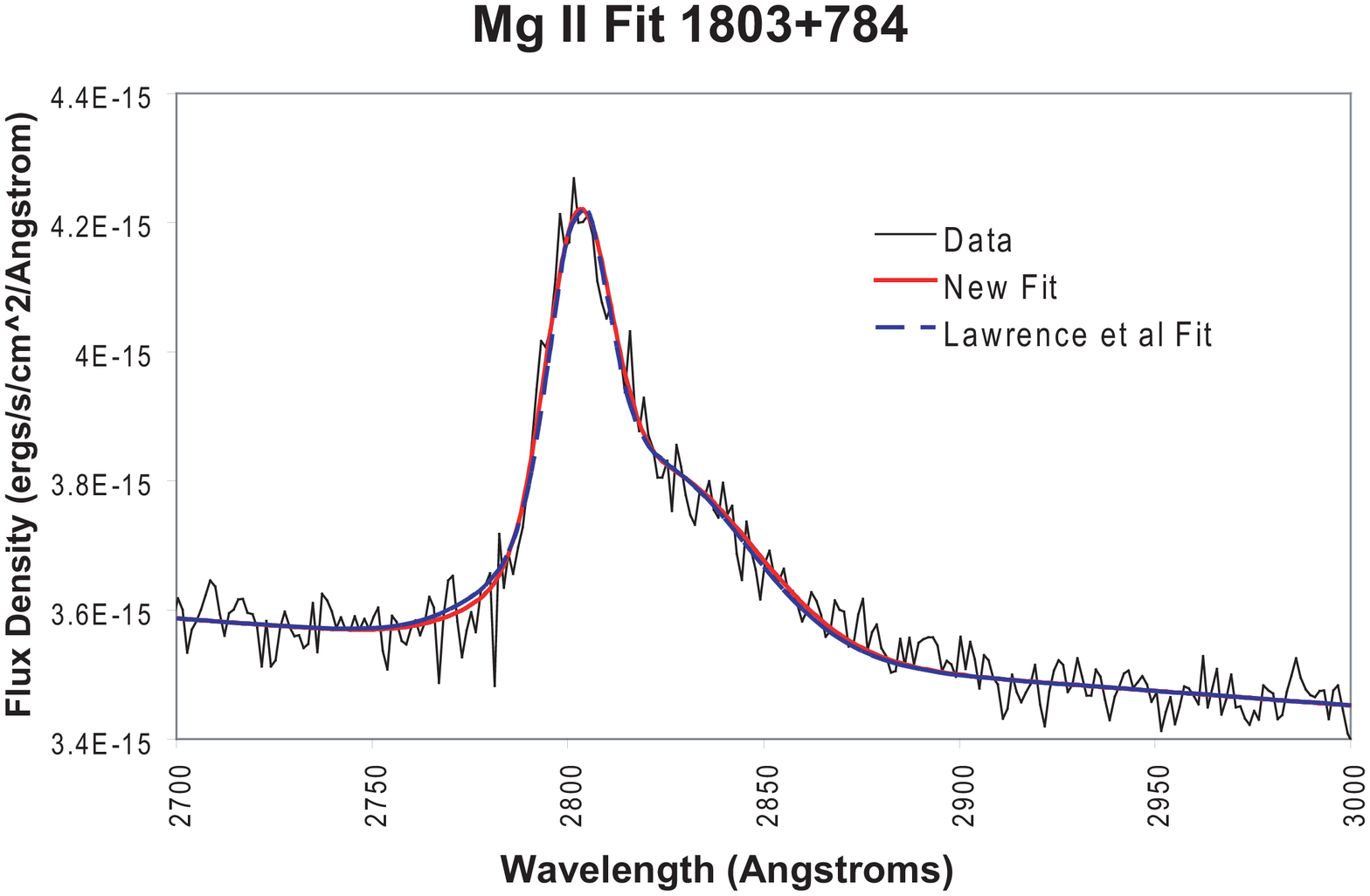}
\caption{The fit to the MgII line of 1803+784 from our analysis is
illustrated in the left hand frame. This is compared to the fit of
\citet{law96} in the right hand frame. The doublet nature of MgII
utilized in this paper made a significant difference in the width of
the narrow BC and a modest difference to the fit to the blue wing of the emission
line.}
\end{center}
\end{figure}
\par As with the other \citet{pea81} source, 0954+556, we refit the line (since the line shape is so unusual) instead
of using the \citet{law96} fit. The main difference in our fit is
that we consider MgII as a doublet. Our fit is in
the left hand frame of Figure 4. The doublet nature of MgII changes
parameters of the ``broad component". In \citet{law96} the
BC had a FWHM = 2450~km~s$^{-1}$, in our doublet fit, the FWHM = 1795~km~s$^{-1}$. We compare our fit to the \citet{law96} fit in the
right hand frame of Figure 4. There is a small difference is in the blue
wing of the MgII line..
\begin{figure}
\begin{center}
\includegraphics[width=70 mm, angle= 0]{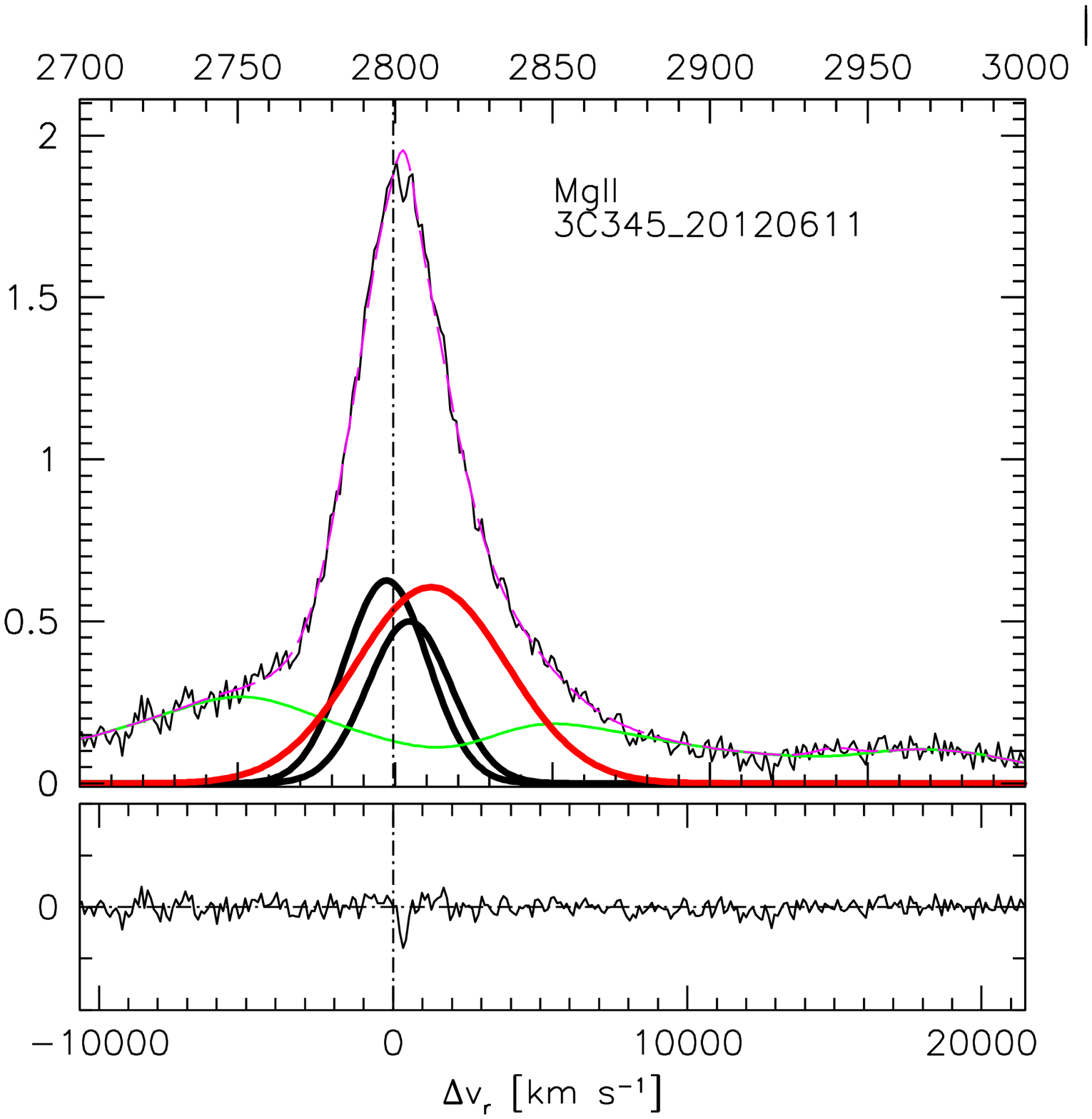}
\includegraphics[width=70 mm, angle= 0]{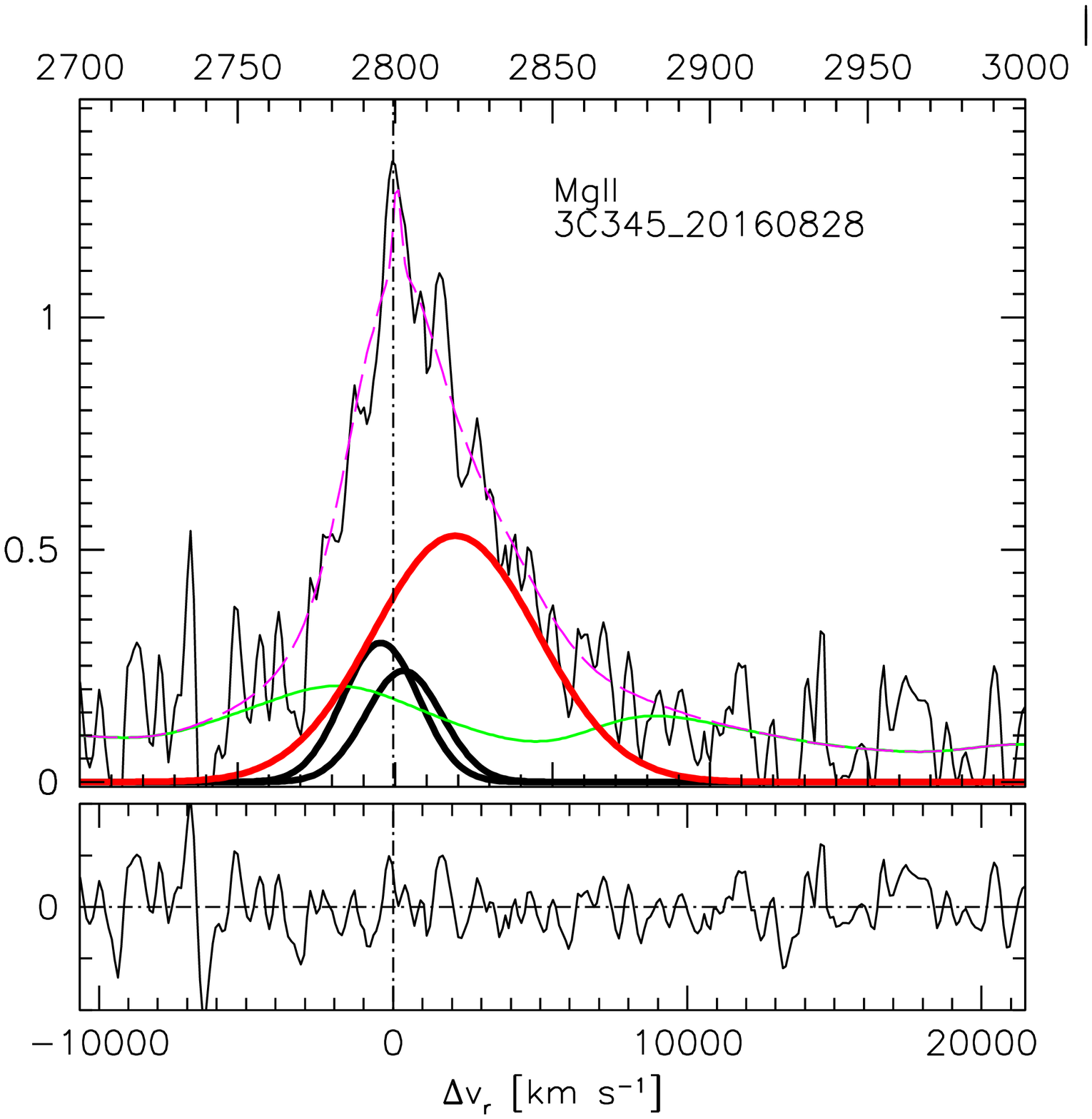}
\caption{The MgII BEL in 3C 345 has recently become more redward
asymmetric asymmetric. The spectrum from 6/11/2012 (left hand frame)
has very high signal to  noise and shows a mildly RA
profile. More recently, the BEL has become weaker. On 8/28/2016 (the
right hand frame), the BEL shows a very pronounced redward
asymmetry. Notice that the VBC strength was steady during the three
months, but the BC luminosity decreased significantly.}
\end{center}
\end{figure}
\par There is no CIV line for comparison of the redshifts of the VBC
for the low and high ionization lines. However, there is an H$\beta$
line that appears more symmetric than the MgII line. Thus, we
consider the fit in \citet{law96} reasonable for our purposes in Tables 2 and 3. One reason that the H$\beta$ line is so important to this analysis is that the MgII line for 1803+784 (as well as 3C 279 and 0954+556) is very asymmetric and the notion of the FWHM is not well quantified in the context of the virial mass estimators. This creates a poorly understood uncertainty with the virial mass estimates computed from highly asymmetric profiles.
\subsection{The Broad Emission Lines of 3C 345} The blazar 3C 345
appears in both the \citet{pea81} complete sample and the 3CR
complete sample. It was included in this study because it has
changed from having a very mildly asymmetric MgII BEL for two
decades to a very asymmetric MgII BEL in the summer of 2016
\citep{ber18}. The emergence of the RA profile is
illustrated in Figure 5. Note that the \citet{law96} fit does not
appear in Table 2 since the quality (spectral resolution and signal
to noise) is rather poor (compared to the other spectra) and not
conducive to accurate fitting. We have also obtained numerous low
signal to noise spectra that cannot be fit convincingly and were not
included in Table 2. The HST spectra were of lower quality and
required smoothing, but these are unique opportunities to
simultaneously observe CIV and MgII.

\begin{table}[htbp]
%\begin{center}
\caption{Derived Properties of Broad Emission Line Fits}
\tiny\begin{tabular}{cccccc} \tableline\rule{0mm}{3mm}
Date &  Line & $L_{\rm{bol}}$ ($10^{45}$ergs/s) & $M_{bh}$ ($10^{8}M_{\odot})$\tablenotemark{a} &  $R_{\rm{Edd}}$ \tablenotemark{b} \\
(1) &  (2)  &  (3) &  (4)  & (5)    \\
\tableline \rule{0mm}{3mm}\\
\hline
\multicolumn{5}{c}{3C279}  \\
\hline
4/8/1994\tablenotemark{c} & H$\beta$ & 3.88 & 22.4/11.6  & 3.09\%  \\
Average Mass & H$\beta$   & ...  &  17.0  & ...  \\
4/9/1992 & MgII & 3.63 & 2.89/5.58 & 2.89\%  \\
4/29/2009 & MgII & 2.37 & 1.59/2.91 & 1.89\%   \\
4/11/2010 & MgII & 2.01 & 1.46/2.80 & 1.60\%   \\
Average Mass & MgII   & ...  &  2.87  &  ... \\
Average Mass & H$\beta$ and MgII   & ...  &  9.94  &  ...  \\
4/8/1992 & CIV & 4.95 & ... & 3.95\%   \\
\hline
\multicolumn{5}{c}{0954+556}   \\
\hline
12/8/1983\tablenotemark{c} & H$\beta$ & 15.0 & 41.0/28.1 & 10.11\%   \\
3/24/2012 & H$\beta$ & 12.6 & 8.39/5.46 & 8.45\%   \\
Average Mass & H$\beta$   & ...  &  20.7  & ...  \\
12/8/1983 & MgII & 5.94 & 2.08/2.70 & 4.00\%   \\
1/12/2003 & MgII & 6.50 & 2.65/3.65 & 4.37\%  \\
3/24/2012 & MgII & 7.24 & 2.58/3.34 & 4.88\%   \\
Average Mass & MgII   & ...  &  2.83  &  ... \\
Average Mass & H$\beta$ and MgII   & ...  &  10.9  &  ...  \\
1/20/1993 & CIV & 12.0 & ... & 8.09\%  \\
\hline
\multicolumn{5}{c}{1803+784}  \\
\hline
7/04/1986 & H$\beta$ & 5.27 & 5.79/3.02 & 12.43\%   \\
Average Mass & H$\beta$   & ...  &  4.40  & ...  \\
7/04/1986 & MgII & 5.15 & 1.97/2.68 & 12.14\%   \\
Average Mass & MgII   & ...  &  2.32  &  ... \\
Average Mass & H$\beta$ and MgII   & ...  &  3.36  &  ...  \\
\hline
\multicolumn{5}{c}{3C 345}  \\
\hline
6/11/2012 & H$\beta$ & 4.56 & 4.43/2.25 & 5.66\%  \\
5/27/2016 & H$\beta$ & 6.80 & 6.54/3.56 & 8.43\%   \\
Average Mass & H$\beta$   & ...  &  4.20  & ...  \\
6/7/1992 & MgII & 25.6 & 10.4/13.6 & 31.7\%  \\
8/20/1995 & MgII & 18.8 & 7.80/10.2 & 23.2\%   \\
3/24/2009 & MgII & 24.3 & 9.78/12.3 & 30.1\%  \\
6/11/2012 & MgII & 11.8 & 5.77/8.32 & 14.6\%  \\
5/30/2016 & MgII & 10.2 & 5.83/9.08 & 12.7\%  \\
8/29/2016 & MgII & 7.43 & 5.84/10.7 & 9.22\%  \\
9/17/2017 & MgII & 6.45 & 4.18/7.10 & 8.00\%   \\
Average Mass & MgII   & ...  &  8.61  &  ... \\
Average Mass & H$\beta$ and MgII   & ...  &  6.40  &  ...  \\
6/7/1992 & CIV & 41.7 & ... & 51.7\%   \\
8/20/1995 & CIV & 23.3 & ...& 27.4\%  \\ \hline
\end{tabular}
 \tablenotetext{a}{\tiny First estimate from \citet{she12}, the second
estimate for MgII  and H$\beta$  are from \citet{tra12} and
\citet{gre05}, respectively. The mass estimates are not adjusted for the blazar line of sight and are likely gross underestimates (See Section 6).}
\tablenotetext{b}{\tiny The Eddington rate is calculated using the mean of the ``average mass" from both H$\beta$ and MgII. $R_{\rm{Edd}}=L_{\rm{bol}}/L_{\rm{Edd}}$, $L_{\rm{Edd}} = 1.26 \times 10^{38} M_{bh}/M_{\odot}$ ergs/sec. These are likely gross overestimates due to the fact that the mass estimate did not include blazar line of sight corrections (see Section 6).}
\tablenotetext{c}{\tiny These measurements are highly uncertain due to noisy spectra. The ``out of family" mass estimates might occur for this reason. The deductions of this analysis do not depend on these values.}
\end{table}
One of the topics of interest is the comparison of the redward
asymmetry of high ionization lines with the low ionization lines. To
this end, we processed the rather noisy HST data. For example, the
1992 MgII data is based on only 3 minutes of observation and 4
minutes for CIV. The 1992 observation was previously discussed in
\citet{wil95} and they found a modest asymmetry of CIV based on
Equation (1), $A=0.22$. They made no attempt to compute the asymmetry
of MgII. We proceeded to fit the HST spectra in 1992 and 1995 after
some smoothing. Unlike the BELs in 3C 279 and 0954+556 (with very
asymmetric BELs), Table 2 does not indicate any significant
difference in the redward shift of the VBC of MgII compared to the
VBC of CIV, but the uncertainties are comparatively large. CIV does
seem to have slightly larger AI and C(1/4) in Table 2 calculated
from Equations (2) and (3), respectively. The observations from 1992
and 1995 represent more symmetric line shapes than we found in 2016
- 2017 and the MgII line luminosity is much larger in 1992 and 1995.

\section{The Jet Power}
The RA quasars are distinguished by their powerful
relativistic radio jets. This is clearly an important aspect of
their peculiar BELs. The most basic things to look at are the line
of sight and the jet power. The latter is the topic of this section.

The are two common methods of estimating the jet power, $Q$, of
quasars. One is based on the low frequency (151 MHz) flux from the
radio lobes on 100 kpc scales, \citet{wil99}, and the other is based
on models of the broadband Doppler boosted synchrotron and inverse
Compton radiation spectra associated with the relativistic parsec
scale jet \citep{ghi10}. Each method has its own advantages. Using
the parsec scale jet, the $Q$ estimate can be approximately
contemporaneous with the observed BELs and this is ideal for our
physical understanding. However, there is a major drawback due to
small line of sight effects and bulk Lorentz factors of 10 -50
\citep{lis16}. The Doppler factor, $\delta$, is given in terms of
$\Gamma$, the Lorentz factor of the outflow; $\beta$, the three
velocity of the outflow and the angle of propagation to the line of
sight, $\theta$; $\delta=1/[\Gamma(1-\beta\cos{\theta})]$
\citep{lin85}. The Doppler factors that are estimated are highly
dependent on the line of sight and most blazar jets are not imaged
as linear structures on parsec scales, but seem to typically swerve
relative to any preferred linear trajectory. Thus, we know that the
small angle of the line of sight to the direction of the observed
structures is variable and the Doppler factor must be changing
significantly for very small lines of sight and ultra-relativistic
motion \citep{lin85}. Therefore an estimate of the Doppler factor
tends to vary significantly from knot to knot in the parsec scale
jet \citep{lis16,kel04}. The observed flux density for an unresolved
ejected component depends on $\delta$ to the fourth power
\citep{lin85}. Consequently, estimates of the jet power can be in
error by two orders of magnitude or more \citep{pun05}. Thus, the 151 MHz method is
generally considered more reliable since it does not involve large
uncertainties due to Doppler beaming \citep{wil99}. A disadvantage
is that it involves long term time averages, $\overline{Q}$, that do
not necessarily reflect the current state of quasar activity.

\par Consequently, in this study we only consider the case $Q \equiv \overline{Q}$. The spectral luminosity at 151 MHz per steradian, $L_{151}$, provides a surrogate for the luminosity of the radio lobes in the method \citep{wil99}.  The spectral index, $\alpha$, of a powerlaw the approximates the radio lobe spectrum is defined in
terms of the spectral luminosity by $L_{\nu} \sim \nu^{-\alpha}$. The estimate assumes a low frequency cut off at 10
MHz to the spectrum of the radio lobes, the jet axis is $60^{\circ}$ to the line of sight, there is no
protonic contribution and 100\% filling factor. The plasma is near minimum energy and a quantity, $\mathcal{F}$,
is introduced to account for deviations of actual radio lobes from these assumptions
as well as energy lost expanding the lobe into the external medium,
back flow from the head of the lobe and kinetic turbulence. $\overline{Q}$ as
a function of $\mathcal{F}$ and $L_{151}$ is plotted in Figure 7 of \citet{wil99},
\begin{equation}
\overline{Q} \approx 3.8\times10^{45} \mathcal{F} L_{151}^{6/7} \rm{ergs/s}\;,
\end{equation}
The quantity $\mathcal{F}$ was estimated to be in the range of $10<\mathcal{F}<20$ for most FRII radio sources
\citep{blu00}. We take $\mathcal{F}=10$ for the lower bound to our estimates of $\overline{Q}$ and $\mathcal{F}=20$ for the upper bound. This
is the source of the error in the estimates in Table 4.

\par We note that method is derived based on classical relaxed double radio sources.
Thus we do not consider the formula relevant for sources with a
linear size of less than 20 kpc which are constrained by the ambient
pressure of the host galaxy. Furthermore, for blazars a classical
relaxed double morphology is never the case. There is a very strong,
dominant Doppler boosted radio core and the kpc scale jet tends to
be Doppler boosted on the approaching side \citep{pun95}. Thus, in
order to make contact with this formula, contributions from Doppler
boosted jets or radio cores are removed. This requires deep radio
images with high dynamic range (the radio core is bright) as discussed in the Appendix B for individual sources.

\section{Distinguishing Characteristics of the RA Blazars}
In this section we look for empirical characteristics of
the 3C 279, 0954+556 and 1803+784 that distinguish them from other
similar blazars. This is crucial for understanding the physics that
is responsible for the asymmetric BEL shapes. In this regard, we
need to compare the blazars to complete samples of similar blazars
and quantify what is meant by ``similar". Based on Table 1, these
are extreme objects even for blazars. Extreme in the standard notion
of blazar means that what we see is a manifestation of an
ultra-relativistic jet combined with a line of sight that is almost
parallel to the direction of jet motion. Thus, we require a parent
sample of objects with
\begin{enumerate}
\item broad emission lines
\item radio jet
\item evidence of an optically thick (due to synchrotron self-absorption, SSA, $\alpha<0.5$) compact radio core
\item evidence of a line of sight that is nearly pole-on
\end{enumerate}
We choose two criteria to indicate point 4, if either is satisfied
then the object is considered to have a line of sight that is similar to 3C
279, 0954+556 and 1803+784. Either the object has high optical
polarization ($>3\%$) or the object displays superluminal motion
above some threshold value. The second criteria is necessary because
high redshift, core dominated quasars rarely show high optical
polarization. Since the UV is red-shifted into the optical band, one
is looking much farther down the synchrotron tail than in low
redshift sources. The optically thin synchrotron tail is much
steeper than the accretion disk emission spectrum per unit
frequency. Thus, accretion disk emission is more pronounced in the
observed optical band of high redshift quasars. The superluminal criterion is an apparent effect due to relativistic motion. The apparent velocity, $v_{\rm{app}}$,
can be expressed in terms of the lines of sight, $\theta$, and the
physical jet speed normalized to the speed of light, $\beta$, as
\begin{equation}
v_{\rm{app}}/c = \beta \sin{\theta}/(1-\beta\cos{\theta}).
\end{equation}
For polar lines of sight this can exceed one. In order to find a $v_{\rm{app}}$ that represents strong blazar activity we consult the 19 years of 15 GHz monitoring of the MOJAVE 1.5 Jy survey with VLBA \citep{lis16}. The median fastest $v_{\rm{app}}$ is approximately 6c for these sources. Thus, we consider this a lower limit for an extreme blazar in criteria 4, above. Our sample might be slightly over-inclusive, there could be a few slightly off angle modest blazars that are mixed in with our extreme blazar sample.

\par In summary, our complete extreme blazar samples are subsamples
of the complete high frequency selected \citep{pea81} sample of
radio sources and the low frequency selected 3CR sample of radio
sources \citep{spi85}. The conditions to define the subsamples are:
\begin{enumerate}
\item they are \citet{pea81} or 3CR sources and they have,
\item broad emission lines
\item radio jet
\item evidence of an optically thick compact radio core
\item either an optical polarization $>3\%$ or $v_{\rm{app}} >6c$.
\end{enumerate}
This is the definition of the extreme blazars in Table 1. Based on
Table 1, the highly RA blazars, 3C 279, 0954+556 and
1803+784, are likely much more extreme blazars than indicated by the
lower cutoffs defined in criteria 5, above. However, in choosing a
complete sample it is much more important to chose a sample that is
overly inclusive than one that is too selective. The blazars are so
extreme that we need a looser criteria for the complete samples
otherwise there would be very few sources.

\begin{figure}[h]
\begin{center}
\includegraphics[width=150 mm, angle= 0]{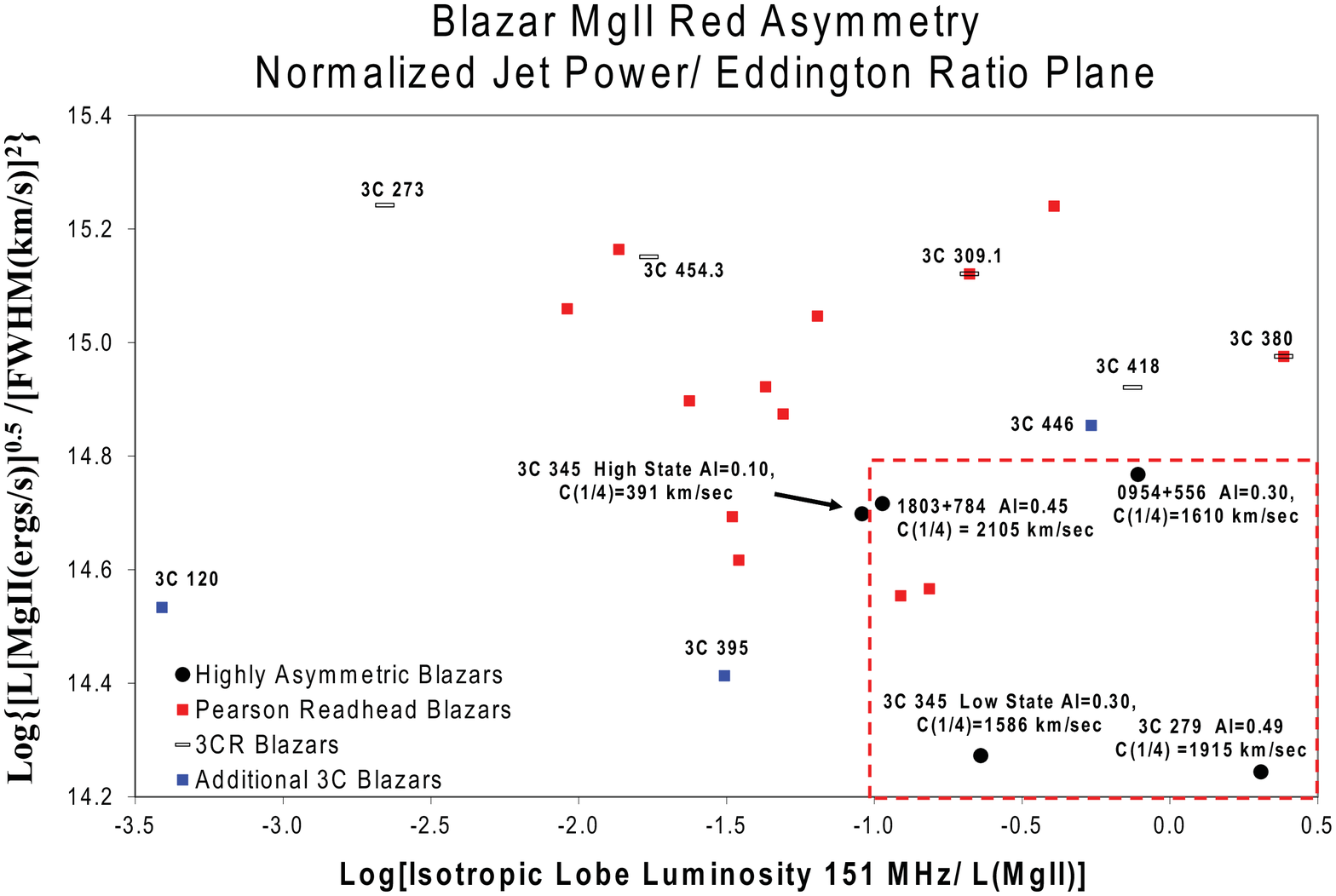}
\caption{The scatter plot is empirical in nature. The redward
asymmetric blazars 3C 279, 0954+556 and 1803+784 are compared with
the with two complete samples, the extreme blazars in the high
frequency selected Pearson-Readhead sample and the low frequency
selected 3CR sample. The horizontal axis is the large scale
isotropic low frequency luminosity, $\nu_{e}L_{\nu}(\nu_{e}=151
\rm{MHz})$ normalized to the MgII BEL luminosity. The vertical axis
is $[L(\rm{MgII})]^{0.5}/\rm{FWHM}^{2}$ which is indicative of the
Eddington ratio based entirely on the MgII line.}
\end{center}
\end{figure}
\begin{figure}[ht]
\begin{center}
\includegraphics[width=85mm, angle= 0]{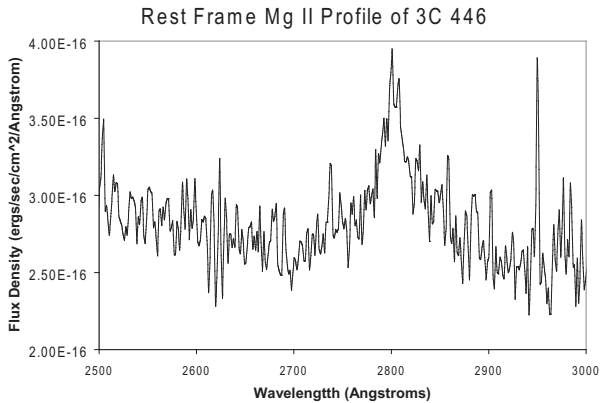}
\caption{The symmetric MgII BEL in the blazar 3C 446. Even though
$\nu_{e}L_{\nu}(\nu_{e}=151 \rm{MHz})/L{\rm{MgII}}$ is large for the isotropic lobe emission, there
is no dramatic redward asymmetry in MgII.}
\end{center}
\end{figure}

\begin{table}
\caption{Jet Power of Broadline Blazars}
\footnotesize
\begin{tabular}{cccccccc} \tableline \rule{0mm}{3mm}
 Quasar & z & Extended Isotropic &$\alpha$ & Isotropic  & $\overline{Q}$  & Reference \\
        &       & Flux Density (mJy) && Flux Density  & $10^{45}$ ergs/s & for  \\
        &       & /Frequency (GHz) && 151 MHz (Jy)  &    & Radio Image   \\
\tableline \rule{0mm}{1mm} \\ \hline
\multicolumn{7}{c}{ Redward    Asymmetric    Blazars } \\
\hline \rule{0mm}{1mm}
3C 279 & 0.536 & 1486/1.67& 1.0 & 16.4 &  $ 9.05 \pm 3.01$ & Appendix B.1  \\
0954+556 & 0.899& 1339/0.408 & 1.0 & 3.62 & $7.25 \pm 2.42$ & Appendix B.2 \\
1803+784 & 0.684 & 67/1.55 & 1.0 & 0.69 &  $ 0.98 \pm 0.33 $ & Appendix B.3  \\
3C 345 & 0.593 & 295/1.55 & 1.0 & 3.03 &  $2.59 \pm 0.86 $ & Appendix B.4 \\
\hline
\multicolumn{7}{c}{ Pearson \& Readhead  (1981)    Sample     Extreme   Blazars } \\
\hline
0016+731 &1.78& 7.8/1.40 & 1.0& 0.07 &  $1.08 \pm 0.36$ & \citet{kha10} \\
0133+476 &0.86& 8.7/1.40 & 1.0& 0.08 &  $0.25\pm 0.08$ & \citet{kha10}  \\
0212+735 & 2.37 & 1.7/1.40 & 1.0 & 0.02 &  $ 0.55 \pm 0.18 $ & \citet{kha10}  \\
0804+499 &1.43& 5.3/1.55& 1.0 & 0.05 &  $ 0.49 \pm 0.16 $ &\citet{mur93} \\
0836+710 & 2.18 & 730/0.408 & 0.96 & 1.90 &  $ 2.73 \pm 0.91 $ & \citet{hum92}  \\
0850+581 & 1.32 & 244/1.40 & 0.8 & 1.45 &  $  7.49 \pm 2.50 $ & \citet{gar91} \\
0859+470 & 1.46 & 222/0.408 & 0.9 & 0.54 &  $  4.00 \pm 1.33  $ & \citet{rei95}  \\
0945+408 & 1.25 & 95/1.40 & 1.0 & 0.88 &  $ 4.35 \pm 1.45  $ & \citet{kha10}  \\
1458+718\tablenotemark{a}  & 0.91 & 344/4.80 & 0.8 & 5.48 &  $  10.47 \pm 3.49   $ & \citet{van84} \\
1633+382 & 1.82 & 32/1.55 & 0.80& 0.21 &  $  2.79 \pm 0.93    $ & \citet{mur93} \\
1637+574 & 0.75 & 150/1.55 & 0.85 & 1.09 &  $  1.76 \pm 0.59    $ & \citet{mur93} \\
1739+522 & 1.38 & 67/0.408 & 1.0 & 0.18 & $  1.38 \pm 0.46    $ & \citet{rei95} \\
1828+487\tablenotemark{b}  & 0.69 & 22310/0.408 & 0.9 & 54.58 &  $  42.66 \pm 14.22 $ & \citet{rei95}\\
\hline
 \multicolumn{7}{c}{ 3CR    Extreme   Blazars } \\
\hline
3C 273  & 0.16 &2328/328& 0.93 & 4.79 &  $ 0.28 \pm 0.09    $ & \citet{pun17}  \\
3C 309.1\tablenotemark{a}  & 0.91 & 344/4.8& 0.8 & 5.48 &  $  10.47 \pm 3.49   $ & \citet{van84} \\
3C 380\tablenotemark{b}   & 0.69 & 22310/0.408 & 0.9 & 54.58 &  $  42.66 \pm 14.22 $ & \citet{rei95}\\
3C 418  & 2.69 & 2466/0.33& 1.25 & 6.55 &  $   45.95 \pm 15.32  $ & \citet{pun18}\\
3C 454.3  & 0.86 & 62/1.40 & 1.0 & 0.59 &  $ 1.36 \pm 0.45  $ & Appendix B.5 \\
\hline \multicolumn{7}{c}{ Additional   3C    Extreme    Blazars } \\
\hline
3C 120  & 0.03 & 300/1.40& 1.0 & 2.78 &  $   0.009 \pm 0.003  $ & \cite{wal87} \\
3C 395  & 0.64 &50/1.60 & 1.0 & 0.53 &  $   0.67 \pm 0.32  $ & \citet{sai90}\\
3C 446  & 1.40 & 340/1.60 & 1.0 & 3.60 &  $   18.63 \pm 6.21 $ & \citet{fej92} \\
\tableline \rule{0mm}{1mm}
\end{tabular}
\tablenotetext{a}{Same quasar}
\tablenotetext{b}{Same quasar}
\end{table}

Table 4 provides estimates of $\overline{Q}$ that were performed
using the methods that were described by means of the detailed
examples in Section 3 and the Appendix. Typically, the estimates were performed with radio images of lower sensitivity and dynamic range than those in Figures 18 and 19 of Appendix B. Column (2) provides the redshift of the extreme blazars. Column (3) is the
detailed data extracted from the radio image that is referenced in
the last column. The estimated isotropic flux (based on the methods
of Section 3) is given along with the frequency of the observation.
The next column is the spectral index that is used to extrapolate
the extended flux to 151 MHz (observer's frame) that is discerned by
the details of the image or in some cases it also utilizes the
details of images at multiple frequency. For low quality images the
default value is $\alpha = 1$. The fifth column is the extrapolated
value of flux density at 151 MHz, $F_{151}$. The next column is $\overline{Q}$ from Equation
(10).

\begin{table}
\caption{Extreme Blazar MgII Broadline Properties}
{\footnotesize\begin{tabular}{cccc} \tableline \rule{0mm}{3mm}
 Quasar & Line Luminosity & FWHM  & Reference  \\
 & ergs/sec  & km~s$^{-1}$ &   \\
\tableline \rule{0mm}{1mm}
        & \citet{pea81} Sample  & Extreme & Blazars  \\
\hline
0016+731 & $7.42\times 10^{43}$  &  4179 &\citet{law96} \\
0133+476 & $1.29\times 10^{43}$  &  2946 & \citet{law96} \\
0212+735 & $8.13\times 10^{43}$  &  2486 & \citet{law96} \\
0804+499 & $1.06\times 10^{44}$  &  3000& \citet{law96}  \\
0836+710 & $2.61\times 10^{44}$  &  3048 & \citet{law96} \\
0850+581 & $1.91\times 10^{44}$  &  6214 & \citet{law96}  \\
0859+470 & $7.33\times 10^{43}$  &  4821 & \citet{law96}  \\
0945+408 & $1.94\times 10^{44}$  &  3536 & \citet{law96}  \\
1458+718\tablenotemark{a}  & $1.65\times 10^{44}$  &  3118 & \citet{law96} \\
1633+382 & $1.73\times 10^{44}$  &  3964 & \citet{law96}\\
1637+574 & $8.76\times 10^{43}$  &  3536 & \citet{law96}  \\
1739+522 & $1.38\times 10^{44}$  & 3857 & \citet{law96}  \\
1828+487\tablenotemark{b}  & $7.48\times 10^{43}$  &  3011 & \citet{law96}  \\
\hline
        & 3CR  & Extreme & Blazars  \\
\hline
3C 273  & $2.26\times 10^{44}$ &   2935 & \citet{tan12} \\
3C 309.1\tablenotemark{a}  & $1.65\times 10^{44}$  &  3118 & \citet{law96} \\
3C 380\tablenotemark{b}  & $7.48\times 10^{43}$ & 3011 & \citet{law96} \\
3C 418  & $2.60\times 10^{44}$  &  4440 & \citet{smi80} \\
3C 454.3  & $1.85\times 10^{44}$  & 3100 & \citet{gup17,leo13} \\
\hline
        & Additional 3C  & Extreme & Blazars  \\
\hline
3C 120  & $2.28\times 10^{43}$ &  3740 & \citet{kon06} \\
3C 395  & $4.50 \times 10^{43}$ &  5090 & \citet{tor12} \\
3C 446  & $1.25\times 10^{44}$ &  3955  & \citet{tan12} \\
\tableline \rule{0mm}{1mm}
\end{tabular}}
\tablenotetext{a}{Same quasar}
\tablenotetext{b}{Same quasar}
\end{table}

The reason for introducing the complete samples of extreme
blazars in Table 1 and Table 4 is to see if there is a
distinguishing characteristic of the RA blazars that
differentiates them from other similar objects - i.e., are they
extreme in some other parameters. In Figure 6, we plot a
suggestive scatter plane. First, we introduce the isotropic lobe
luminosity (computed form Table 4) in the rest frame of the quasar,
$\nu_{e}L_{\nu}(\nu_{e}=151~\rm{MHz})$, where the subscript ``e"
means emitted. From Equation (10) this translates directly
to the estimate of $\overline{Q}$. Consider the bolometric
correction in Equation (4) that is derived from the MgII line
strength. One can therefore define an empirical quantity that is a
measure of a normalized jet power, $\nu_{e}L_{\nu}(\nu_{e}=151
\rm{MHz})/L({\rm{MgII}})$; related to the ratio of estimators,
$\overline{Q}/L_{\rm{bol}}$. This is the horizontal coordinate of
the scatter plane in Figure 6. Similarly, based on the virial estimator in Equation (7), combined with the bolometric correction in Equation (4), we
introduce the empirical vertical coordinate in Figure 6. The
empirical quantity, $[L(\rm{MgII})]^{0.5}/\rm{FWHM}^{2}$ relates
directly to estimates of the Eddington ratio, $L_{\rm{bol}}/L_{\rm{Edd}}$.
\begin{figure}[ht]
\begin{center}
\includegraphics[width=150 mm, angle= 0]{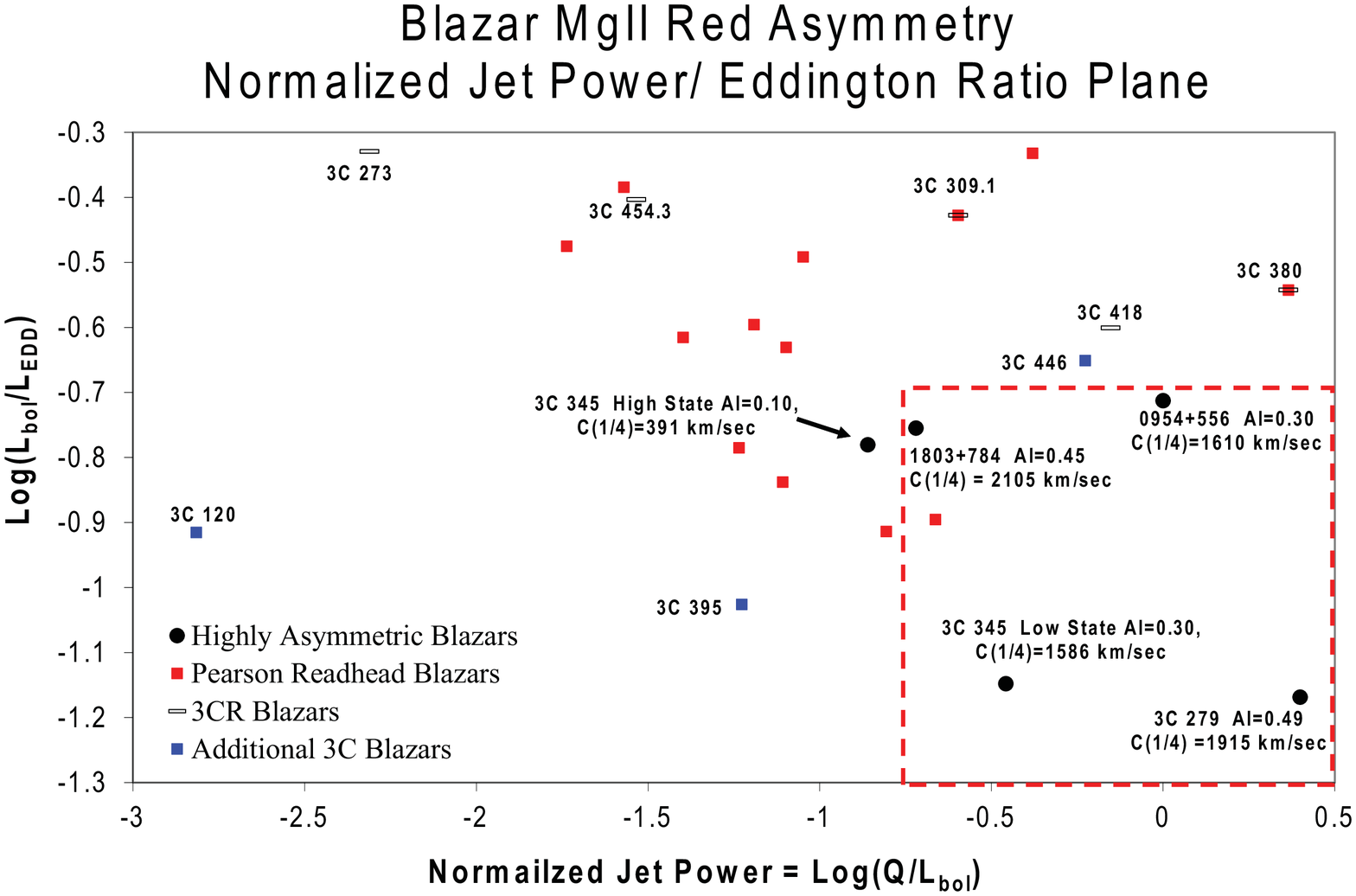}
\caption{This scatter plot is the same as Figure 6 except that the
observed quantities are converted to estimators of physical
parameters by means of Equations (4), (6), (7) and (10). The
RA blazars occur in the region of low Eddington rate and
large jet power relative to the accretion luminosity.}
\end{center}
\end{figure}

\begin{figure}[ht]
\begin{center}
\includegraphics[width=115 mm, angle= 0]{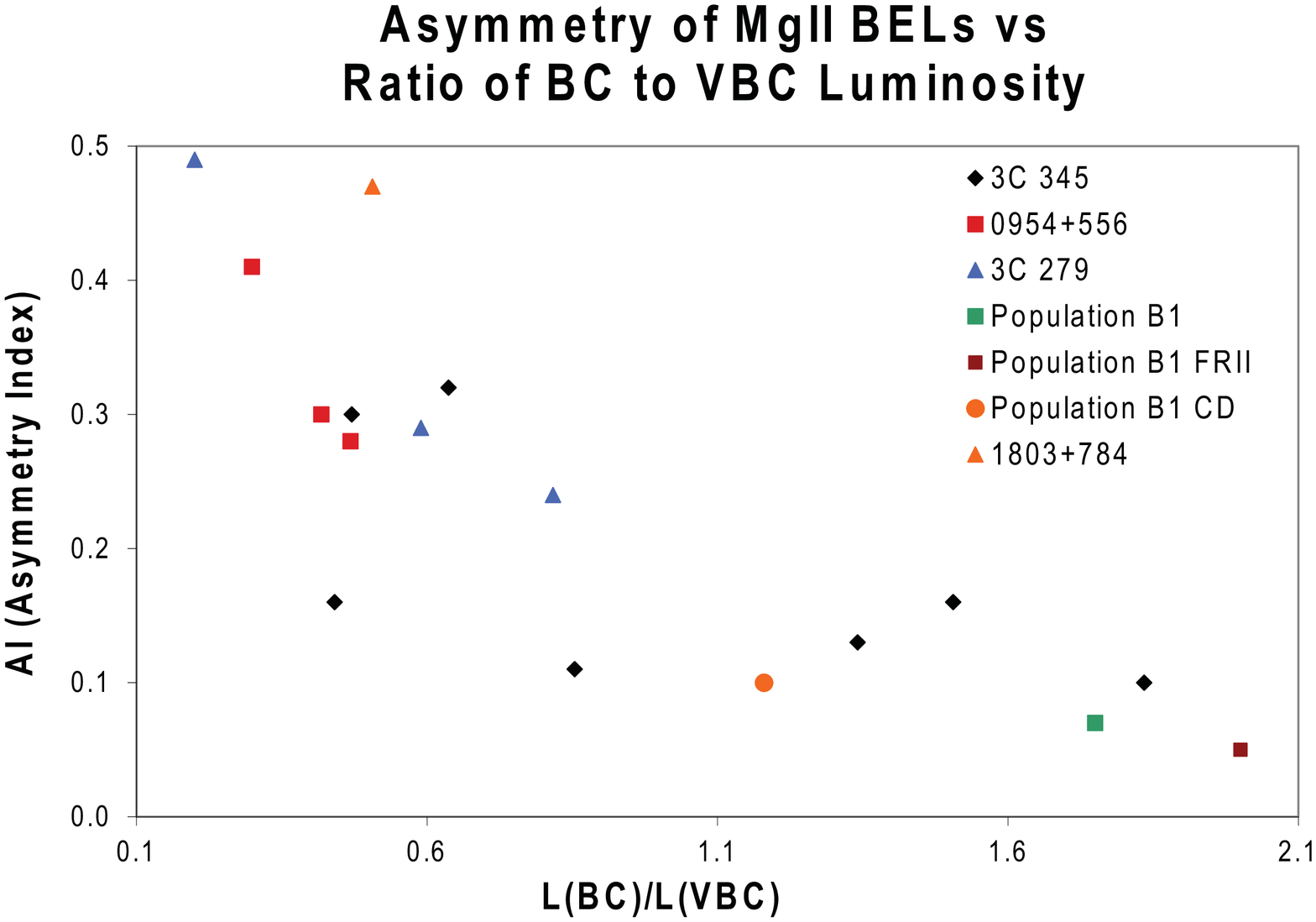}
\caption{The scatter plot shows the dependence of C(1/4) on the
ratio of the broad component luminosity to the VBC luminosity.  The
very RA blazars have a small BC to VBC luminosity
ratio.}
\end{center}
\end{figure}

The MgII lines used for 3C 279 and 0954+556 are from the highest
signal to noise spectra that were shown in Figure 1. The high 3C 345
state is from 1995, simply because the signal to  noise was a little
better than 1992, which lies in a similar region of the scatter
plot. The low state for 3C 345 is the high signal to noise spectrum
with the large red asymmetry from August 28 2016 shown in Figure 5.
There is a definite pattern, the RA blazars are in
the lower right hand corner of the scatter plot inside the dashed
red rectangle. None of the other quasars in the scatter plot have a
pronounced redwatd asymmetry in their MgII profile. The MgII line
properties for the other extreme blazars is tabulated in Table 5.
In particular, all of the quasars in Table 5 have MgII BEL profiles with $AI \sim 0.10$ or less.
\par Based on Equations (7) and (10), the location in the
scatter plot of the RA blazars has a natural
physical interpretation, $\overline{Q}/L_{\rm{bol}}$ is large and
$L_{\rm{bol}}/L_{\rm{Edd}}$ is small. A large value of
$\nu_{e}L_{\nu}(\nu_{e}=151 \rm{MHz})/L({\rm{MgII}})$ (an estimate
for the physical quantity $\overline{Q}/L_{\rm{bol}}$) is not
sufficient to be associated with a large redward asymmetry. The
quasars, 3C 446, 3C 418 and 3C 380 have very large
$\nu_{e}L_{\nu}(\nu_{e}=151 \rm{MHz})/L{\rm{MgII}}$ (as large or
larger than the RA blazars). The asymmetry of the 3C
418 broad line is difficult to assess due to some strong narrow
absorption features, but does not show any obvious signs of redward
asymmetry \citep{smi80,pun18}. The MgII emission line in 3C 380 is
extremely well fit by a symmetric Gaussian \citep{law96}. Figure 7
is the spectrum of 3C446 near MgII generously provided by Michael
Brotherton. There were two observations that were combined. There
was the McDonald Observatory 2.7 meter LCS spectrum from September
1991, as well as a McDonald Observatory 2.1 meter ES2 spectrum from
October 1991.  The spectra were consistent with each other, with
similar SNR.  Data were not photometric and they were all scaled to
match the flux level of the HST spectrum from September 11 1991.
In summary, the driver of the redward asymmetry in blazar BELs is not the jet power,
alone.
\par The other characteristic of the zone of highly asymmetric BELs in the
scatter plane of Figure 6 is the low value of
$[L(\rm{MgII})]^{0.5}/\rm{FWHM}^{2}$ (an estimate of the physical
quantity, Eddington ratio, $L_{\rm{bol}}/L_{\rm{Edd}}$). Is this
alone sufficient to drive a redward asymmetry in the MgII BEL? The
scatter plot is missing sources in the lower left hand corner that
could answer this question. This is a selection effect. Sources in
the lower left hand corner are radio quiet Seyferts and would not appear in
the complete samples of extreme blazars. Numerous Seyfert galaxies with observed MgII
exist in this region and none of them have pronounced redward
asymmetry in their emission lines \citep{eva04,kin91}. Thus, our
studies of complete samples indicate that the red dashed region is
the preferred region for quasars with RA emission lines.

Even though estimates of the physical quantities have inherent
uncertainty, it is useful to convert Figure 6 to physical quantities
with the caveat that additional uncertainty has been introduced. The
estimate for $L_{\rm{bol}}$ is from Equation (4). The estimate for
$M_{bh}$ is the average of estimates in Equations (6) and (7). The
estimate for $\overline{Q}$ is from Table 4. The results are plotted in Figure 8.
We note there is one source in the preferred RA blazar sector of parameter space (the dashed red rectangle in the lower right hand corner) that is not an RA blazar. It is the radio source 0859+470. We note that the lobe flux (and therefore $\overline{Q}$) estimate is a bit uncertain based on the radio images \citep{rei95,mur93}. There is a sub-arcsecond jet to the north that is likely Doppler boosted. The diffuse emission at 408 MHz is dominated by contours that are at the same level as the image noise and the lobe flux is deemed uncertain by the observers \citep{rei95}. They hoped that L-band observations with VLA would clarify the situation. The most sensitive image is from \citet{mur93}, but there are very strong negative contours near the source and the diffuse features tend not to line up with those at 408 MHz. It might be that it does not belong inside the RA blazar region of the scatter plane if better L-band VLA data is obtained.

\section{The Phenomenology that Creates RA
Emission Lines} In the last section, we explored the physical
aspects of the quasar that are associated with these extremely
RA BELs. In this section, we explore the
substructure of the BELs that creates the unusual asymmetric shape.
Fortunately, \citet{mar13}, provides a detailed study of the nature
of MgII emission line profiles in quasars. We can use this as a
large control sample. The analysis of \citet{mar13} relies on
decomposing the quasar population into subsamples, Population A and
Population B \citep{sul00}. Population A are quasars with an
H$\beta$ $\rm{FWHM} < 4000~\rm{km/s}$. Population B are quasars with
an H$\beta$ $\rm{FWHM} > 4000~\rm{km/s}$. RQQs can reside in either
Population A or B, but RLQs are almost always Population B. In order
to separate out extremely wide profiles sub-samples of Population
B are defined. In particular, Population B1 quasars are defined by
an H$\beta$ $4000~\rm{km/s} <\rm{FWHM} < 8000~\rm{km/s}$. All of the
RA sources in Table 1 are Population B1 quasars per
Table 2.

The most obvious line characteristic to suspect as the origin
of the RA MgII BELs is the large redward peak shifts
of the VBC in Table 2. These shifts are in the range of 1500 km/s -
2100 km/s. This is the range of VBC shifts used in the fits
of the Population B1 sources in \citet{mar13}, Table 2. However, by contrast, their table
also indicates that these quasars have rather symmetric profiles
with the quantities from Equations (2) and (3) being
\begin{equation}
\rm{AI} =0.07\pm 0.03\;, \quad \rm{C(1/4)} = 230 \pm 140
\rm{km/s}\;,
\end{equation}
Comparing these numbers to Table 2 indicates (surprisingly) that the
1500 km/s - 2100 km/s redward shift of the VBC is not the origin of
the RA profiles in these extreme blazars.

A clarification of the origin of the redward asymmetry is indicated
in the scatter plot in Figure 9. The plot shows the ratio of the
luminosity of the BC component to the luminosity of the VBC. Notice that the BC is actually more
luminous than the VBC in Population B1 sources as indicated in Table
2 of \citet{mar13} and shown pictorially in Figure 9. This is true
even if we look at the subsamples from \citet{mar13}, B1 FRII, the
FRII (steep spectrum Fanaroff-Riley type II RLQs) or B1 CD (core
dominated RLQs). In line with this scatter plot is Figure 5. When 3C
345 has a symmetric profile the BC is much stronger than when it has
an asymmetric profile. Consistent with the scatter plot in Figure 9, a strong redwing develops when the BC drops in luminosity.
Figure 9 indicates that it is a weak BC relative to
the VBC that produces a redward asymmetry in the MgII BEL in these
extreme blazars.
\par The remainder of this paper is devoted to understanding why the
BC is relatively weak compared to the VBC in these RA blazars. In
particular, based on the results of Section 4, we explore the
creation of BELs in the context of
\begin{itemize}
\item A low Eddington rate
\item A strong jet relative to $L_{\rm{bol}}$
\item A polar line of sight
\end{itemize}

\section{The Line of Sight Adjusted Eddington Ratio} One of the distinguishing characteristics of the broad emission line blazars with RA BELs is their low Eddington ratios in the scatter planes of Figures 6 and 8. Even so, these Eddington ratio estimates from Table 3 are actually grossly over-estimated as a consequence of the blazar polar line of sight. The line of sight reduces the observed FWHM of the BEL and therefore reduces the estimates of $M_{bh}$ in Equations (6)-(9). In this section, we recompute the Eddington ratio, $R_{\rm{Edd}}$, from the last column in Table 3 after correcting for a polar line of sight.
\par The low ionization BEL gas is generally considered to be distributed in an equatorial disk, orthogonal
to the jet axis, with a random velocity, $v_{r}$, superimposed on an
equatorial velocity, $v_{p}$, that is predominantly bulk motion from
Keplerian rotation \citep{bro86} For example, \citet{jar06}, use the relation
\begin{eqnarray}
&& [\rm{FWHM}]^{2} \simeq
4v_{p}^{2}\left[\frac{v_{r}^{2}}{v_{p}^{2}} +
\sin^{2}{\theta}\right] \;.
\end{eqnarray}
It has been estimated that $\frac{v_{r}^{2}}{v_{p}^{2}}\approx 0.024$
for the H$\beta$ BEL \citep{bro86,cor98}.  The average
line of sight (LOS) to the radio jet in a radio loud quasar has been
estimated as $31^{\circ}$ and the maximum line of sight has been
estimated as $\sim 45^{\circ}$ \citep{bar89}. A blazar line of sight would be $\lesssim 5^{\circ}$ to the disk normal \citep{jor17}. According to Equation (13), if the gas were viewed within 5 degrees of the jet axis, the FWHM would appear smaller than that of an average (steep spectrum) radio loud quasar by a factor
$[\rm{FWHM}(\theta=5^{\circ})]^{2}/[\rm{FWHM}(\theta=31^{\circ})]^{2} = (0.024+0.008)/(0.023+0.265) =0.11$. This can be expressed as
\begin{equation}
3[\rm{FWHM(blazar\, LOS)}]\approx \rm{FWHM(steep\, spectrum\, quasar\, LOS)}\;.
\end{equation}
This was corroborated in the study of \citet{dec11} in which the host galaxy bulge luminosity estimates of $M_{bh}$ were compared with virial mass estimates in order to find \begin{equation}
\rm{FWHM(blazar\, LOS)}/\rm{FWHM(normal\, quasar\, LOS)}\approx 5.6/2=2.8\;.
\end{equation}
where ``normal" quasars are a mix of radio quiet and steep spectrum quasars. Thus, we multiply the H$\beta$ and the MgII FWHM in Table 2 by 2.8 and recompute Table 3 with Equations (6)-(8) and present the results in Table 6.
\par Consider the estimates of $M_{bh}$ for 3C 279 in Table 6. In \citet{nil09}, using the empirical
correlation between bulge luminosity and $M_{bh}$, they found
$\log\left(\frac{M_{bh}}{M_{\odot}}\right)=8.9\pm0.5$. Using the same data \citet{dec11} estimated $\log\left(\frac{M_{bh}}{M_{\odot}}\right)=9.4$. Both estimates are consistent with our estimates in Table 6 $\log\left(\frac{M_{bh}}{M_{\odot}}\right)=9.27$. The estimates in Table 6 are very crude considering the many assumptions and parameters that are not directly observed. However, the results indicate a basic conclusion, the Eddington ratios for the RA  blazars are very low for quasars, probably $\sim 1\%-2\%$.

\begin{table}
\caption{Estimated Line of Sight Adjusted Properties of Broad
Emission Lines} {\footnotesize\begin{tabular}{cccccc} \tableline\rule{0mm}{3mm}
1 &  2  &  3 &  4  & 5    \\
 Date &  Line & $L_{\rm{bol}}$ ($10^{45}$ergs/s) & $M_{bh}$ ($10^{9}M_{\odot})$\tablenotemark{a} &  $R_{\rm{Edd}}$ \tablenotemark{b} \\
\tableline \rule{0mm}{3mm}
3C279 &    &   &    &    \\
\hline
4/8/1994\tablenotemark{c} & H$\beta$ & 3.88 & 16.8/8.75  & 1.63\%  \\
4/9/1992 & MgII & 3.63 & 1.20/4.38 & 1.52\%  \\
4/29/2009 & MgII & 2.37 & 0.66/2.88 & 0.99\%   \\
4/11/2010 & MgII & 2.01 & 0.60/2.20 & 0.84\%   \\
Average Mass & MgII   & ...  &  1.89  &  ... \\
4/8/1992 & CIV & 4.95 & ... & 2.08\%   \\
\hline
0954+556 &    &   &    &   \\
\hline
12/8/1983\tablenotemark{c} & H$\beta$ & 15.0 & 30.8/22.0 & 3.37\%   \\
3/24/2012 & H$\beta$ & 12.6 & 6.31/4.28 & 2.82\%   \\
Average Mass & H$\beta$   & ...  &  5.29  & ...  \\
12/8/1983 & MgII & 5.94 & 0.86/2.11 & 1.34\%   \\
1/12/2003 & MgII & 6.50 & 1.10/2.86 & 1.46\%  \\
3/24/2012 & MgII & 7.24 & 1.07/2.62 & 1.63\%   \\
Average Mass & MgII   & ...  &  1.77  &  ... \\
Average Mass & H$\beta$ and MgII   & ...  &  3.53  &  ...  \\
1/20/1993 & CIV & 12.0 & ... & 2.70\%  \\
\hline
1803+784 &    &   &    &   \\
7/04/1986 & H$\beta$ & 5.27 & 4.37/2.39 & 1.93\%   \\
Average Mass & H$\beta$   & ...  &  3.38  & ...  \\
7/04/1986 & MgII & 5.15 & 0.82/2.10 & 1.86\%   \\
Average Mass & MgII   & ...  &  1.02  &  ... \\
Average Mass & H$\beta$ and MgII   & ...  &  2.20  &  ...  \\
\hline
3C 345 &    &   &    &   \\
\hline
6/11/2012 & H$\beta$ & 4.56 & 4.92/2.79 & 1.25\%  \\
5/27/2016 & H$\beta$ & 6.80 & 3.33/1.77 & 0.84\%   \\
Average Mass & H$\beta$   & ...  &  3.20  & ...  \\
6/7/1992 & MgII & 25.6 & 4.32/10.3 & 4.67\%  \\
8/20/1995 & MgII & 18.8 & 3.24/7.97 & 3.44\%   \\
3/24/2009 & MgII & 24.3 & 4.06/9.67 & 4.46\%  \\
6/11/2012 & MgII & 11.8 & 2.40/6.52 & 2.17\%  \\
5/30/2016 & MgII & 10.2 & 2.42/7.12 & 1.87\%  \\
8/29/2016 & MgII & 7.43 & 2.42/8.40 & 1.37\%  \\
9/17/2017 & MgII & 6.45 & 1.74/5.57 & 1.19\%   \\
Average Mass & MgII   & ...  &  5.44  &  ... \\
Average Mass & H$\beta$ and MgII   & ...  &  4.32  &  ...  \\
6/7/1992 & CIV & 41.7 & ... & 7.66\%   \\
8/20/1995 & CIV & 23.3 & ...& 4.05\%  \\
\end{tabular}}
\tablenotetext{a}{First estimate from \citet{she12}, the second
estimate for MgII  and H$\beta$  are from \citet{tra12} and
\citet{gre05}, respectively.}\tablenotetext{b}{The Eddington rate is
calculated using the mean of the ``average mass" from both H$\beta$ and MgII. Note that the average mass does not include H$\beta$ for 3C 279 per table note c.}
\tablenotetext{c}{These measurements are highly uncertain due to noisy spectra. These observations not used in average mass.}
\end{table}
\section{The Kinematics of the Redshift of the VBC}
The proposed explanation in \citet{cor97} of the very redward shifted VBC profiles seen in some BELs
was Keplerian motion and gravitational redshift of gas in a thin
wedge. In order to produce line shapes a model of the gas
distribution had to be made. There is no self-consistent modeling of
the ionizing flux, gas density and commensurate radiative transfer
solution. This model is adhoc, so it is the preference here to
kinematically constrain the approximate centroid of the gas
distribution comprising the VBC without a specific model of the gas
distribution. Namely, consider an element of ionized gas in a
circular Keplerian orbit near a black hole that is emitting UV
recombination emission lines at a rest frame frequency
$\nu_{\mathrm{BEL}}$. The line of sight of the observer makes an
angle $\theta$ to the normal to the plane of the Keplerian orbit.
The UV line emission seen at earth is at a frequency, $\nu_{\rm{earth}}$, that
depends on where in the orbit the element of the line emitting gas
is located. We define the frequency in the following calculation as that which is evaluated in the cosmological rest frame of the quasar, $\nu$, and is related to the that which is observed by, $\nu = (1+z)\nu_{\rm{earth}}$. The maximum observed frequency, $\nu_{max}$, is when the element is approaching earth (maximum blueshift) at maximal
velocity. Similarly, the minimum observed frequency (maximum
redshift), $\nu_{min}$, is when the element is an orbital location
such that it is receding from earth with maximal speed. The
approximate peak of the gas distribution is defined by the frequency
\begin{eqnarray}
&& \nu_{\mathrm{peak}}  =0.5\mathrm{c}(\nu_{max} + \nu_{min})\;, \nonumber\\
&& \nu_{\mathrm{earth}}(\rm{peak})  =[0.5\mathrm{c}(\nu_{max} + \nu_{min})]/(1+z)\;.
\end{eqnarray}
In order to determine $\nu_{min}$ and $\nu_{max}$, we must solve for
the combined frequency shift from the kinematic Doppler shift and
the gravitational redshift. Since, the BEL gas resides far from the
black hole, spin corrections to the metric that describe the geometry external to the central black hole are negligible,
justifying the use of the Schwarzschild metric. Let $K^{\mu}$ be the
momentum four vector of a photon emitted by the BEL gas at
coordinate r. Let $U^{\mu}_{\mathrm{BEL}}$, $U^{\mu}_{o}$,
$U^{\mu}_{\mathrm{quasar}}$ be the four velocity of an element of BEL
emitting gas in the three reference frames, the BEL gas rest frame,
the local static frame and the cosmological rest frame of quasar, respectively. All quantities in the local static frame
will be designated by a subscript "o" throughout. Note that the local static frame is defined by (the following parallels problem 15.8 of Lightman et al 1975),
\begin{equation}
U^{\mu}_{\mathrm{quasar}} = U^{\mu}_{o}\sqrt{1- \frac{2M}{r}} \;.
\end{equation}
In Equation (17), we have introduced geometrized units, $M\equiv GM_{bh}/c^{2}$ and c=1.
With these definitions, one has
\begin{equation}
\frac{\nu_{o}}{\nu_{\mathrm{BEL}}} =\frac{U_{o}\cdot K}{U_{BEL}\cdot
K} = \frac{1}{\Gamma (1 - \beta \cos{\phi})}\;.
\end{equation}
In Equation (18), the angle $\phi$ is the angle between the local
velocity vector and the line of sight to earth, $\beta = v/c $ is
the three-velocity of the BEL gas as measured in the static frames
and $ \Gamma = 1/\sqrt{1 - \beta^{2}} $. Equation (18) is the
standard kinematical Doppler shift \citep{lin85}. The transformation
from the BEL rest frame to the local orthonormal static frames is equivalent to a
local special relativistic boost. Similarly, one has the following transformation of the
frequency from the local static frame to the quasar cosmological frame (at asymptotic infinity in the Schwarzschild metric),
\begin{equation}
\frac{\nu}{\nu_{o}} =\frac{U_{\mathrm{quasar}}\cdot
K}{U_{o}\cdot K} = \sqrt{1- \frac{2M}{r}}\;,
\end{equation}
This constitutes the gravitational redshift. The Keplerian velocity,
$c \beta_{\mathrm{kep}}$, in the static frames is, \citep{lig75},
\begin{equation}
c \beta_{\mathrm{kep}} = \frac{d\phi_{o}}{dt_{o}}= \frac{r
d\phi}{\sqrt{1- \frac{2M}{r}}dt}=\sqrt{\frac{M}{r -2M}}\;.
\end{equation}
Note that $d\phi_{o}$ has units of distance since the static frames are orthonormal.
Thus, combining equations (18) - (20) at any point on the orbit, the
frequency in the cosmological frame of reference of the quasar is given by
\begin{equation}
\frac{\nu}{\nu_{BEL}}=\frac{\sqrt{1- \frac{2M}{r}}}{\Gamma (1 -
\beta_{\mathrm{kep}} \cos{\phi})}\;.
\end{equation}
\begin{figure}[ht]
\begin{center}
\includegraphics[width=145 mm, angle= 0]{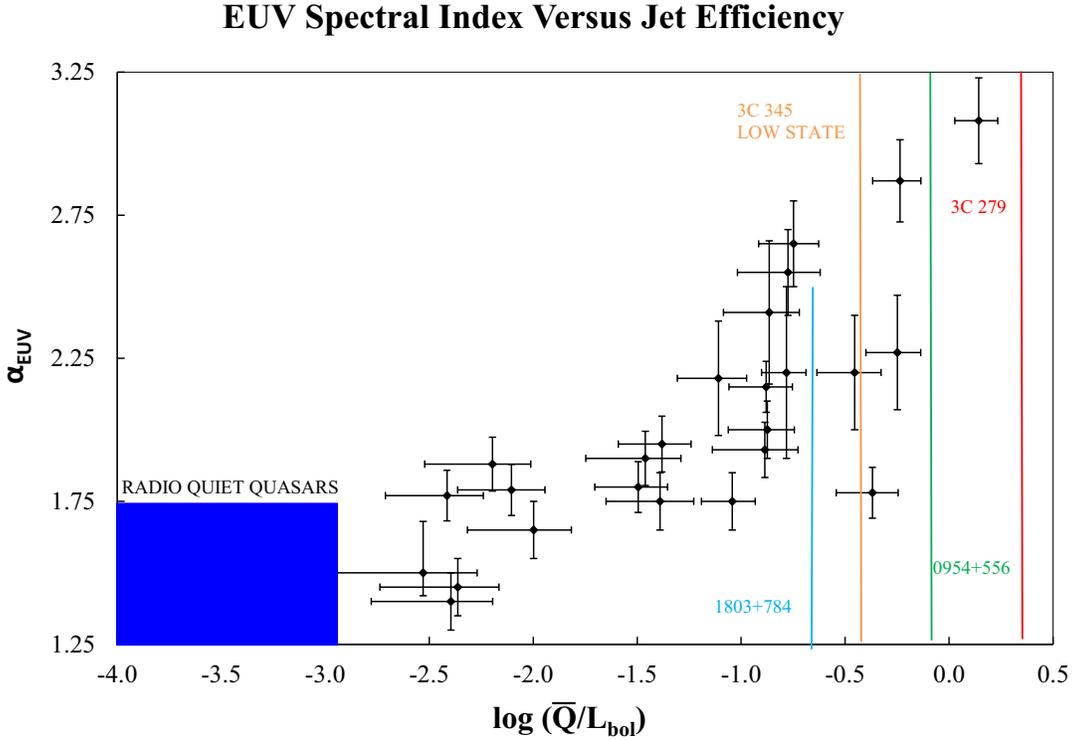}
\caption{This scatter plot shows the EUV deficit of RLQs is originally from \citep{pun17}. We superimpose the locations of the x-variable ($\overline{Q}/L_{\rm{bol}}$) for the RA blazars for comparison purposes. The $\overline{Q}/L_{\rm{bol}}$ estimates for the RA blazars are from Tables 3 and 4.}
\end{center}
\end{figure}
Recall that $\theta$ is defined as the angle between the line of sight and the normal of the Keplerian plane. The maximum value of frequency in Equation (21) occurs when $\beta_{\mathrm{kep}} \cos{\phi} = \beta_{\mathrm{kep}} \sin{\theta}$ corresponding to motion toward the line of sight. The minimum value of frequency in Equation (21) occurs when $\beta_{\mathrm{kep}}\cos{\phi} = -\beta_{\mathrm{kep}} \sin{\theta}$ corresponding to motion away from the
line of sight. Equation (21), therefore implies that
\begin{eqnarray}
&& \frac{\nu_{max}}{\nu_{BEL}}= \frac{\sqrt{1-
\frac{2M}{r}}}{\Gamma (1 - \beta_{\mathrm{kep}}\sin{\theta})}\;,\\
&& \frac{\nu_{min}}{\nu_{BEL}}= \frac{\sqrt{1- \frac{2M}{r}}}{\Gamma (1 +
\beta_{\mathrm{kep}}\sin{\theta})}\;.
\end{eqnarray}
From Equations (16), (22) and (23), the offset of the peak of the VBC from the systemic
velocity of the quasar, $v_{\mathrm{peak} \; \mathrm{offset} }$,  does not depend on the line of sight,
\begin{equation}
\frac{v_{\mathrm{peak} \; \mathrm{offset} }}{c}=  \frac{\nu_{\rm{peak}}}{\nu_{BEL}} =1 - (\Gamma)^{-1}\sqrt{1-\frac{2M}{r}}
\approx \frac{M}{r}+\frac{1}{2}\beta_{\mathrm{kep}}^{2}\approx \frac{3M}{2r}\;.
\end{equation}
The first term on the right hand side of equation (24) is the gravitational redshift
and the second term is the transverse Doppler shift from orbital motion. The fact that the result is independent of the line of sight
naturally explains the fact that the Population B1 quasars in \citet{mar13} have very similar MgII VBC redshifts to the nearly polar line of sight blazars in Table 2, $v_{\mathrm{peak} \; \mathrm{offset} } \sim 2000$~km~s$^{-1}$. This is compelling evidence that this is the correct explanation of the redshift as opposed to a directional outflow or inflow. If one assumes approximately Keplerian motion of the BEL gas, Equation (24) can be solved for r. A VBC peak offset of 2000~km~s$^{-1}$ occurs at r = 225 M by Equation (24). For the CIV VBC redshift of 3C 279 in Table 2, $v_{\mathrm{peak} \; \mathrm{offset} } \sim 5000$~km~s$^{-1}$ and by Equation (24), r = 90 M\footnote{In cgs units this would be $r = 90[M_{bh}/(10^{9}M_{\odot})]1.45\times 10^{14}$cm}. This leads to the natural question how can this BEL gas reside so close to the black hole and not be too over-ionized to radiate efficiently?
\section{The Ionization State of the VBC} The results of Sections 4, 5 and 7 indicate that the strong redwing in these extreme blazars is emitted from VBC gas deep in the gravitational potential at $r \sim 100 -200$ M. This section is an analysis of the ionization state of gas this close to the central quasar. We verify the expected trivial result that gas this close to the photo-ionizing source  will be too ionized to radiate BELs efficiently in most quasars. We explore the conditions necessary for this region to be a bright source of BELs using Figures 6 and 8 as a guide. The low Eddington rate estimated at $\sim 1$\% in Table 6 is an obvious factor to lower the strength of the ionizing source. However, it does not explain any connection to the powerful radio jet which also appears to be relevant based on Figures 6 and 8. This section explores a possible connection to the radio jet that is based on the interaction of the radio jet launching mechanism and the innermost regions of the quasar accretion flow, a region that is the source of the extreme ultraviolet (EUV) continuum \citep{zhe97,tel02}. The difference imposed on the quasar spectrum associated with the jet launching mechanism was quantified as an ``EUV deficit" of radio loud quasars that increases with the strength of the jet \citep{pun15}. This is particularly relevant because these are the bulk of the photons that ionize the BEL gas. In order for gas to emit MgII BELs it must be singly ionized which requires a photon of energy $> 15$ eV. Thus, photons with a wavelength $<825\AA$ are required to ionize the gas in order for MgII BELs to occur due to photo-ionization by the accretion flow radiation.

\begin{figure}[ht]
\begin{center}
\includegraphics[width=145 mm, angle= 0]{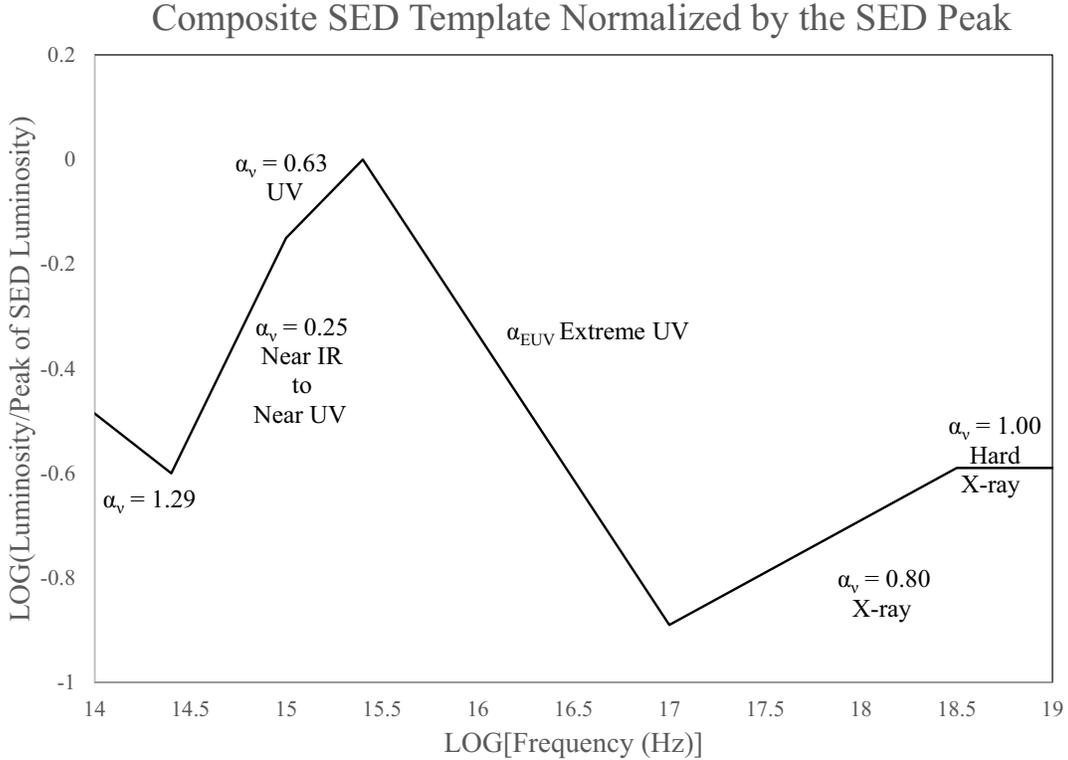}
\caption{The quasar composite SED described in the text is plotted above. We use names for the various regions (such as UV) to make contact with Table 7. The spectral indices represent power laws for the spectral luminosity $L_{\nu}\propto \nu^{-\alpha_{\nu}}$. The $\alpha_{\nu}$ values from this template are used to bridge the gaps between regions of the SED which are constrained by observation. A prescription is necessary since no SED that we use has full simultaneous frequency coverage.}
\end{center}
\end{figure}
This is the EUV region of the quasar spectrum. The EUV is more
important than the soft X-ray band in the photo-ionization of BEL
clouds at a distance, $R_{BC}$, from the quasar, since it is the number of ionizing photons that are relevant
to the degree of ionization and the number spectrum, $N_{\nu} \equiv
L_{\nu}/(4\pi R_{BC}^{2}c\rm{h}\nu)$ (where $L_{\nu}$ is the
spectral luminosity) is very steep in the EUV region of quasar
spectra \citep{tel02}. Thus, most of the ionizing photons are
emitted just below $825\AA$. Consider the plot in Figure 10. This
plot was originally developed in \citet{pun15} and the version in
Figure 10 was presented in \citep{pun17} with some additional
sources\footnote{Note the minor change to the definition of $\overline{Q}$ in Equation (10). This induces an $\sim 10\%$ change from the formulae used in \citet{pun15,pun17}. The reasons for the minor change are described in \citet{pun19}, yet the results of Figure 10 are not noticeably changed by the modified formula.} . The spectral luminosity in the EUV is defined by a power law, $L_{\nu} \sim \nu^{-\alpha_{\rm{EUV}}}$. In Figure 10, $\alpha_{\rm{EUV}}$ is estimated in the region $1100\AA$ to $700\AA$. We do not know the extent of the power law in general, but we assume that a continuous power law approximation is valid short-ward of $1100\AA$ in the following. This continuum is the main source of ionizing photons. Based on Figure 10, radio loud quasars will produce fewer ionizing photons than radio quiet quasars with the same near UV luminosity. The EUV deficit becomes much larger for the radio loud quasars with very strong jets, such as 3C 279.

\section{A CLOUDY Analysis of the Redward Asymmetry}
In this section, we describe a single zone CLOUDY 13 analysis of the ionization state associated with the photo-ionizing continuum of the reward asymmetric blazars \citep{fer13}. We want to explore the roles of low Eddington rate and jet power. However, we note that radio quiet Seyfert 1 galaxies typically have $R_{Edd} = 0.01 -0.1$
\citep{wan99}. The low Eddington states of radio quiet Seyfert
galaxies have accretion rates similar to the estimated line of sight
corrected values of $R_{Edd}$ for the RA blazars in Table 6. Yet, their
is no evidence of extreme RA MgII emission lines in
radio quiet Seyfert galaxies, even in low states \citep{eva04,kin91}. However, these objects tend to have much harder EUV continua than the
more luminous radio quiet quasars \citep{sco04,ste14,ste06}. Thus, our analysis must also include spectral shape changes associated with different classes of objects as well as the Eddington ratio and the jet power.
\subsection{The Design of the CLOUDY Experiment}The CLOUDY numerical experiment will be designed to test three hypotheses.
\begin{description}
\item[Experiment 1] Do the SEDs associated with the low Eddington rates in Table 6 for the RA blazars enhance a highly redshifted VBC compared to the VBC induced by the SED of quasars with more typical Eddington rates: $\sim 10\% - 20\%$ \citep{sun89}?
\item[Experiment 2] Does the steep EUV continuum indicated in Figure 10 for these strong jet sources enhance a highly redshifted VBC compared to the VBC excited by a broad line radio quiet AGN SED with similarly low Eddington rates: $\sim 1\%$?
\item[Experiment 3] Is the large redward asymmetry in the BELs a manifestation of a strong VBC as opposed to a weak BC?
\end{description}
The second experiment shows that a low accretion rate is not a sufficient condition. It highlights the importance of the change in spectral shape in the EUV as the radio jet gets stronger. The first and third experiments will compare the line emissivity as a function of distance from the central black hole (as a function of redshift through Equation (24)) induced by two different SEDs. First, a representative RA blazar (Eddington ratio $\sim 1\%$) SED and, second, the average Population B1 quasar SED in \citet{mar13} with a $\sim 12\%$ Eddington ratio. The second experiment will compare the line emissivity as a function of distance from the central black hole (as a function of redshift) induced by the same $\sim 1\%$ Eddington ratio blazar SED with that induced by a low state Seyfert 1 galaxy SED also with $\sim 1\%$ Eddington ratio.
\subsection{Input SEDs} CLOUDY requires an input photo-ionizing spectrum in order to execute the computation of line strengths originating from different BEL densities and distances from the photo-ionizing source. None of our three input SEDs have complete, simultaneous observations of the ionizing continuum, EUV, soft X-ray and hard X-ray. This is far from ideal, but it is not deleterious to our efforts since we are looking at the effect of gross changes to the ionizing spectrum. Thus, coarse approximations to the photo-ionizing continuum will be adequate to demonstrate large changes in the VBC luminosity. The most poorly constrained is 3C 279 for which the entire SED from the accretion flow is masked by the jet emission. However, we can indirectly estimate $L_{\rm{bol}}$ from the BEL line strengths through Equations (4) and (5) and Tables 3 and 6. This sets an overall normalization for the SED and Figure 10 constrains the EUV. In order to implement these two pieces of information we introduce a quasar SED template in Figure 11 that is based on composite spectra. The template will not be the exact SED shape at the time of observation. However, for the example of 3C 279, the normalization estimates from Equations (4) and (5) and the estimate of $\alpha_{\rm{EUV}}$ from Figure 10 will give us the rough estimate of the ionizing continuum that we need for the experiment.
\begin{figure}[ht]
\begin{center}
\includegraphics[width=80 mm, angle= 0]{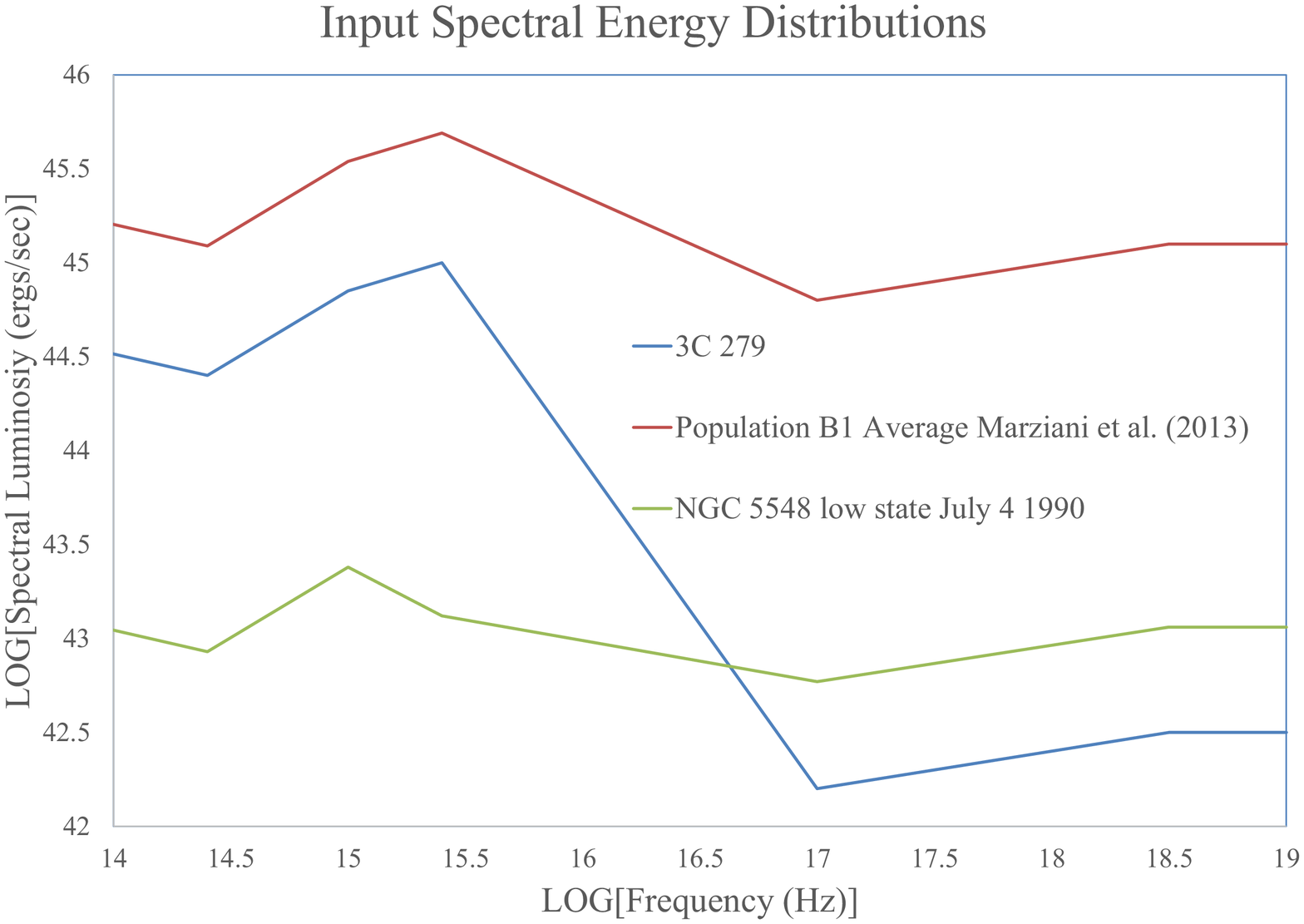}
\includegraphics[width=80 mm, angle= 0]{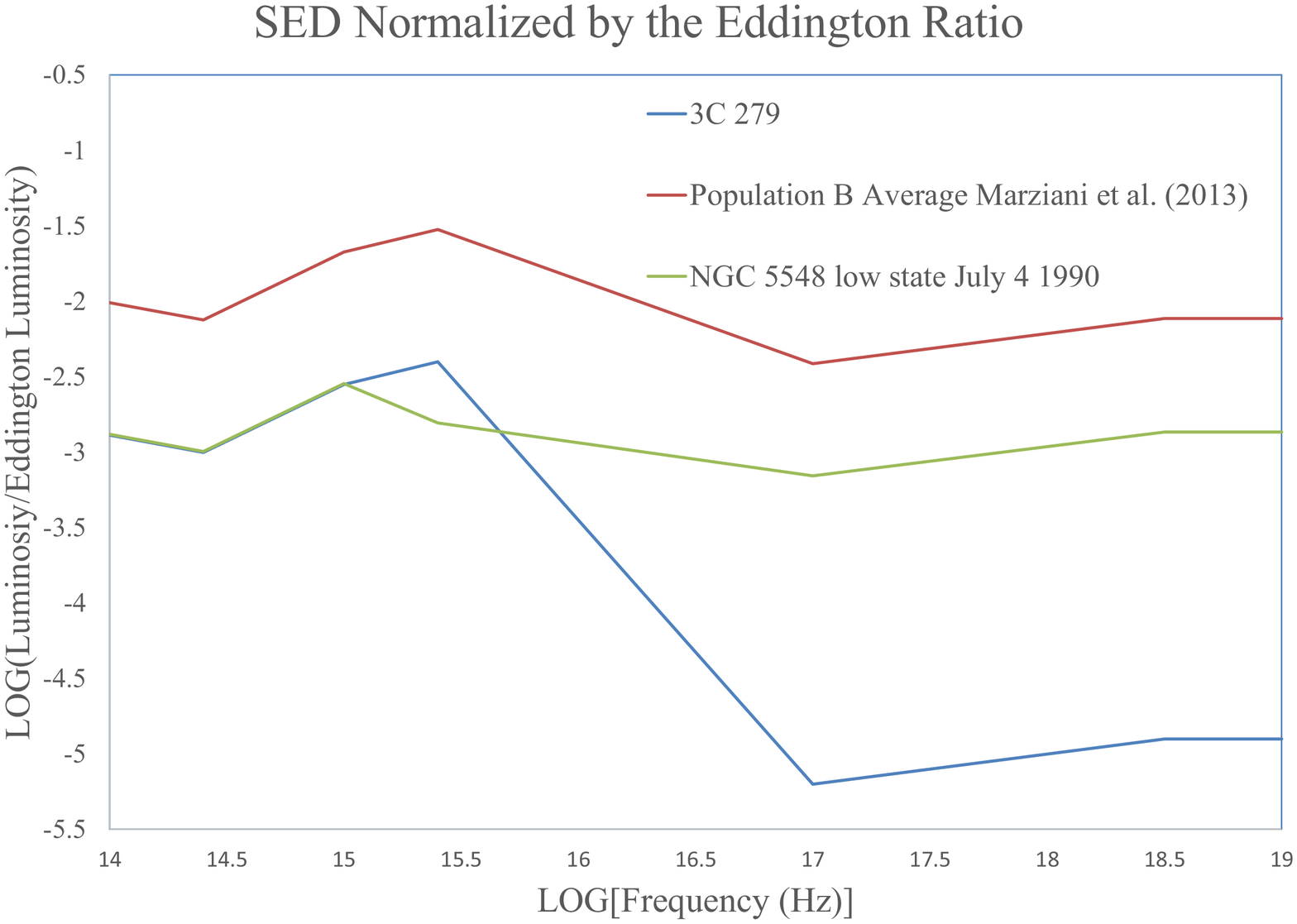}
\caption{The three input SEDs for our CLOUDY numerical experiment as described in Section 9.2. The left plot is based on luminosity and the right plot scales the SEDs in terms of Eddington luminosity. The redwing asymmetric blazars are represented by the most extreme SED, 3C 279. It is very low Eddington rate and by Figure 10 is very steep in the EUV. The Population B1 SED provides a contrast between more ``typical" quasars and 3C 279. The NGC 5548 low state SED provides a contrast between a low Eddington rate radio quiet broad line AGN and the RA blazars.}
\end{center}
\end{figure}
\subsubsection{The Quasar SED Template}Figure 11 is our chosen quasar SED template. The SED is associated with emission from the disk, not distant molecular clouds or the relativistic jet. It is directly related to the estimates of $L_{\rm{bol}}$. The SED functions to provide a crude estimate of the spectral index in regions where there are gaps in our observational measurements. The Figure is based on the quasar composite spectrum in \citet{lao97} that was predicated primarily on the HST composite spectrum \citep{zhe97}. Our modification is to use newer XMM quasar X-ray spectra in the 0.5 keV -12 keV regime and the hard X-ray data (20keV -100 keV) of broad line AGN from INTEGRAL \citep{pic05,bec09}. The junction between the EUV and the X-ray is chosen to be $\sim0.4$ keV ($10^{17}$ Hz). This region is not well defined because of the existence of a soft X-ray excess in some quasars, usually higher Eddington rate quasars \citep{pic05}. \citet{lao97} has the junction (somewhat arbitrarily) at 2 keV. But, the amount of soft X-ray excess in our lower Eddington rate quasars might not be very large and we ignore it to zeroth order in our approximation of the ionizing continuum.

\begin{table}[ht]
\caption{Input SEDs for CLOUDY Experiment }
{\footnotesize\begin{tabular}{cccccccc} \tableline \rule{0mm}{3mm}
 SED & Normalization &Near IR & UV\tablenotemark{a} & $\alpha_{\rm{EUV}}$\tablenotemark{a}  & X-ray\tablenotemark{a}  & Hard\tablenotemark{a} \\
        &       & to Near UV\tablenotemark{a}  &&   &  & X-ray  \\
\tableline \rule{0mm}{1mm}
3C 279 & $R_{\rm{Edd}}$, $M_{bh}$ & $\alpha_{\nu}=0.25$\tablenotemark{b}& $\alpha_{\nu}=0.63$\tablenotemark{b} & $\alpha_{\rm{EUV}}=2.75$  & $\alpha_{\nu}=0.80$\tablenotemark{c}& $\alpha_{\nu}=1.00$\tablenotemark{c} \\
 & Table 6 & &  & from Figure 10 &   &   \\
Population B1 & $\lambda L_{\lambda}(3000\,\AA)$& $\alpha_{\nu}=0.25$\tablenotemark{d} & $\alpha_{\nu}=0.63$\tablenotemark{d} & $\alpha_{\rm{EUV}}=1.55$\tablenotemark{e} & $\alpha_{\nu}=0.80$\tablenotemark{c} & $\alpha_{\nu}=1.00$\tablenotemark{c} \\
NGC 5548 & $\lambda L_{\lambda}(1350\,\AA)$, & $\alpha_{\nu}=0.25$\tablenotemark{c} & \tablenotemark{f} & $\alpha_{\rm{EUV}}=1.22$\tablenotemark{c} & $\alpha_{\nu}=0.80$ \tablenotemark{g} & $\alpha_{\nu}=1.00$\tablenotemark{c} \\
&  $\lambda L_{\lambda}(2670\,\AA)$  & &  & &   &
\end{tabular}}
\tablenotetext{a}{Regions of the SED are defined in Figure 11.}
\tablenotetext{b}{The spectral index is from, Figure 11, the SED template. Luminosity set by Eddington rate and mass from Table 6.}
\tablenotetext{c}{The spectral index in Figure 11 SED template is used to extrapolate SED form the adjacent frequency band.}
\tablenotetext{d}{SED template spectral index from Figure 11. Luminosity set by median value in \citep{mar13}.}
\tablenotetext{e}{Value from quasar composite of similar luminosity \citep{tel02}.}
\tablenotetext{f}{Spectrum from \citep{cla92}.}
\tablenotetext{g}{Spectral index from Figure 11 and normalization from \citep{cla92}.}
\end{table}
\subsubsection{The 3C279 SED} The SED of the accretion flow is completely obscured by the bright synchrotron power law from the relativistic jet. Our only ability to peer into the hidden disk emission is through the luminosity of the emission lines. This was the basis of the $R_{\rm{Edd}}$ estimates in Tables 3 and 6. The BEL luminosity for the various epochs seem to indicate $R_{\rm{Edd}} \sim 0.01$. We choose $R_{\rm{Edd}} = 0.01$ and a black hole mass estimate of $ M_{bh}\approx 2 \times 10^{9}M_{\odot}$ from Table 6 and the bulge luminosity estimate in \citet{dec11} in order to scale the SED template in Figure 11. Furthermore, a value of $\alpha_{\rm{EUV}}=2.75$ is estimated from the normalized jet power in Figure 10. This allows us to crudely estimate the SED in Figure 12. It is loosely constrained, but it does contain the essential features of the hidden SED: an extremely low $R_{\rm{Edd}}$ for a quasar and an extremely steep EUV continuum for a quasar. These features are sufficient for us to perform our CLOUDY numerical experiment.
\begin{figure}[ht]
\begin{center}
\includegraphics[width=80 mm, angle= 0]{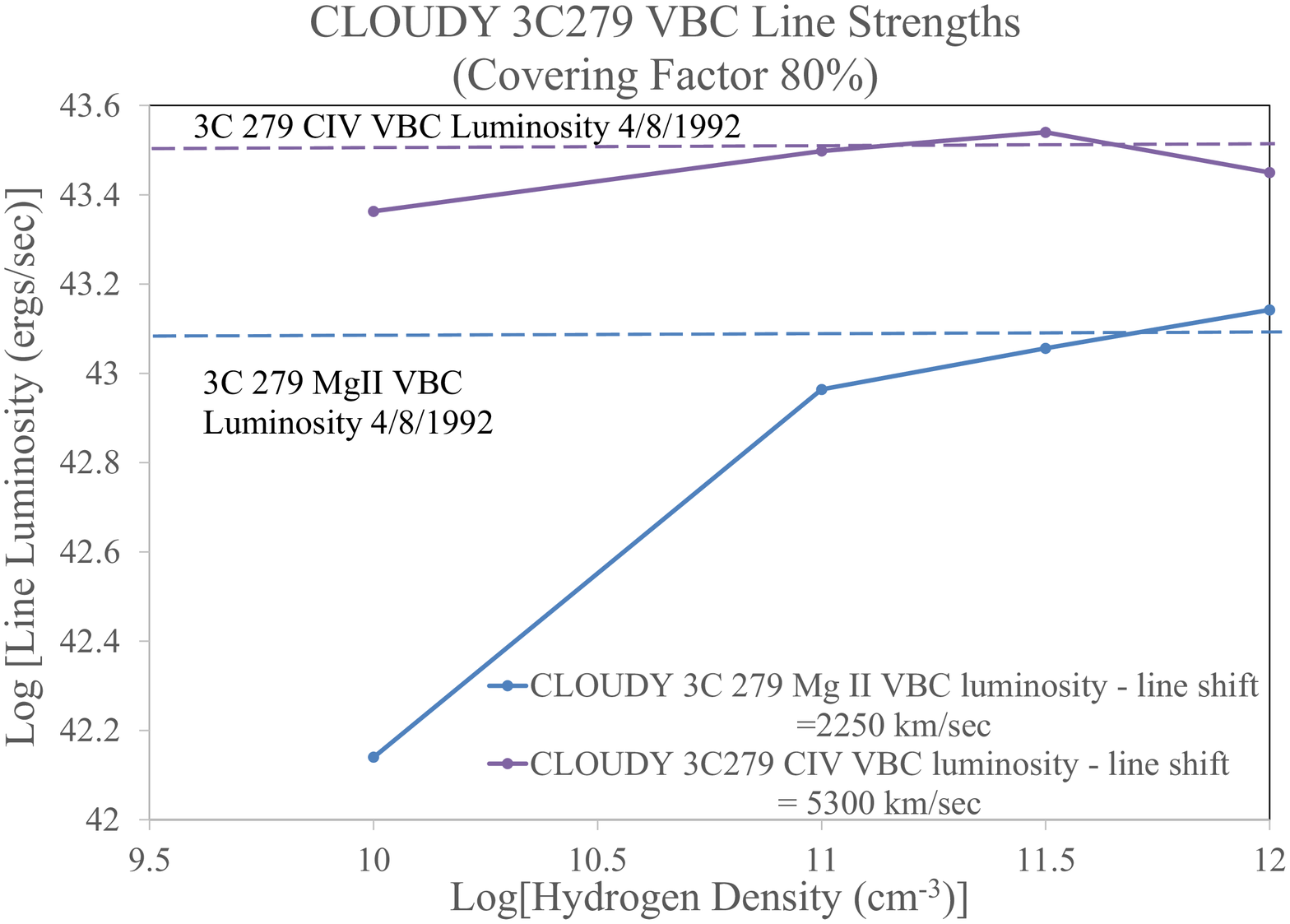}
\includegraphics[width=80 mm, angle= 0]{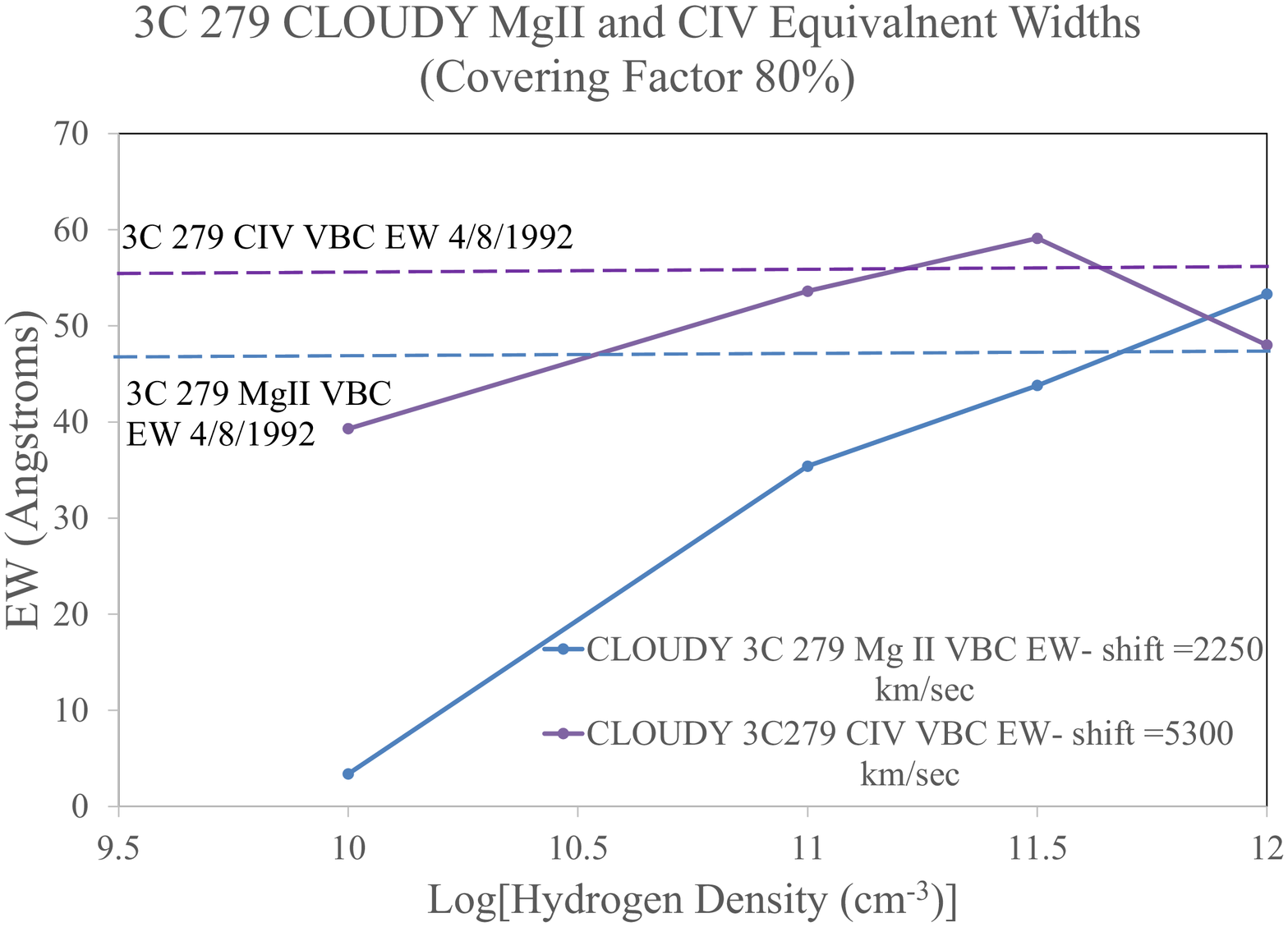}
\caption{The CIV and MgII VBC luminosity (left) and the rest frame EW (right) as a function of hydrogen density computed from the 3C 279 SED in Figure 12. The dashed horizontal lines are the fitted values for the VBC from April 1992 in Table 2. The emission line (red) shifts chosen from the CLOUDY runs are the fitted line shifts in Table 2. The results show that the 3C 279 SED in Figure 12 efficiently excites a dense BEL gas, near the black hole, if $\log{n} =$ 11.5 - 12 and the line strength is consistent with the observed values.}
\end{center}
\end{figure}

Table 7 is provided to capture the logic of how each of the SEDs was completed across the entire range of frequencies in Figures 11 and 12. The first column is the object/objects used to determined the approximate SED. The first piece of information is the normalization in the second column. Then, in the next five columns, we show how all the regions of the template in Figure 11 were determined in Figure 12. The details of how the observational data was implemented in these columns is given in Sections 9.2.2-9.2.4. Figure 12 has two plots. The left plot compares the SEDs of our three types of sources. The right plot has these SEDs normalized by the Eddington luminosity of the central black hole. This gives us a scale based on $M$ which is relevant for gauging the strength of the photo-ionizing source in the gravitational potential well. We proceed to describe the origins of the other two SEDs.

\subsubsection{The Population B1 SED} The Population B1 quasars discussed in Section 5, were similar to the RA blazars in many respects. They are associated with large central black hole masses and small or modest $R_{\rm{Edd}}$. Their properties are discussed in Tables 2-4 of \citet{mar13} with median values. We use these to construct a median SED in Figure 12. From Table 4 of \citet{mar13} $ M_{bh}\approx 1.3 \times 10^{9}M_{\odot}$. The first column of Table 3 is their estimate of $L_{\rm{bol}}$ which is derived from the $3000\,\AA$ luminosity ($L_{3000}\equiv \lambda L_{\lambda}(3000\,\AA)$) in their first column,
\begin{equation}
\log{L_{\rm{bol}}} = (9.24\pm 0.77)+(0.81 \pm 0.02)\log{L_{3000}}\;.
\end{equation}
We invert Equation(27) with $\log{L_{\rm{bol}}}=46.13$ for the Population B1 quasars to find $\log{L_{3000}}=45.54$. We use this point to scale the template SED in Figure 11 in order to get the SED in Figure 12. Most of the Population B1 quasars in \citet{mar13} are radio quiet, 82\%. Thus, we pick $\alpha_{\rm{EUV}}=1.55$ based on the radio quiet HST EUV composite spectra with $\log{L_{\rm{bol}}}\gtrsim 46$ \citep{tel02}. The resultant SED has $R_{\rm{Edd}}=0.12$.
\begin{figure}[ht]
\begin{center}
\includegraphics[width=80 mm, angle= 0]{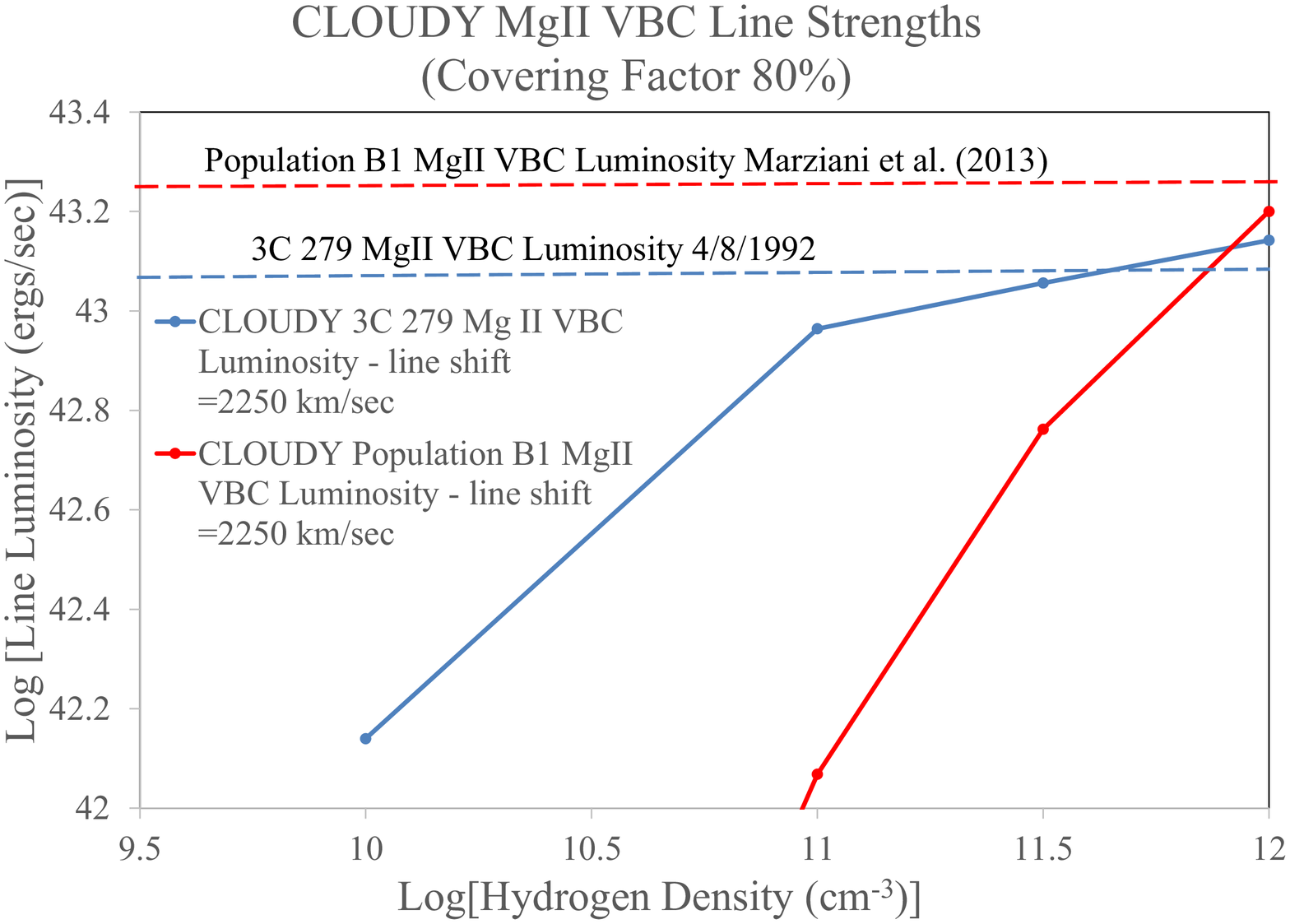}
\includegraphics[width=80 mm, angle= 0]{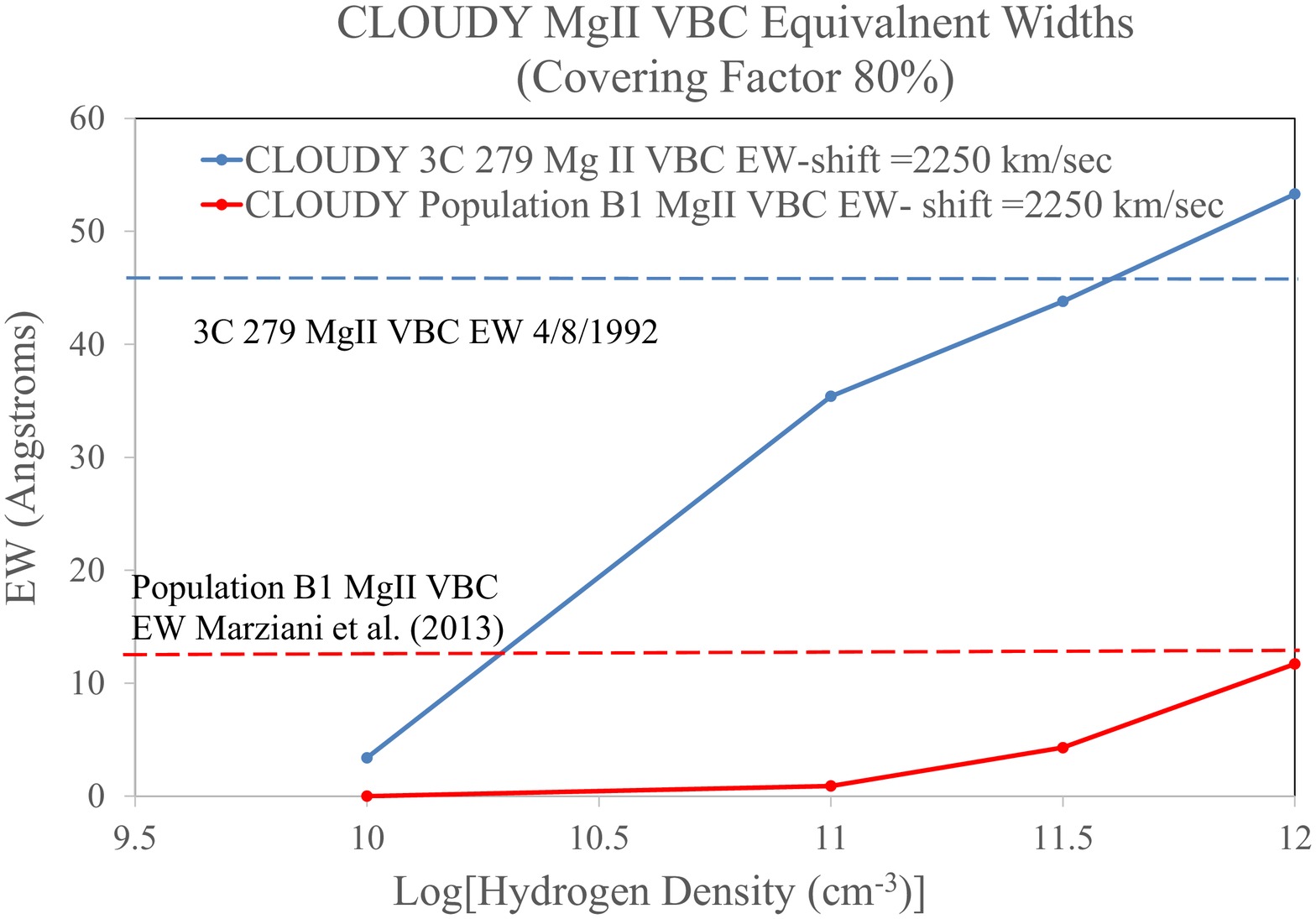}
\caption{A comparison of the MgII VBC luminosity (left) and the rest frame EW (right) as a function of hydrogen density computed from the 3C 279 SED and the Population B1 SED in Figure 12. The blue dashed horizontal line is the fitted value for the VBC of 3C 279 from April 1992 in Table 2. The red dashed horizontal line is the median value from the \citet{mar13} Population B1 sample. The emission line (red) shift from \citet{mar13} and for 3C 279 in 1992 are both $\gtrsim$ 2100~km~s$^{-1}$. The right hand panel shows that the 3C 279 SED in Figure 12 efficiently excites a dense BEL gas, near the black hole, if $\log{n} =$ 11.5 - 12, yet the Population B1 SED seems to over-ionize the VBC region for efficient MgII line emission.}
\end{center}
\end{figure}
\subsubsection{The SED of NGC 5548 in a Low State}We were fortunate to find a low state of NGC 5548 that was observed simultaneously in the UV and X-ray on July 4 1990 \citep{cla92}. It is not the lowest state of NGC 5448 ever observed, but it was the lowest with both UV and X-ray coverage. The simultaneity is very important since there is so much variability in this source \citep{cla92}. There are tabulated continuum flux densities at $2670\, \AA$, $1840\, \AA$ and $1350\, \AA$ from IUE observations and the 2-10 keV flux from Ginga. We get a strong constraint on the EUV by extrapolating the X-ray spectrum to 0.4 keV and the far UV continuum to just shortward of Ly$\alpha$, then connecting these with a power law. It is not exact, but it should be close. The resultant EUV spectral index is $\alpha_{\rm{EUV}} =1.22$. This is a relatively hard spectrum for a broadline AGN, but it is consistent from our expectations for a radio quiet, low luminosity, broad line AGN \citep{tel02,ste14,sco04,ste06}.
\par We choose $M_{bh} = 6.7 \times 10^{7} M_{\odot}$ based on \citep{pet04,ben09,woo10}. The input SED for NGC 5548 in Figure 12 implies that $R_{\rm{Edd}}\approx 0.013$. Thus, when contrasted with the input 3C 279 SED, it provides a good test of the impact of the EUV deficit of radio loud quasars on the VBC in our CLOUDY experiment.
\begin{figure}[ht]
\begin{center}
\includegraphics[width=100 mm, angle= 0]{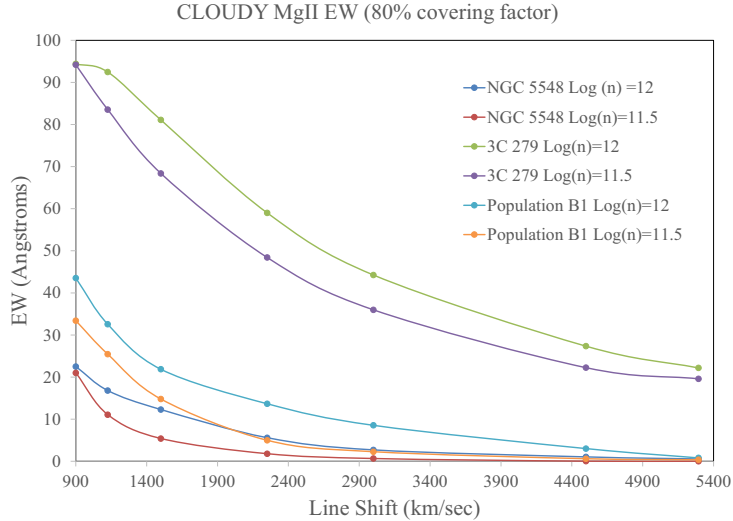}
\caption{A comparison of the MgII VBC rest frame EW as a function of line (red) shift computed from the 3C 279 SED, the Population B1 SED and the low state NGC 5548 SED in Figure 12. The Figure shows that the 3C 279 SED in Figure 12 efficiently excites a dense BEL gas at large redshift, yet both the Population B1 SED and the low state NGC 5548 SED seem to over-ionize the VBC region for efficient MgII line emission to a similar degree. The NGC 5548 contrast with 3C 279 along with the input SEDs in Figure 12, show that the steep EUV continuum associated with a powerful jet makes a profound difference on the VBC strength. }
\end{center}
\end{figure}
\subsection{CLOUDY Analysis of the 3C 279 Ionizing Continuum} We ran CLOUDY for hydrogen number density, $n$, in units of $\rm{cm}^{-3}$ of $\log(n)=$ 10, 11, 11.5, 12. We chose five times solar metalicity as is commonly assumed in BEL studies of quasars \citep{kor97}. As a test of our technique we take advantage of the quasi-simultaneous April 1992 observations of two BELs of 3C 279. We have both a low ionization line MgII and the CIV line (see Table 2 and Figure 2). The VBC line shifts and luminosities are fitted in Table 2. Based on our CLOUDY experiment, the simple one zone model shows agreement of the VBC line shifts of Equation (24) and the VBC luminosity for both lines at $\log(n)=11.5 - 12$ and an 80\% covering fraction as evidenced by the top left hand frame of Figure 13. For direct comparison of the VBC radiative efficiency from object to object, we convert this to a rest frame equivalent width, EW, using the local continuum from the SED in Figure 12. The CLOUDY generated EWs are plotted in the right hand frame of Figure 13. The dashed lines are the fitted values computed from Table 2 and the local SED continuum. We see that the preference for $\log(n)=11.5 - 12$ and an 80\% covering fraction in order to be consistent with both VBCs.
\subsection{Experiment 1: Comparing the BELs from the 3C 279 and Population B1 Ionizing Continua} Figure 13 validates that our CLOUDY analysis seems to provide reasonable results. Thus, we apply our methods to Experiment 1 from the introduction to Section 9. We compare the effect of the ionizing continuum of 3C 279 to that of quasars with conventional Eddington rates ($\sim 10\% - 20\%$) as represented by the Population B1 SED in Figure 12. The left hand panel of Figure 14 shows that parameters similar to those used in Figure 13, $\log(n)\approx 12$ and an 80\% covering fraction reproduce the VBC line shift and luminosity of MgII for both 3C 279 and the values for the Population B1 quasars \citep{mar13}. The right hand panel of Figure 14 makes the conversion to EW by dividing the luminosity by the local continuum flux density from the SED in Figure 12. It is very reassuring that the single zone model with a small spread in density and covering fraction is consistent with two types of quasars (Figure 14) and two ionization states of very broad emission lines (Figure 13). Thus, we proceed to interpret the results further. The right hand panel clearly shows that the VBC region of the Population B1 quasars is over-ionized for efficient MgII line emission as proposed in Section 8. This is not the case for 3C 279. The low Eddington rate and a steep EUV (associated with a powerful jet) in 3C 279 does not over-ionize the VBC near the black hole and it radiates MgII BELs efficiently.
\begin{figure}[ht]
\begin{center}
\includegraphics[width=100 mm, angle= 0]{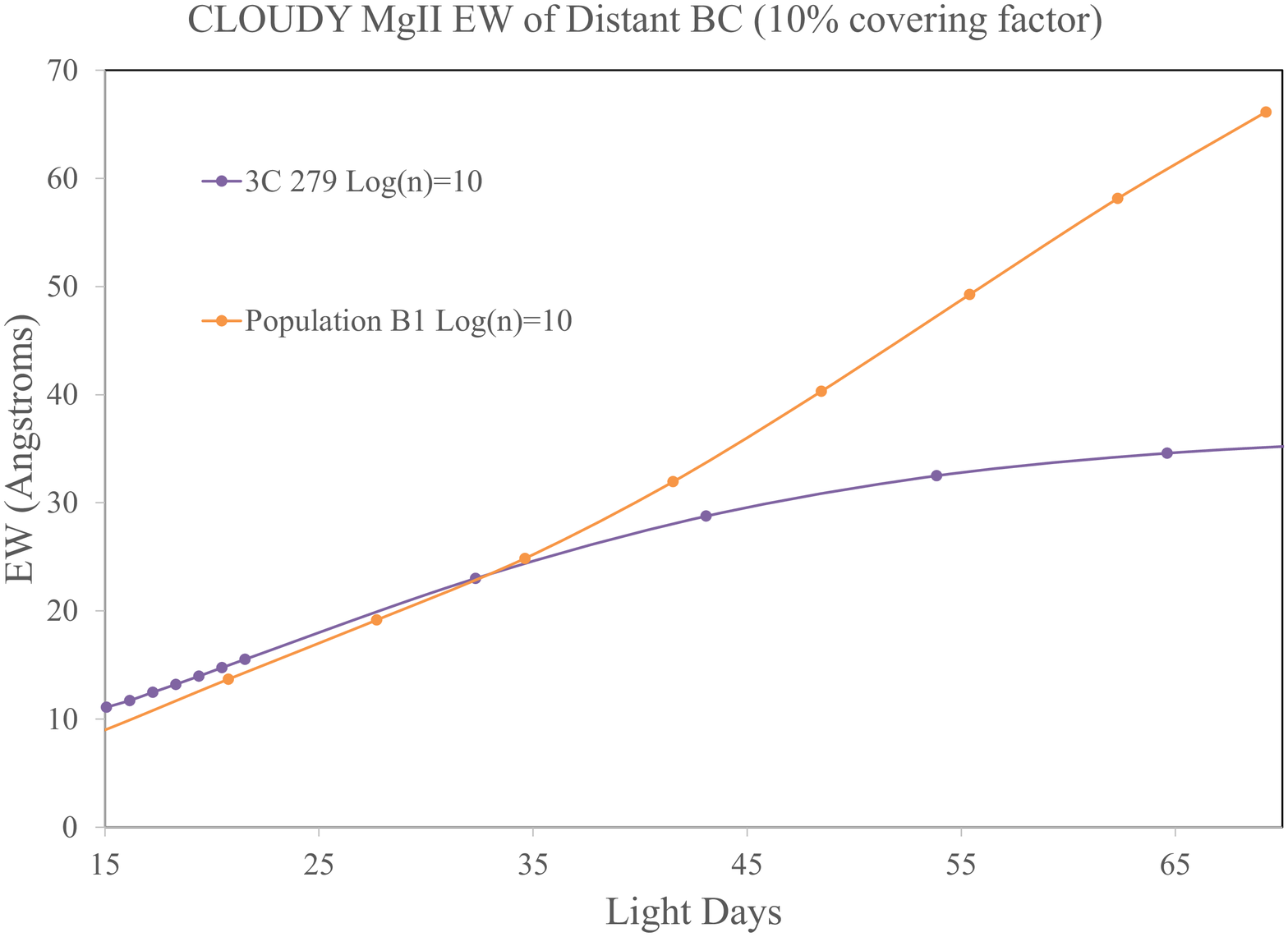}
\caption{A comparison of the MgII BC EW as a function of the number of light days from the quasar computed from the 3C 279 SED and the Population B1 SED in Figure 12. From Equation (26), we estimate that the MgII BC is $\sim70$ lt-days from the quasar in 3C 279. The Population B1 SED induces much more efficient MgII emssion than the 3C 279 SED $\sim70$ lt-days from the quasar. Apparently, the weak ionizing continuum of 3C 279 can under-ionize the distant BC gas for efficient MgII emission.}
\end{center}
\end{figure}

\subsection{Experiment 2: Comparing the BELs from the 3C 279 and a Low luminosity Radio Quiet Seyfert Galaxy}Experiment 1 combined two control variables. It contrasted quasars with relatively weak or weak jets and a modest accretion rate with 3C279, a quasar with a very strong jet and a low accretion rate. Thus, it is not clear if one or both of these differences created the much stronger VBC emission in 3C 279. In formulating Experiment 2, we note there are many low Eddington rate broad line radio quiet AGN, Seyfert 1 galaxies. They do not show highly RA UV BELs \citep{pun10,kin91,eva04}. Experiment 2 explores the origin of this phenomenon with only one control variable being altered. In particular, we contrast the CLOUDY VBCs induced by the SED of 3C 279 with those induced by the low Eddington state SED of NGC 5548 in Figure 12. Figure 15 plots the MgII VBC EW as a function of line shift from the CLOUDY experiment for the NGC 5548 low state, the Population B1 quasars and 3C 279. The EWs are similar to the Population B1 quasars, but even smaller. Apparently, the very hard ionizing continuum compensates for the low Eddington rate and over-ionizes the potential VBC emitting gas near the central black hole, with a line shift $>900$ km/s. In this experiment, the EUV deficit of radio loud quasars is the sole cause of the large luminosity increase of the highly redshifted VBC of 3C 279. The experiment emphasizes the importance of the EUV deficit of radio loud quasars in producing the RA UV BELs.

As a qualification to this experiment, it should be noted that the mean escape probability used by CLOUDY is a crude approximation to model the radiation transfer in the partially ionized zone where X-rays are absorbed. MgII is expected to be enhanced in the partially ionized zone, so the CLOUDY predictions might be less reliable for MgII than for a high ionization line. The ionizing continuum of NGC 5548 has a large X-ray contribution in Figure 12. Thus, we expect that this is the least accurate of the CLOUDY experiments. However, the over-ionization effect is so large in Figure 15 that we do not expect this shortcoming of our calculational method to alter the conclusion of the experiment.

\subsection{Experiment 3: Is the Distant BC Under-Ionized by the 3C 279 Continuum?} A second possible explanation of the large RA seen in the UV BELs of 3C 279 is that the BC is actually weak due to the fact that the ionizing continuum is of insufficient strength to photo-ionize distant gas, sufficiently. We do not know exactly where the BC is located due to the lack of a reverberation experiment. There are some estimates of the radial distance of the MgII BLR from the quasar, $R_{BLR}$, based on $L_{3000}$. For the SED of Figure 12 for 3C 279  $L_{3000} = 7.1 \times 10^{44}$ ergs/s. The estimator based on the SDSS sample with $L_{3000} < 10^{46}$ ergs/s of \citet{tra12} is the most relevant to 3C 279:
\begin{eqnarray}
&& \log{\left[\frac{R_{BLR}}{\rm{lt-days}}\right]} = (1.400\pm0.019)+(0.531\pm 0.015)\log{\left[\frac{L_{3000}}{10^{44}\rm{ergs/s}}\right]}=\nonumber\\
&& R_{BLR} = 71\pm 5 \, \rm{lt-days}\;.
\end{eqnarray}
It is not clear how appropriate Equation (26) is to the strangely shaped BELs in Figure 1. So the results of our CLOUDY analysis that implement this value of $R_{BLR}$ need to be viewed with caution. Using Equation (26) for our purposes is tantamount to assuming that the VBC is negligibly small in the BELs considered in SDSS sample with $L_{3000} < 10^{46}$ ergs/s \citep{tra12}. For 3C 279, 71 lt-days corresponds to a Keplerian velocity $\sim 3500$~km~s$^{-1}$.
\par We look at the CLOUDY results in the region around 70 lt-days from the quasar in 3C 279. These are compared to the MgII EW for the Population B1 quasars in the same region. Figure 16 shows the results of our experiment. A major source of uncertainty in this experiment is the hydrogen number density, $n$. We choose $n=10^{10}\rm{cm}^{-3}$, a ``standard" value for the MgII BEL that has been used in the past \citep{kor97}. The Figure shows that the 3C 279 SED begins to underproduce MgII BEL emission relative to the Population B1 SED $>45$ lt-days from the quasar. Thus, this is a suggestive experiment that the weak ionizing continuum of 3C 279 under produces the narrower BC components relative to more luminous radio quiet quasars.
\subsection{Summary of the CLOUDY Experiments}The simple single zone CLOUDY experiment established that the weak ionizing continuum of the blazars with highly RA UV BELs successfully explains their existence and the properties of the VBC. In particular, we showed
\begin{enumerate}
\item In Section 9.3 and Figure 13, it was shown that the 3C 279 SED explains the large redshifts of the VBC of both MgII and CIV as well as the large EWs.
\item In Section 9.4 and Figure 14, it was shown that the 3C 279 SED and the Population B1 SED explain the relative EWs of the MgII VBC in these objects. The combination of the low $R_{\rm{Edd}}$ and the EUV deficit associated with the powerful jet in 3C 279 conspire to lower the ionizing continuum so that the nearby VBC gas does not get over-ionized for efficient MgII emission. But, the experiment does not elucidate the relative importance of the two causative agents.
\item In Section 9.5 and Figure 15, it was shown that the 3C 279 SED and the NGC 5548 SED in a low state explain why the VBC is weak in low luminosity radio quiet Seyfert 1 galaxies even though it is strong in 3C 279 that has a similarly low value of $R_{\rm{Edd}}$. The EUV deficit associated with the powerful jet that diminishes the ionizing continuum allowing the nearby VBC to not be over-ionized in 3C 279. This experiment highlights the importance of the strong jet in the production of a strong VBC.
\item In Section 9.6 and Figure 16, it was shown that the 3C 279 SED not only creates a strong VBC that is highly redshifted, but the low ionizing flux also likely under-ionizes and therefore under-produces the conventional BC emission far way. The strong VBC is required for a highly RA BEL. The weak BC is a second order effect that can accentuate the asymmetry, significantly (see for example Figure 5).
\end{enumerate}

\par We note that the crude input SEDs in Figure 12 are not tightly constrained by observation and one might question if the inaccuracy in our SED approximations has led to a false conclusion in item 2. The driver of the redward asymmetry in the CLOUDY experiment is a depressed ionizing continuum. This cannot be observed directly in 3C 279 because of the dominant high frequency synchrotron tail of the jet. Considering the uncertainty in the SED for 3C 279, the CLOUDY experiment is not proof that this is the cause of the highly RA BELs. We note that the input SED for 3C 279 can reproduce the velocity shift and strength of the VBC for both MgII and CIV as indicated in Figure 13. But, without directly observing the SED of the accretion flow, we cannot rule out that this is coincidental. At a minimum, the CLOUDY experiments are a consistency check of the ingredients required for highly RA UV BELs indicated in Figures 6, 8 and 10.

\par In Appendix A, we show that the VBC profiles can be described by the accretion disk model of \citet{chenhalpern89}, if we restrict the emitting region to being close to the central black hole as indicated in the CLOUDY experiments.

\section{Comparison to Other Posited Explanations of RA UV Broad Lines}
Various explanations have been offered as to the origin of highly RA BELs. We compare and contrast these with our analysis in this section.
\begin{enumerate}
\item Model 1: Developing an idea proposed by \citet{pop95}, \citet{cor97,cor98}
suggested that the redward asymmetry arises from the VBC. It is a planar distribution of
optically thin gas that is located "near" the black hole,
$ r \sim$ 100 M -200 M. For a nearly face on orientation, the component of the Keplerian
velocity along the line of sight is small and the gravitational redshift is
comparatively large at $ r \sim$ 100 M -200 M, thereby shifting the VBC
emission significantly towards the red in blazars. This is the scenario that we have developed in this paper.
\item Model 2: The BEL emission has a large component emanating from an axial flow of
optically thick clouds. The unilluminated side of the clouds is
dusty gas and is not a BEL emitter. The illuminated side is ionized
and is a BEL emitter. Thus, we are viewing the backside of the
clouds from the axial outflow on the far side of the quasar. The
near side axial outflow is undetected \citep{net95,wbr96}.
\item Model 3: There is an optically thin axial inflow in RLQs that is a powerful source of BEL emission \citep{net95}.
\item Model 4: Optically thick gas at the inner edge of the
dusty torus radiates Fe II emission just redward of the BEL. These Fe II lines
are the actual source of the redward emission in the BEL
broad wing. Low inclination views are shielded from the Fe II
emitting gas on the near side by the dusty torus in the standard
model \citep{ant93}. For polar lines of sight all of the Fe II emitting
gas is visible, hence the stronger redwings in polar (blazar)
RLQs \citep{jac91,wbr96}.
\item Model 5: The BEL clouds are confined by a magnetized accretion disk wind. The blue BEL emission originates from the near side, approaching, faces of the BEL clouds. The red side of the observed BEL is light from the inner faces of the BEL clouds that is seen in reflection. The BEL emission from the inner face scatters off of electrons in the accretion disk corona back towards the observer \citep{emm92}. If the scattering is primarily at small $r$ then the red side can be redshifted significantly as discussed in Section 7 and might cause RA BELs.
\end{enumerate}

\begin{table*}[h]
\footnotesize{\scriptsize}
\caption{Compliance Matrix for the Explanations of Highly RA UV BELs}
\begin{tabular}{lcccccc}
\hline\hline\
Explanation    & Polar LOS & Strong Jet &   Low $R_{Edd}$ & VBC velocity  & Can red side of  \\
               & Required for & Required for & Required for & shift of peak & BEL be much more  \\
               & Strong RA & Strong RA &    Strong RA       & same as & luminous than blue   \\
               &            &          &                    & Pop. B quasars & side of BEL?  \\
\hline
Model 1   & Yes   &  Yes & Yes  &  Yes &  Yes\\
Model 2   & Yes  &  No & No & No  &  Yes \\  		
Model 3 & Yes  &  No & No & No  &  Yes \\
Model 4 & Yes  &  No & No & No  &  Yes \\
Model 5   & No   &  No & No & Yes  &  No \\

\hline
\end{tabular}
\tablenotetext{}{Yes means compliant. No means non-compliant}
\end{table*}
\par Table 8 is a direct comparison of the desired properties of a putative explanation of highly RA BELs for the various models. This is the logic table that allowed us to down select to Model 1. Columns (2) - (6) are the properties that we motivated in this paper as being the constraints on a plausible explanation of highly RA BELs as discussed in Section 5. Column (2) is motivated by the fact that all the known sources with highly RA UV BELs are blazars \citep{wbr96,pun10}. Columns (3) and (4) are motivated by Figures 6 and 8. The requirement in column (5) is from \citet{mar13} and Table 2 as discussed in Section 5. The requirement in column (6) is motivated by the BEL line fits in Table 2 and the line profiles in Figures 1 - 4. Only Model 1 complies with all these requirements. Models 2 and 3 do not require a strong jet or low $R_{Edd}$. The velocity of the VBC peak shifts vary with the cosine of the line of sight to the axial flow, so it is larger ($\sim 20\%$) for blazars compared to Population B quasars. This violation is minor, so in of itself, it is not determinant. Model 4 has no dependence on jet strength or $R_{Edd}$. There is also no reason to believe that the VBC peak shift would be the same if different regions of the FeII BLR are exposed by different lines of sight (i.e., $\sim 30^{\circ}$ for Population B quasars versus a polar line of sight). Model 5 would have difficulty explaining the much stronger red wings than blue wings in UV BELs in column (6). It also has no obvious dependence on jet strength, $R_{Edd}$ or the LOS. Based on Table 8, Model 1 would appear to be the preferred explanation, but it does not categorically rule out all other explanations.
\section{Discussion and Concluding Remarks}Previous research has shown that RA UV BELs are associated with radio loud quasars and extreme asymmetry is associated with blazars in particular \citep{wbr96,pun10}. The motivating principle of this study is that the highly RA BELs are paradoxically found in blazars (objects known for possessing the most extreme blue shifts in the Universe), and this paradox is a consequence of a fundamental physical process associated with relativistic jet launching and is not a coincidence.
\par Our investigation began empirically. In Section 4 (Figures 6 and 8), we found that the BEL blazars with the most RA BELs were those that had both an extremely low $R_{\rm{Edd}}$ and a very large ratio of long term, time-averaged jet power to accretion flow thermal luminosity, $\overline{Q}/L_{\rm{bol}}$. In Section 5, we summarized the elements required for the highly asymmetric lines:
\begin{enumerate}
\item A low Eddington rate
\item A strong jet relative to $L_{\rm{bol}}$
\item A polar line of sight
\end{enumerate}
\par In Section 7, we used the enormous redshifts of the VBC in these asymmetric lines to locate the CIV VBC at $\approx 90$M and the MgII VBC $\approx200$M from the central black hole in 3C 279. In Section 8, we noted that this requires a very weak photo-ionizing continuum, otherwise the gas will be over-ionized near the black hole the and the VBCs will be weak. So, the investigation becomes one in which we try to understand what makes a quasar continuum a weak photo-ionizing source. Clearly a lower $R_{\rm{Edd}}$ will produce a weaker photo-ionizing field for a given $M$. We also noted a phenomenon know as the EUV deficit of radio loud quasars \citep{pun14,pun15}. The larger $\overline{Q}/L_{\rm{bol}}$, the steeper (weaker) the EUV continuum as evidenced by the plot in Figure 10. 3C 279 has one of (if not the) largest known values of $\overline{Q}/L_{\rm{bol}}$. We verified that it is these two elements that will produce the RA BELs (a luminous, highly redshifted VBC) with a series of CLOUDY numerical experiments in Section 9. Equivalently, we have identified two elements that create a weak photo-ionizing continuum and a luminous VBC at large redshift:
\begin{enumerate}
\item A low Eddington rate
\item A strong jet relative to $L_{\rm{bol}}$
\end{enumerate}
This is almost identical to the empirical properties of the RA blazars that are listed in the paragraph just above, but it does not include the last item: the polar line of sight. We argue that this becomes relevant because it creates a selection effect. The low $R_{\rm{Edd}}$ broad line radio loud AGN have enormous Doppler enhancement of the radio flux due to the polar line of sight. Thus, some rather nondescript sources appear as very bright and therefore qualify for flux limited samples of strong radio sources. If they were viewed off axis they would appear as modest lobe dominated radio sources with a weak optical core. The synchrotron spectrum would be faint and an observer at Earth would detect the accretion flow flux in the optical band with negligible contamination from the jet emission. Based on our crude SEDs in Figures 11 and 12, we estimate a Johnson $m_{v}=19.5$ ($m_{v}=20.3$) for 3C 279 (0954+556) if it were viewed off axis (i.e. $\sim 30^{\circ}$). Furthermore, if the BEL gas is flattened distribution of gas that rotates in the gravitational potential, for a typical quasar line of sight, from Equation (14),
\begin{equation}
3\,[\rm{FWHM(blazar\, LOS)}]\approx \rm{FWHM(steep\, spectrum\, quasar\, LOS)}\;.
\end{equation}
Since the VBC flux is independent of the line of sight, the amplitude of the VBC Gaussian will be reduced by a factor $\approx 3$ in order to accommodate the $\approx 3$ times larger FWHM. The VBC FWHM would be $\sim$ 15,000~km~s$^{-1}$. In summary, these quasars would appear with very faint (but blue) optical cores and very squat BEL profiles. These spread out BELs would make a small contrast relative to the underlying continuum and the FeII emission. A good example of such a wide MgII BEL in a lobe dominated quasar is 3C 68.1 \citep{smi80,aar05}. Considering this example and noting the weak optical core for these blazars, they might be identified as radio galaxies if viewed off axis. However, in practice, the diminished prominence of the BEL would make a VBC/BC decomposition impractical. It would be extremely challenging to get an observation good enough to reliably determine the degree of redward asymmetry, AI and C(1/4) from Equations (2) and (3).

This discussion allows us to understand the nature of the selection effects that associate the most RA UV BELs with the polar line of sight. There are two cases.

\begin{itemize}
\item There are Fanaroff-Riley II, FRII, (see \citet{fr74} for a definition) steep spectrum quasars with modest RA UV BELs \citep{pun10,bar90}. However, the asymmetry is not as strong as in Figure 1. For these to be known quasars, $R_{\rm{Edd}}$ is typically at least an order of magnitude larger than that of 3C 279. Thus, the redward asymmetry is only driven by one of the two factors, the EUV deficit of radio loud quasars.
\item The other possibility are the FRII quasars with $R_{\rm{Edd}}\sim 0.01$. However, as noted above if viewed off axis they might be considered radio galaxies. Even if the squat broad lines were identified, it would not likely be possible to do more than a crude single Gaussian fit with no asymmetry information. We note that since the highly RA BELs are associated with quasars at the very high end of the $\overline{Q}/L_{\rm{bol}}$ distribution, they are rare \citep{pun07}. We searched the literary archives for $m_{v}\sim20$, 4C and 6C radio sources in order to try and find an object with squat RA BELs, but have been unsuccessful.
\end{itemize}

\appendix

\section{Relativistic accretion disk model fits to the blazar CIV and MgII VBC}\label{ad}
{The relativistic line shifts computed in Section 7 and the CLOUDY results in Section 9 indicate a mutually consistent dynamic in which the CIV (MgII) VBC emission originates from very close to the central black hole, $r \sim 90$M ($r \sim 225$M). The accretion disk itself (or a portion thereof) is such a distribution of gas capable of line emission if illuminated by ionizing radiation. Thus, it is of interest to explore a parametric model of the accretion disk as the source of the VBC \citep{chenhalpern89}. It is beyond the scope of this work to justify if the model results from a physically consistent distribution of gas, ionization state and emissivity. The model, through its parametrization, does determine an emissivity profile that is consistent with the VBC profiles in Table 2 and Figure 2, 3 and 5.}

\par The panels of Figure \ref{fig:ad} shows the CIV and MgII {total BEL (in black) and VBC (in red)} profiles of the extreme blazars 3C 279, 0954+556, 3C345 and 1803+784 with superimposed model profiles (in orange) of a relativistic, geometrically thin accretion disk  \citep{chenhalpern89}.  The model parameters are reported in Table \ref{tab:ad} that lists in the following order: source, line fitted, exponent of power-law representing radial disk emissivity ($\epsilon \propto r^{-a}$), local broadening parameter $q$, reference wavelength, inner radius $r_\mathrm{in}$\ {of the emitting region} and outer radius $r_\mathrm{out}$ {of the emitting region}\ (both in units of gravitational radii), and inclination of the accretion disk axis with respect to the line-of-sight.

The {chosen parametric disk model} profiles reproduce the empirical VBC {very accurately in all instances}. The models can fully account for the extended red wing typical of the extreme blazars (Fig. \ref{fig:ad}) for both CIV and MgII. From Table \ref{tab:ad} we derive that:

\begin{itemize}
    \item Orientation values are always less or equal 5 degrees. This result is consistent with the blazar nature of the sources, and all blazars are believed to be oriented with the jet/disk axis close to the line of sight \citep{urrypadovani95}.
    \item The inner radius of the emitting {portion of the} disk has to be very small, less than 100 gravitational radii. The velocity field of a geometrically thin, optically thick accretion disk is expected to remain almost Keplerian down to the innermost stable orbit. {From Equation (24), $r_\mathrm{in}<100 $M for the CIV VBC and $r_\mathrm{in}<225 $M for the MgII VBC in 3C 279 and 0954+556 based on kinematic constraints. This was also corroborated by the CLOUDY results in Figure 13. The values of $r_\mathrm{in}$ from the disk models in Table 8 are consistent with these bounds}.
    \item To account for a large {\em} peak shift, the $r_\mathrm{out}$ should be relatively small, or the emissivity extremely steep ($a\gg1$). The small inclination ensures that the blue Doppler boosted peak and the red peak are blended together.
    \item The $q$\ local broadening parameter is relatively large with respect to the original value derived by \citet{chenhalpern89} for Arp 102B. This ensures a smooth, almost Gaussian-like appearance for the disk model.
\end{itemize}

\begin{table*}[h]
\tabletypesize{\scriptsize}
\caption{Accretion disk parameters \label{tab:ad}}
\begin{tabular}{lcccccccc}
\hline\hline\
Source     & Line & $q$\tablenotemark{a} &    $\sigma$\tablenotemark{b} & $\lambda_0$ & $r_\mathrm{in}$ & $r_\mathrm{out}$ & $i$ \\
\hline
3C 279   & CIV   &  3.5 & 8.0E-3 & 1549. &  60& 10000&  5 \\
3C 279   & MgII  &  2.0 & 7.5E-3 & 2800  &  80&  1100&  4 \\  		
0954+556 & CIV   &  3.5 & 5.0E-3 & 1549  &  60& 10000&  5 \\
0954+556 & MgII  &  1.5 & 9.0E-3 & 1549  & 100&   800&  5 \\
3C 345   & CIV   &  3.5 & 8.0E-3 & 1549  &  80& 10000&  5 \\
3C 345   & MgII  &  1.3 & 9.0E-3 & 2800  &  80&   800&  2 \\
1803+784 & MgII  &  2.0 & 5.0E-3 & 2800  &  60&   700&  2 \\
\hline
\end{tabular}
\tablenotetext{a}{Exponent of the radial emissivity law $\epsilon(r) \propto r^{-q}$.}
\tablenotetext{b}{~Local broadening parameter defined by \citet{chenhalpern89}.}
\end{table*}

\begin{figure}[htp!]
\begin{center}
\includegraphics[angle= 0,scale=0.4]{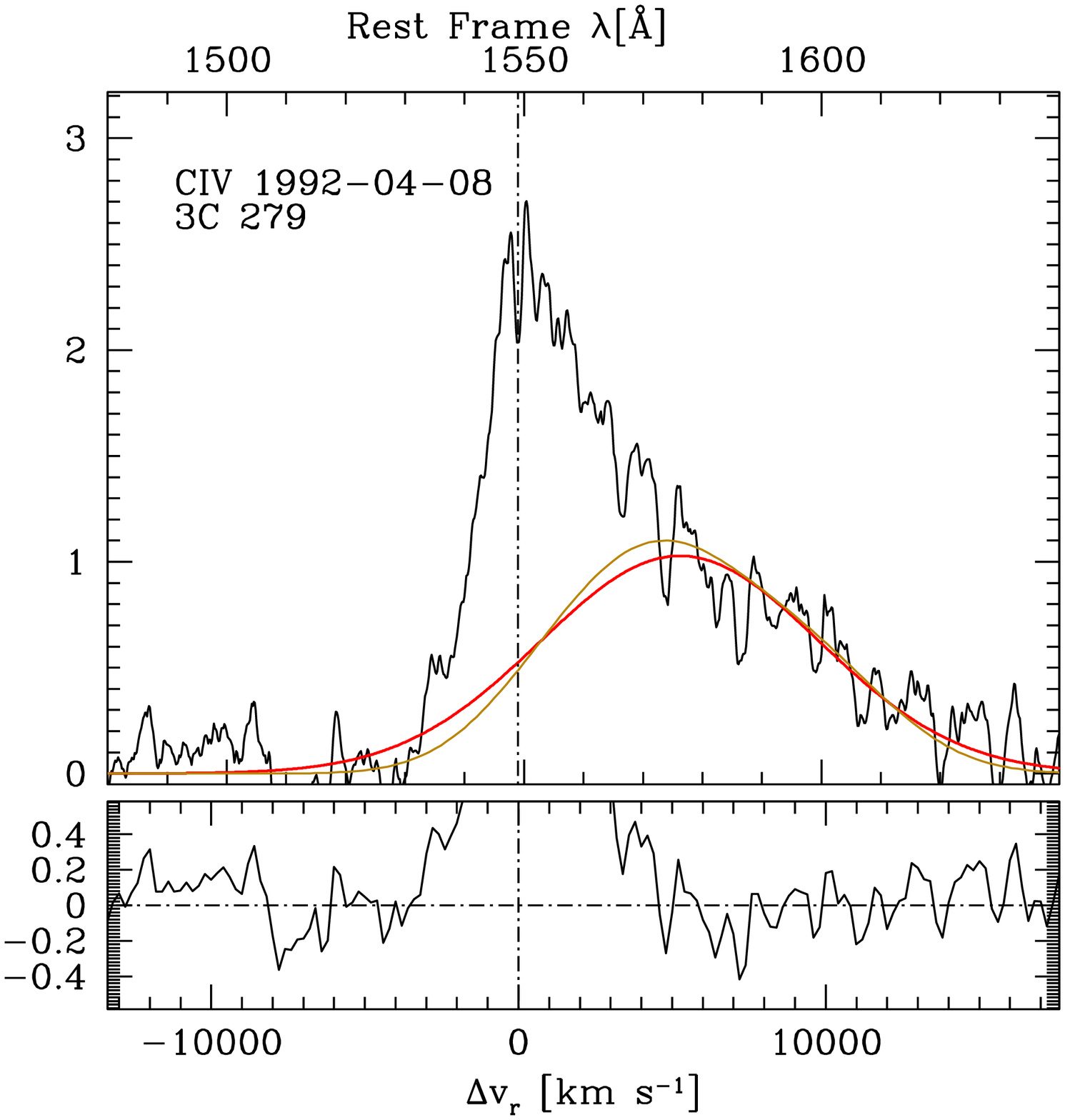}
\includegraphics[angle=0,scale=0.4]{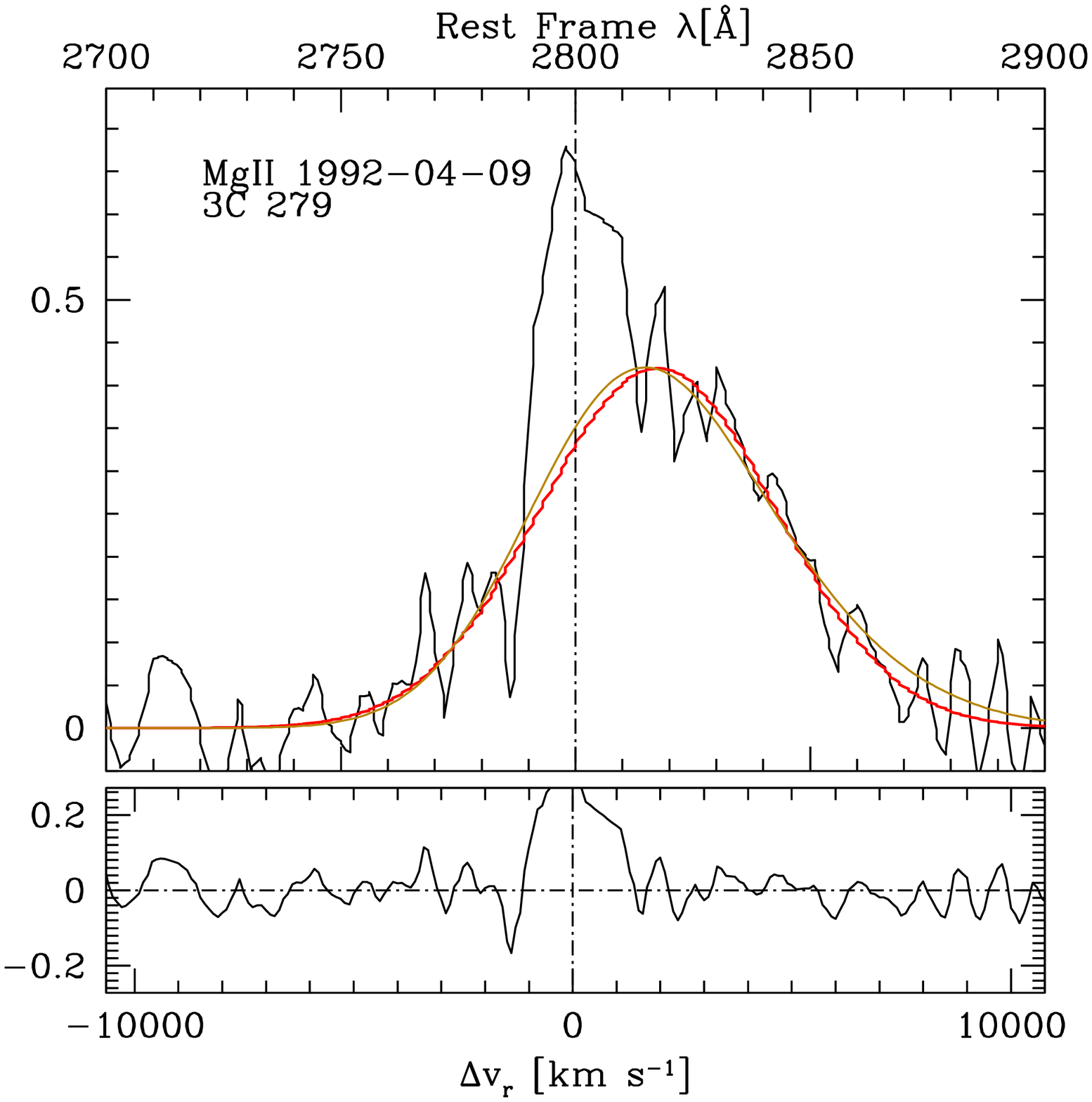}\\
\includegraphics[angle= 0,scale=0.4]{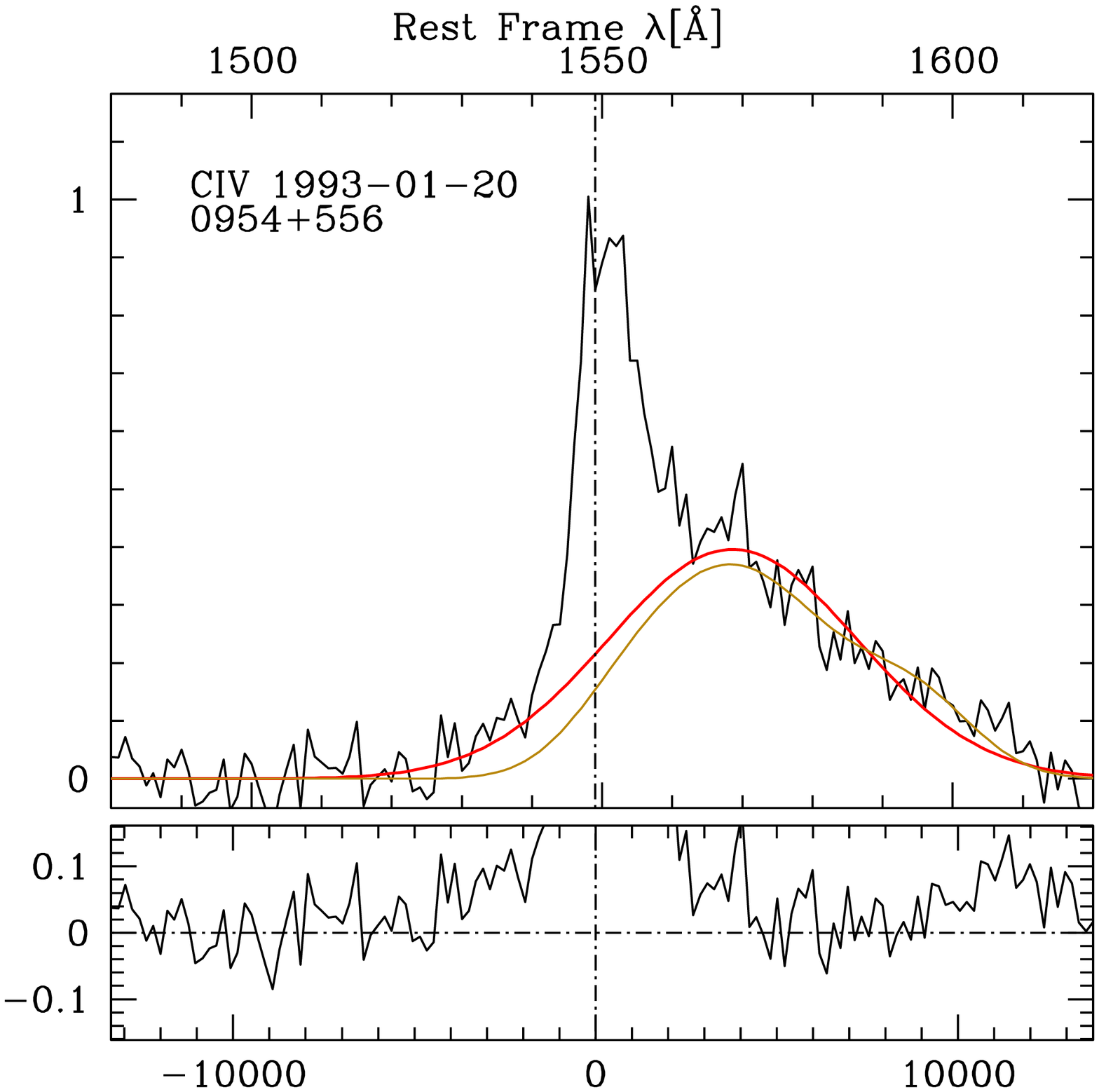}
\includegraphics[angle=0,scale=0.4]{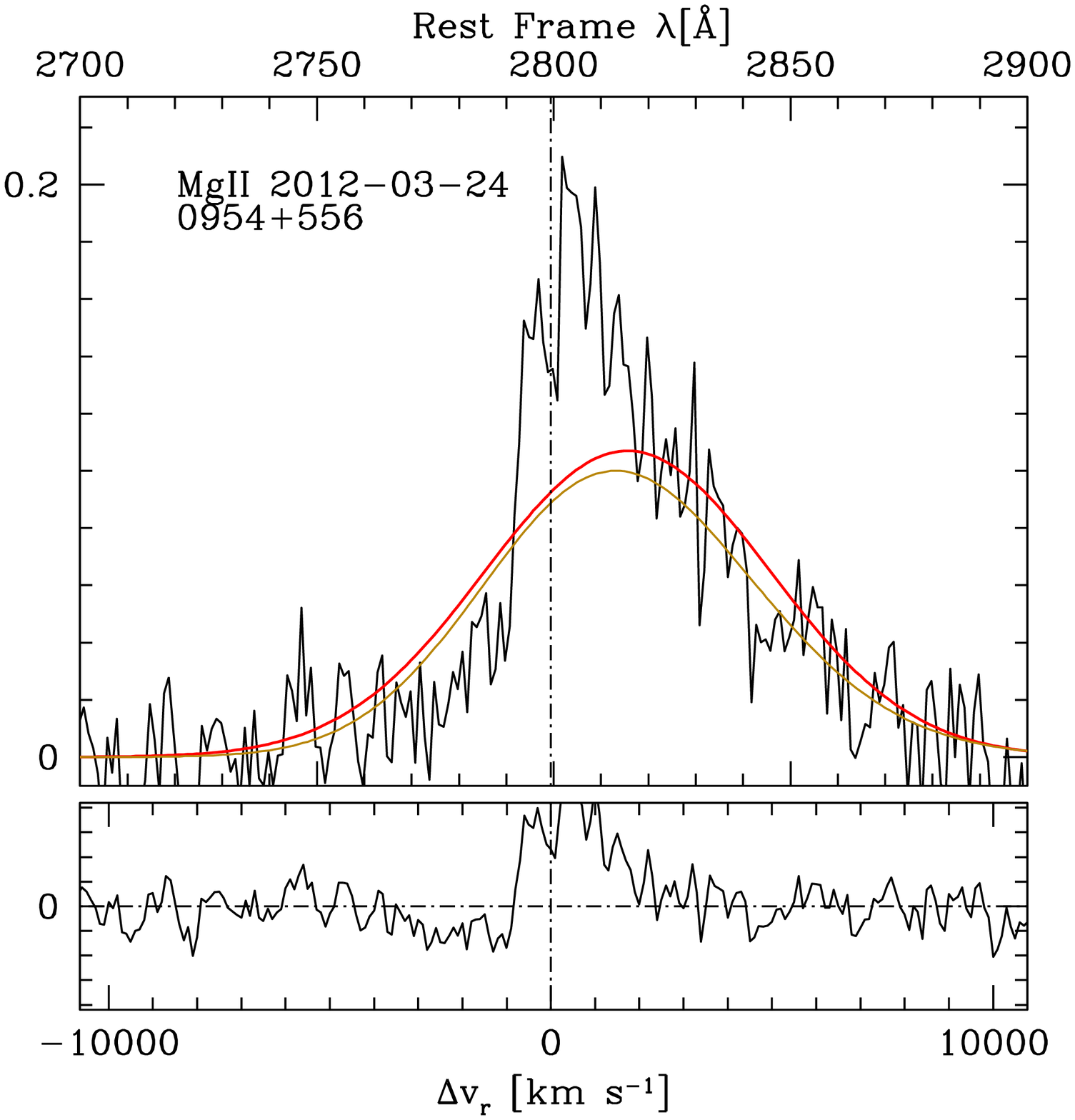}\\
\caption{Relativistic accretion disk model profiles and empirical very broad components for CIV (left column) and MgII (right). The red line traces the VBC derived from the multi-component line fit in Table 2 and the orange profile is the accretion disk model fit whose parameters are reported in Table \ref{tab:ad}. \label{fig:ad}}
\end{center}
\end{figure}

\addtocounter{figure}{-1}

\begin{figure}[htp!]
\begin{center}
\includegraphics[angle= 0,scale=0.4]{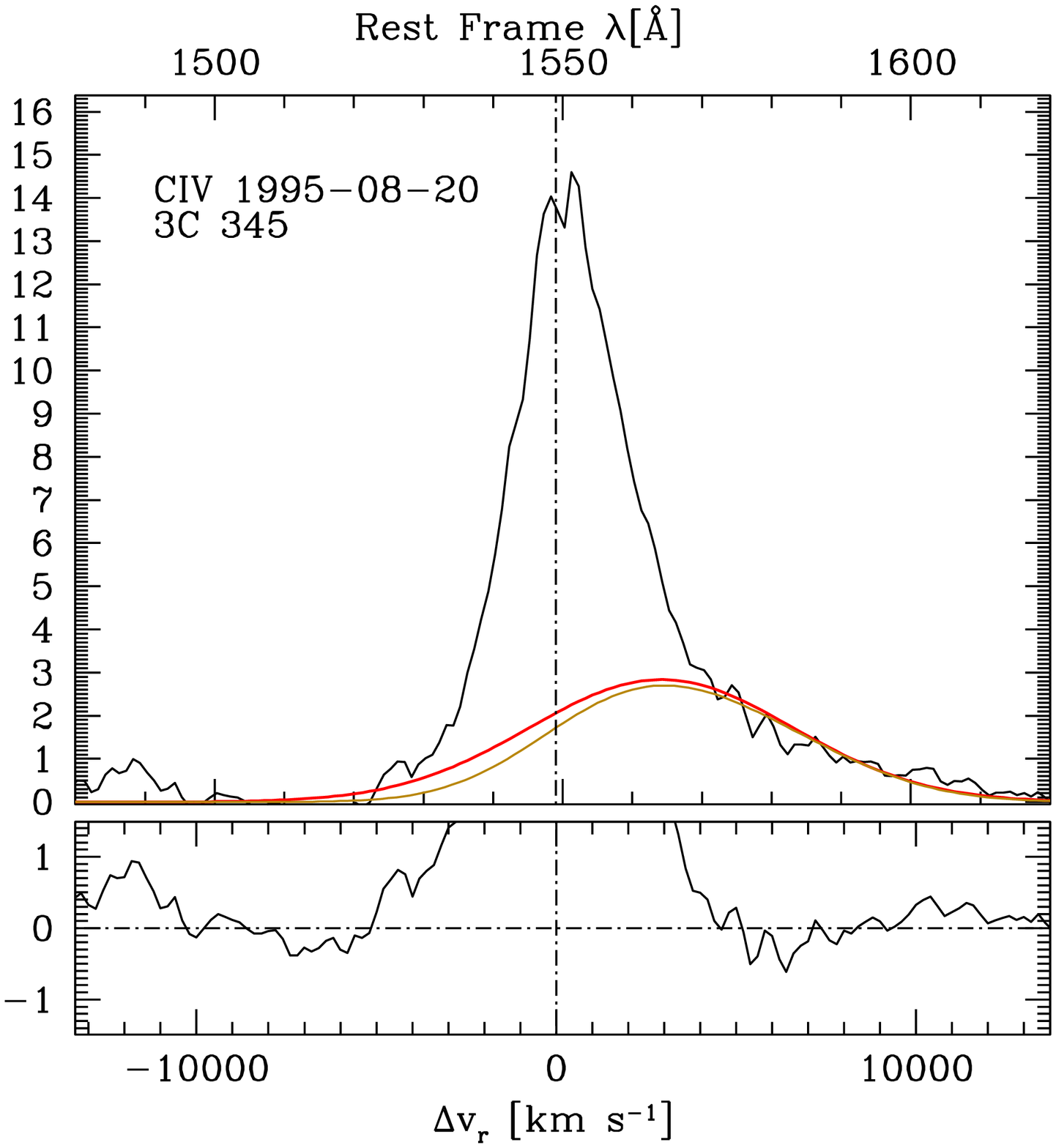}
\includegraphics[angle= 0,scale=0.4]{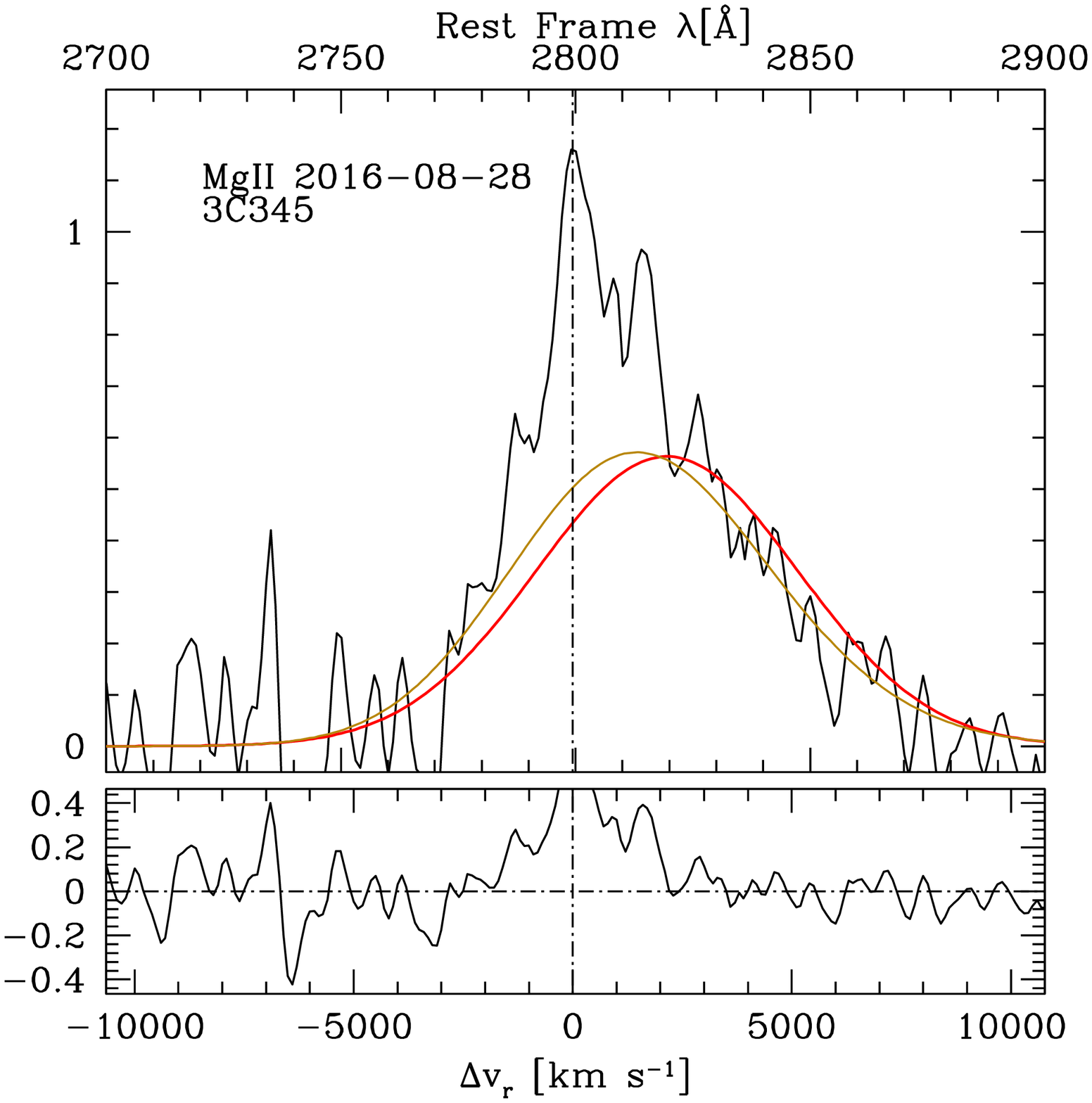}\\
\includegraphics[angle= 0,scale=0.4]{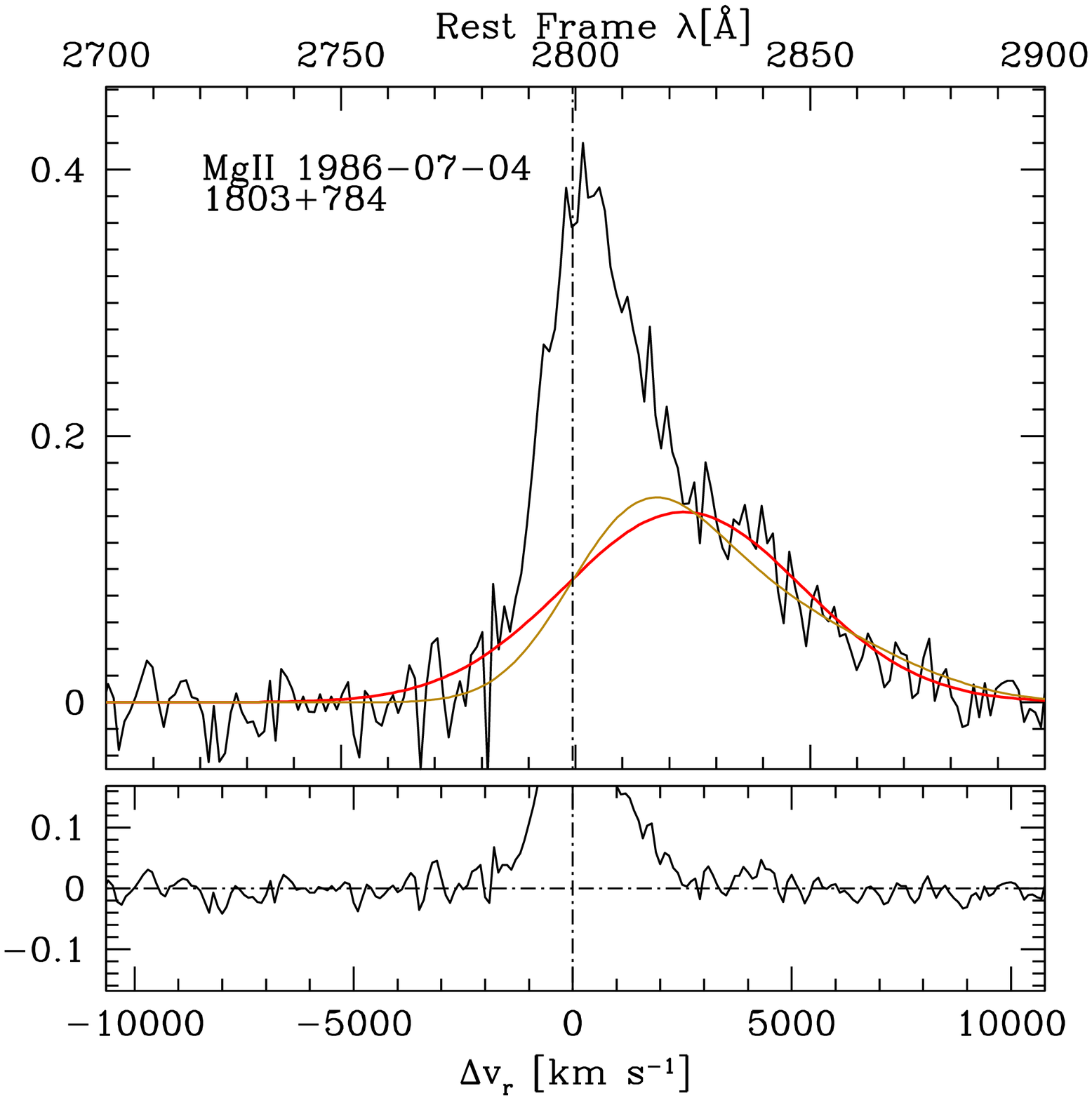}\\
\caption{Cont.}
\end{center}
\end{figure}

\pagebreak

\section{Detailed Lobe Luminosity Estimates for Individual Sources}

This Appendix provides new radio images and data analyses in order to supplement the jet power estimates in Section 3.

\subsection{The Extended Diffuse Flux of 3C279} Even though 3C279 is
perhaps the best known example of the blazar phenomenon, no deep
image of the extended emission has ever been published. The best
image that we have is a C-band image with the early Very Large
Array (VLA) \citep{pat83}. Due to the steep spectrum nature of extended
flux, C-band ($\sim5$~GHz) images tend to resolve-out diffuse emission and require
higher dynamic range than L-band ($\sim1.4$~GHz) images (the core being flat
spectrum contributes more to the total flux density at high
frequency). Equation (10) requires deep images for the
proper extraction of the isotropic diffuse extended flux. To this
end, we present the first deep L-band image of 3C 279 in Figure 18.
The observation was with the VLA in A-array on January 11, 2001.
The observing frequency was 1.665 GHz and the image was restored
with a 1.5$\arcsec$ circular beam.

\begin{figure}
\centerline{
\includegraphics[width=10cm]{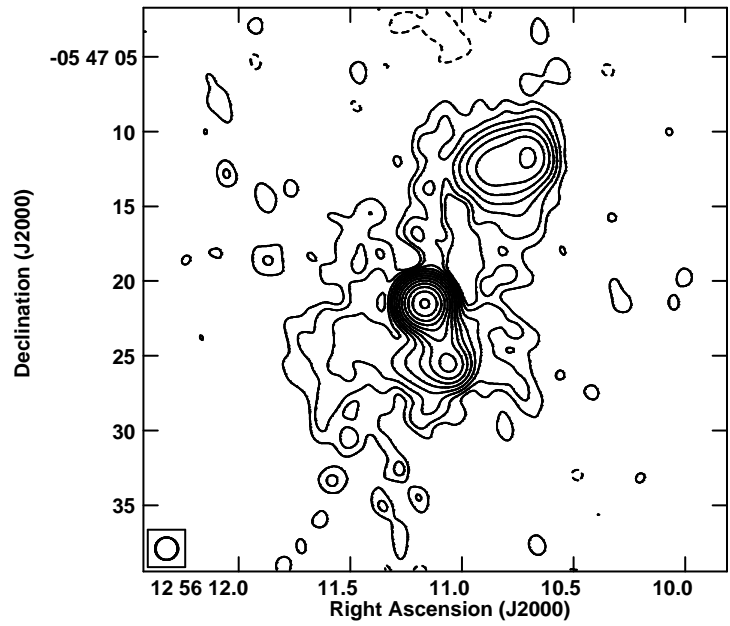}}
\caption{The 1665 MHz image of 3C 279. The radio contours are in percentage of peak surface brightness (= 10.2~Jy~beam$^{-1}$) and increase in steps of 2, with the lowest contour being $\pm0.02$\%. The jet points toward the
south. Note the luminous radio lobe to the north and the diffuse
halo type emission.}
\end{figure}

\par The image is useful for extracting the isotropic extended
emission. The beamed emission originates in the core and possibly in
the kpc jet to the south. A conservative estimate of the isotropic
diffuse flux is therefore obtained by subtracting the core plus jet
flux density (11.11 Jy) from the total flux (12.60 Jy) in order to
yield an extended isotropic flux density of 1.49 Jy. Figure 18 shows
a large amount of diffuse structure. Thus, the A-array observation
has undoubtedly resolved-out some diffuse emission. To properly
constrain the total lobe flux density would require a P-band
observation with the new correlator of the Jansky VLA. An L-band
B-array observation would also be helpful. For now, we must work
with the L-band image in Figure 18 in order to extrapolate the
emission to 151 MHz. Considering the large diffuse cloud of emission
and the conservative estimate above, a spectral index of $\alpha=1$
seems reasonable for the extrapolation from 1.665 GHz \citep{kel69}.
This yields 16.4 Jy at 151 MHz for the isotropic emission. From
Equation (10) with $10<\mathcal{F}<20$, this equates to an estimate of the long term
time averaged jet power of $\overline{Q} = 9.05 \pm 3.01 \times
10^{45} \rm{ergs/s}$. Based on the deep observations of blazars in
\citet{mur93}, 3C279 has one of the largest $\overline{Q}$ of any
known blazar.
\subsection{The Extended Diffuse Flux of 0954+556} Fortunately,
excellent images of this blazar exist in the literature. The 408 MHz
MERLIN observation in \citet{rei95} is of closely matched resolution
with the \citet{mur93} 1.552 GHz VLA observation. This allows us to
compute a spectral index of 0.95 for the extended emission. The
object appears to be a triple that is viewed along the jet axis
making the object appear much more compact than it actually is. The
symmetry of the object indicates that there is very little if any
Doppler beaming on kpc scales. Extrapolating the extended 408 MHz
flux density of 1.339 Jy to 151 MHz with a spectral index of 1
yields a flux density of 3.26 Jy. From Equation (10) with $10<\mathcal{F}<20$, this equates to an estimate of the long term
time averaged jet power of $\overline{Q} = 7.25 \pm 2.42 \times
10^{45} \rm{ergs/s}$. As for 3C279, this is an extremely large time averaged jet power for a blazar.

\subsection{The Extended Diffuse Flux of 1803+784} The source
1803+784 does not have the large isotropic flux that occurs in the
other three RA blazars that are discussed in this section. It has a
very distant secondary 45" away that has been shown to be connected
to the powerful radio core \citep{bri05,ant85}. There is a large
halo of very diffuse radio emission that has never been detected by
the VLA \citep{bri05}. The distant secondary is resolved and the
measured flux density in L-band depends on the beam size and u-v
coverage used with the VLA. The flux density of the secondary was
found in \citet{mur93} in the combined A and B array observations
which were optimized for u-v coverage to be 67 mJy at 1.552 GHz. No
halo emission was detected. We looked at archival P-band VLA
observations in order to see evidence of the halo. The observations
were filled with artifacts and were not useable. We conclude that a
deep P-band image with the new correlator of the Jansky VLA is
required to in order to detect the large faint halo emission. Since
the halo flux was undetected, the flux density of the secondary at
1.552 GHz will yield a conservative estimate of the isotropic flux
density at 151 MHz. Thus, we choose a steep spectral index of
$\alpha =1$ for the extrapolation to 0.69Jy of isotropic flux
density at 151 MHz.  From
Equation (10) with $10<\mathcal{F}<20$, this equates to an estimate of the long term
time averaged jet power of $\overline{Q} = 9.81 \pm 3.27 \times
10^{44} \rm{ergs/s}$.
\subsection{The Extended Diffuse Flux of 3C 345} A deep L-band VLA image
has been previously published \citep{mur93}. However, the data
reduction does not separate the isotropic extended flux from the
dominant one sided jet (which is likely beamed). Thus, we use the
archival observation A-array observation at 1.552 GHz from December 23 1984 in order to
estimate the isotropic extended flux. The image is restored with a
1.5" circular beam and is presented in Figure 19.

\begin{figure}[ht]
\begin{center}
\includegraphics[width=10cm]{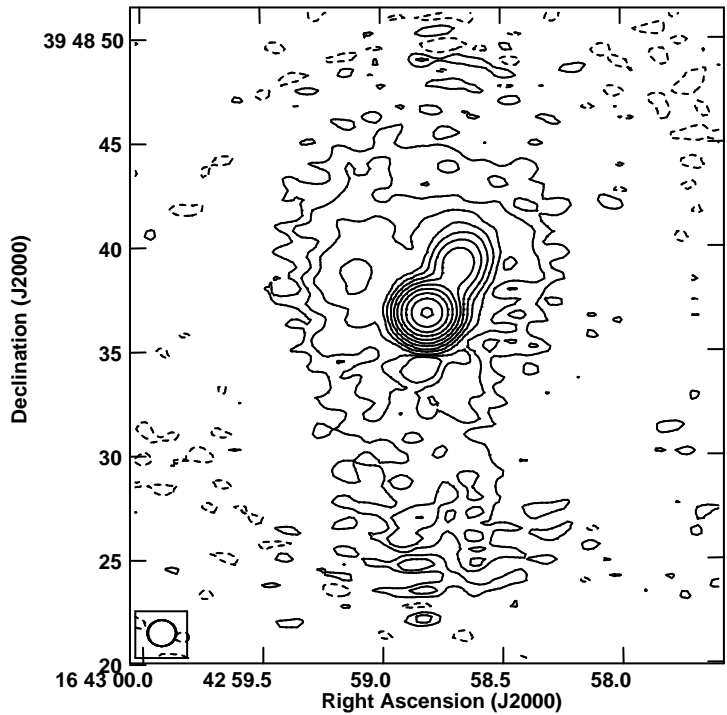}
\caption{The 1552 MHz image of 3C 345. The radio contours are in percentage of peak surface brightness (= 7.7~Jy~beam$^{-1}$) and increase in steps of 2, with the lowest contour being $\pm0.042$\%. Note the diffuse halo type
emission that surrounds the core-jet structure.}
\end{center}
\end{figure}

The total flux density in Figure 12 is 8.934 Jy. The flux density of
the (Core + Jet/Northern Hotspot) is 8.639 Jy. The extended
isotropic flux density is estimated as Total - (Core + Jet/Hotspot)
is 0.295 mJy. This is primarily very diffuse halo type emission and
is likely very steep spectrum. It is likely that more of this flux
would be detected in B and C array. Thus, we choose $\alpha =1$ to
extrapolate to the isotropic flux density to 151 MHz. Clearly a much
better estimate could be achieved with the aid of a new P-Band
Jansky VLA image. Based on the current image and data reduction we
find that $F_{151} =3.03 \rm{Jy}$. From
Equation (10) with $10<\mathcal{F}<20$, this equates to an estimate of the long term
time averaged jet power of $\overline{Q} = 2.59 \pm 0.86 \times
10^{45} \rm{ergs/s}$.
\subsection{The Extended Diffuse Flux of 3C 454.3}
\par One object that did not have an existing deep radio image that
revealed the diffuse isotropic extended emission is the famous
blazar, 3C 454.3. Even the excellent observations of \citet{mur93}
only reveal a core-jet morphology with the jet directed towards
Earth with a strong possibility of significant Doppler beaming on
kpc scales \citep{pun95}. In Figure 20, never before published images
form the VLA in A-array observation on August 10, 1999 at L-band are
presented. We show images at slightly different frequencies (1435
MHz and 1365 MHz), in order to explore possible artifacts in diffuse
structure.

\begin{figure}
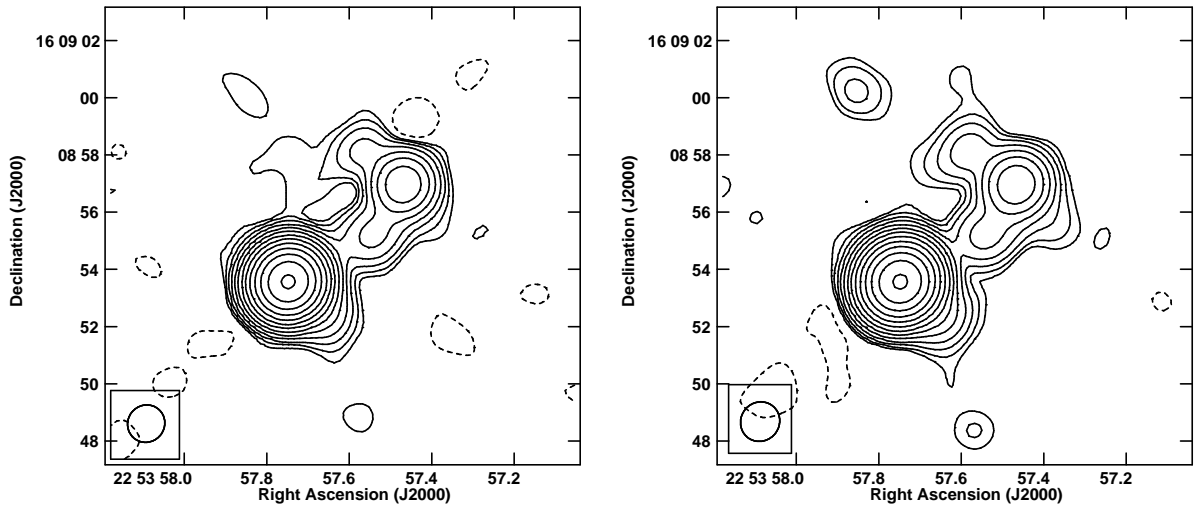

\begin{center}
\includegraphics[width=8cm]{f35.eps}
\includegraphics[width=8cm]{f36.eps}
\caption{1365~MHz (left) and 1435~MHz (right) images of 3C 454.3. The radio contours are in percentage of peak surface brightness (left: 13.35~Jy~beam$^{-1}$, right: 13.32~Jy~beam$^{-1}$) and increase in steps of 2, with the lowest contour being $\pm0.02$\%. Comparing the two panels
indicates that there is an artifact to the northeast of the hot spot in the left image.}
\end{center}
\end{figure}
\par Compare the left hand panel in Figure 20 at 1365 MHz with the right hand panel. There
is a prominent feature to the northeast of the jet terminus that is
much weaker in the lower 1435 MHz image. This is considered an
artifact. By contrast, the excess emission surrounding the jet
terminus which is pronounced towards the east is apparent at both
frequencies is concluded to be inherent to 3C 454.3. We estimate a
flux density of 51 mJy at 1.4 GHz for this feature. There is also a
feature the protrudes to the west of the core (a possible
counter-jet) that appears in both images and is likely associated
with 3C 454.3. We estimate a flux density of 11 mJy at 1.4 GHz for
this feature. The combined extended flux density is 62 mJy at 1.4
GHz. From
Equation (10) with $10<\mathcal{F}<20$, this equates to an estimate of the long term
time averaged jet power of $\overline{Q} = 1.36 \pm 0.45 \times
10^{45} \rm{ergs/s}$.

\begin{thebibliography}{}
\bibitem[Aars et al.(2005)]{aar05} Aars, C. E.; Hough, D. H.; Yu, L. H.; Linick, J. P.; Beyer, P. J.; Vermeulen, R. C.; Readhead, A. C. S.. 2005, AJ 130 23
\bibitem[Abdo et al.(2009)]{abd09} Abdo, A. et al. 2009, ApJ 700 597
\bibitem[Antonucci(1993)]{ant93} Antonucci, R.J. 1993, Annu. Rev. Astron. Astrophys. 31 473
\bibitem[Antonucci and Ulvestad(1985)]{ant85} Antonucci R., Ulvestad J., 1985, ApJ 294, 158
\bibitem[Balmaverde and Capetti(2014)]{bal14}Balmaverde, B. and Capetti, A. 2014 A\&A 563 119
\bibitem[Barthel (1989)]{bar89} Barthel, P. 1989, ApJ 336 606
\bibitem[Barthel (1990)]{bar90} Barthel, P., Tytler, D., Thomson, B. 1990, A\&A Supp. 82 339
\bibitem[Beckmann et al.(2009)]{bec09} Beckmann, V.; Soldi, S.; Ricci, C. et al. 2009, A\&A 505 417
\bibitem[Bentz et al.(2009)]{ben09} Bentz, M., Walsh, B., Barth, A. et al. 2004, ApJ 705 199
\bibitem[Berton et al.(2018)]{ber18} Berton, M., Liao, N., La Mura, G. et al, 2018, A\& A 614 148
\bibitem[Best et al.(1999)]{bes99}Best, P., Rottgering, H., Lehnert, M. 1999 MNRAS
310 223
\bibitem[Bicknell et al.(1998)]{bic98}Bicknell, G., Dopita, M., Tsvetanov, Z., Sutherland,
R. 1998 ApJ 495 680
\bibitem[Britzen et al (2005)]{bri05} Britzen, S., Krichbaum, T., Strom, R. et al, 2005
444 443
\bibitem[Blundell and Rawlings(2000)]{blu00} Blundell, K., Rawlings, S. 2000 AJ 119 1111
\bibitem[Brotherton (1996)]{bro96}Brotherton, M. 1996, ApJS 102
1
\bibitem[Brotherton et al(1994)]{bro94}Brotherton, M., Wills B., Steidel, C., Sargent, W. 1994, ApJ 430
131
\bibitem[Bruhweiler and Verner(2008)]{bru08}Bruhweiler, F. and Verner, E. 2008, ApJ
675 83
\bibitem[Chen \& Halpern(1989)]{chenhalpern89} Chen, K., \& Halpern, J.~P.\ 1989, \apj, 344, 115
\bibitem[Clavel et al.(1992)]{cla92}Clavel, J., Nandra K., Makino, F., et al. 1992, ApJ 393 113
\bibitem[Corbin (1991)]{cor91} Corbin, M. 1991, ApJ 375 503
\bibitem[Corbin (1997a)]{cor97} Corbin, M. 1997, ApJ 485 517
\bibitem[Corbin (1997b)]{cor98} Corbin, M. 1997, ApJS 113 245
\bibitem[Corbin and Francis (1994)]{cor94} Corbin, M. and Francis, P. 1994, AJ 108 2016
\bibitem[Corbin and Boroson (1996)]{cor96} Corbin, M. and Borosn, T. 1996, ApJS 107
69
\bibitem[Davis and Laor(2011)]{dav11}Davis, S., Laor, A. 2011, ApJ 728 98
\bibitem[Denney et al.(2009)]{den09} Denney, K. 2009, to appear in ApJL http://xxx.lanl.gov/abs/0908.0327
\bibitem[DeCarli et al.(2011)]{dec11}DeCarli, R. Dotti, M> Treves, A. 2011, MNRAS
413 39
\bibitem[Elvis et al.(1994)]{elv94} Elvis, M., Wilkes, B., McDowell, J. et al. 1994, ApJS 95
1.
\bibitem[Emmering et al.(1992)]{emm92} Emmering, R., Blandford, R., Shlosman, I. 1992 ApJ 385 460
\bibitem[Evans and Koratkar(2004)]{eva04} Evans, I. and Koratkar 2004, ApJS 150 73
\bibitem[Fanaroff and Riley(1974)]{fr74}Fanaroff, B.~L.; Riley, J. ~M., 1974, MNRAS, 167, 31P
\bibitem[Fejes et al.(1992)]{fej92}Fejes, I., Porcas, R., Akujor, C. 1992,
A\& A 257 459
\bibitem[Ferland et al.(2013)]{fer13}Ferland, G., Porter, R., van Hoof, P. et al. 2013 Revista Mexicana de Astronomía y Astrofísica 49 137
\bibitem[Fricke et al.(1983)]{fri83} Fricke, K., Kollatschny, W., Witzel, A. 1983 A \& A 117
60
\bibitem[Garrington et al.(1991)]{gar91} Garrington, S., Conway, R., Leahy, J. 1991 MNRAS
250 171
\bibitem[Ghisellini et al.(1998)]{ghi98} Ghisellini, G. et al 1998 301 451
\bibitem[Ghisellini et al.(2009)]{ghi09} Ghisellini, G. et al 2009 to appear in MNRAS http://xxx.lanl.gov/abs/0909.0016
\bibitem[Ghisellini et al.(2010)]{ghi10} Ghisellini, G., Tavecchio, F., Foschini, L., Ghirlanda, G., Maraschi, L., Celotti, A. 2010
MNRAS 402 497
\bibitem[Greene and Ho(2005)]{gre05} Greene, J. E., \& Ho, L. C. 2005, ApJ, 630, 122
\bibitem[Gupta et al.(2017)]{gup17} Gupta, A., Mangalam, A., Wiita, P. 2017, MNRAS,
472 788
\bibitem[Hummel et al.(1992)]{hum92}Hummel, C., Muxlow, T., Krichbaum, T., et al. 1992, Astronomy and Astrophysics, 266, 93
\bibitem[Impey and Tapia (1990)]{imp90} Impey, C. and Tapia, S., 1990 ApJ 354
124
\bibitem[Impey et al (1991)]{imp91} Impey, C., Lawrence, C. and Tapia, S., 1991 ApJ
375 46
\bibitem[Jackson and Browne(1991)]{jac91} Jackson, N., Browne, I.W.A. 1991 MNRAS 250
422
\bibitem[Jarvis and McLure(2006)]{jar06}Jarvis, M., McLure, R. 2006
MNRAS 369 182
\bibitem [Jorstad et al.(2017)]{jor17}Jorstad, S., Marscher, A., Morozova, D., et al. 2007, ApJ, 846, 98
\bibitem [Kellermann et al (1969)]{kel69}Kellermann, K. I., Pauliny-Toth, I. I. K., Williams, P. J. S. 1969 ApJ 157 1
\bibitem[Kellermann et al.(2004)]{kel04} Kellerman, K.I. et al 2004 ApJ 609
539
\bibitem[Kharb et al.(2010)]{kha10} Kharb, P., Lister, M., Cooper, N.  2010,
ApJ 710, 764
\bibitem[Kinney et al.(1991)]{kin91} Kinney, A., Bohlin, R., Blades, J., York, D. 1991,
ApJS 75, 645
\bibitem[Kong et al.(2006)]{kon06} Kong, M.-Z., Wu, X.-B., Wang, R., Han, J.-L. 2006 Chin.J.Astron.Astrophys. 6:396-410
\bibitem [Korista et al.(1997)]{kor97}Korista, K. T., Baldwin, J., Ferland, G., Verner, D. 1997, ApJS, 108, 401
\bibitem[Laor(1991)]{lao91} Laor, A. (1991) ApJ 376 90
\bibitem[Laor et al.(1997)]{lao97} Laor, A. Fiore, F., Elvis, M., Wilkes, B., McDowell, J. (1997) ApJ 477
93
\bibitem[Laor and Davis(2014)]{lao14}Laor, A., Davis, S. 2014 ApJ 428 3024
\bibitem[Lawrence et al.(1996)]{law96} Lawrence, C. et al. 1996, ApJS 107
541
\bibitem[Leon-Tavares et al.(2013)]{leo13} Leon-Tavares, J., Chavushyan, V., Patino-Alvarez, V. et al. 2013, ApJ
763 36
\bibitem[Lightman et al.(1975)]{lig75}Lightman, A., Press, W., Price, R. \& Teukolsky, S. 1975, \emph{Problem Book in Relativity and Gravitation}(Princeton University Press, Princeton)
\bibitem[Lind and Blandford(1985)]{lin85}Lind, K., Blandford, R.
1985, ApJ 295 358
\bibitem[Lister et al.(2013)]{lis13}Lister, M. L., Aller, M. F., Aller, H. D., et al. 2013, AJ, 146, 120
\bibitem[Lister et al.(2016)]{lis16}Lister, M., Aller, M. and Aller, H. et al. 2016 AJ
152 12
\bibitem[Lister et al(2009)]{lis09}Lister, M. et al 2009, ApJL 696 22
\bibitem[Malkan(1983)]{mal83} Malkan, M. 1983, ApJ 268, 582
\bibitem[Malkan and Moore(1986)]{mal86}Malkan, M. and Moore, R. 1986, ApJ 300 216
\bibitem[Marziani et al.(1996)]{mar96}Marziani, P., Sulentic, J., Dultzin-Hacyan, D., Calvani, M., Moles, M. 1996, ApJS 104 37
\bibitem[Marziani et al.(2013)]{mar13}Marziani, P., Sulentic, J., Plauchu-Frayn, I., del Olmo, A. 2013 A\&A 555 89
\bibitem[McCarthy(1993)]{mcc93}McCarhty, P. 1993  Annu. Rev. Astron. Astrophys. 31 639
\bibitem[Murphy et al.(1993)]{mur93} Murphy, D., Browne, I.W.A., Perley, R. 1993, MNRAS 264 298
\bibitem[Murray et al(1995)]{mur95} Murray, N. Chiang, J., Grossman, S., Voit, G. 1995, ApJ 451 498
\bibitem[Netzer et al.(1994)]{net95} Netzer, H., Kazanas, D., Baldwin, J., Ferland, G., Browne, I.W.A. 1995, ApJ 430
191
\bibitem[Nilsson et al.(2009)]{nil09} Nilsson, K.; Pursimo, T.; Villforth, C.; Lindfors, E.; Takalo, L. O.. 2009, A\&A
505 601
\bibitem[de Pater and Perley(1983)]{pat83} de Pater, I., Perley, R. 1983, ApJ 273 64
\bibitem[Pearson and Readhead(1981)]{pea81} Pearson, T., Readhead, A. 1981, ApJ 248
61
\bibitem[Peterson et al.(2004)]{pet04} Peterson, B. M., Ferrarese, L., Gilbert K. et al. 2004, ApJ, 613, 682
\bibitem[Piconcelli(2005)]{pic05} Piconcelli, E.; Jimenez-Bailón, E.; Guainazzi, M. et al. 2005, A \& A 432 15
\bibitem[Popovic et al(1995)]{pop95} Popovic, L., Vince, I., Atanackkovic-Vukmanovic, O., Kubicela, A,
 1995, A \& A 293 309
\bibitem[Punsly(1995)]{pun95}Punsly, B. 1995 AJ 109 1555
\bibitem[Punsly and Tingay(2005)]{pun05}Punsly, B. and Tingay, S. 2005 ApJL 633 89
\bibitem[Punsly(2007)]{pun07} Punsly, B. 2007, MNRAS Lett. 374 10
\bibitem[Punsly(2010)]{pun10} Punsly, B. 2010, ApJ 713 232
\bibitem[Punsly(2012)]{pun12} Punsly, B. 2012, ApJL 762 25
\bibitem[Punsly(2014)]{pun14}Punsly, B. 2014 ApJL 797 33
\bibitem[Punsly(2015)]{pun15}Punsly, B. 2015 ApJ 806 47
\bibitem[Punsly et al.(2016)]{pun16}Punsly, B., Marziani, P., Zhang, S., Muzahid, S., O'Dea, C. 2016, ApJ
830, 104
\bibitem[Punsly and Kharb(2016)]{pun17}Punsly, B., Kharb, P. 2016,
ApJ 833 57
\bibitem[Punsly and Kharb(2017)]{pun18}Punsly, B., Kharb, P. 2017,
MNRAS Letters 468 72
\bibitem[Punsly et al.(2018)]{pun19}Punsly, B., Tramacere, P., Kharb, P., Marziani, P. 2018 ApJ 869 164
\bibitem[Punsly and Zhang(2011)]{pun11}Punsly, B., Zhang, S,. 2011,
ApJL 735 3
\bibitem[Reid et al.(1995)]{rei95}Reid, A., Shone, D., Akujor, C., et al. 1995, A\%AS.
110 213
\bibitem[Richards et al.(2002)]{ric02}Richards, G. et al 2002, AJ 124 1
\bibitem[Rumnoe et al.(2013)]{rum13}Rumnoe, J., Brotherton, M., Shang, Z. Wills, B., DiPompeo. M. 2013, MNRAS
429 135.
\bibitem[Saikia et al.(1990)]{sai90} Saikia, D. , Muxlow, T., Junor, W. 1990,
MNRAS 245 503
\bibitem[Scott et al.(2004)]{sco04} Scott, J., Kriss, G., Brotherton, M., Green, R., Hutchings, J. Shull, M., Zheng, W. 2004,
ApJ 615 135
\bibitem[Shen and Liu(2012)]{she12} Shen, Y., \& Liu, X.\ 2012, ApJ 753 125
\bibitem[Smith and Spinrad(1980)]{smi80}Smith, H., Spinrad, H. 1980,
ApJ 236, 419
\bibitem[Spinrad et al.(1985)]{spi85}Spinrad H., Djorgovski., S., Marr J., Aguilar, L. 1985,
PASP 97 932
\bibitem[Steffen et al.(2006)]{ste06}Steffen, A., Strateva, I., Brandt, W. N., Alexander, D., Koekemoer, A., Lehmer, B., et al.
2006 Astron. J. 131, 2826
\bibitem[Stevans et al.(2014)]{ste14}Stevans, M., Shull, M., Danforth, C., Tilton, E. 2014 ApJ 794 75
\bibitem[Sulentic et al.(2000)]{sul00}Sulentic, J., Marziani, P., and Dultzin-Hacyan, D. 2000 ARA\& A 38, 521
\bibitem[Sun and Malkan(1989)]{sun89}Sun, W.-H., and Malkan, M. A 1989, ApJ 346 68
\bibitem[Tang et al.(2012)]{tan12}Tang B., Shang Z., Gu Q., Brotherton M. S., Runnoe J. C., 2012, ApJ
201 38
\bibitem[Torrealba et al.(2012)]{tor12}Torrealba, J., Chavushyan, V., Cruz-Gonzalez, I., Arshakian, T.
G., Bertone, E., Rosa-Gonzalez, D. 2012, RMxAA 48 9
\bibitem[Telfer et al.(2002)]{tel02}Telfer, R., Zheng, W., Kriss, G., Davidsen, A. 2002, ApJ 565 773
\bibitem[Trakhtenbrot \& Netzer(2012)]{tra12} Trakhtenbrot, B. and Netzer, H.\ 2012, MNRAS  427 3081
\bibitem[Urry \& Padovani(1995)]{urrypadovani95} Urry, C.~M., \& Padovani, P.\ 1995, \pasp, 107, 803
\bibitem[van Breugel et al.(1984)]{van84}van Breugel, W., Miley, G., Heckman, T. 1984, AJ
189 5
\bibitem[Vestergaard and Peterson(2006)]{ves06}Vestergaard, M. and Peterson, B. 2006, ApJ 641
689
\bibitem[Vestergaard and Wilkes(2001)]{ves01}Vestergaard, M. and Wilkes, B. 2001, ApJS 134
1
\bibitem[Walker et al.(1987)]{wal87}Walker, R., Benson, J., Unwin, S. 1987 ApJ
316 546
\bibitem[Wandel et al.(1999)]{wan99}Wandel A., Peterson , B, Malkan, M. 1999 ApJ
526 579
\bibitem[Wang et al.(2004)]{wan04}Wang, J.-M., Luo, B, Ho, L. 2004 ApJL 615 9
\bibitem[Willott et al.(1999)]{wil99}Willott, C., Rawlings, S., Blundell, K., Lacy, M. 1999, MNRAS 309 1017
\bibitem[Wills et al.(1992)]{wil92} Wills, B., Wills, D., Breger, M., Antonucci, R.,  Barvainis, R. 1992 ApJ
398 454
\bibitem[Wills et al.(1993)]{wil93} Wills, B. et al 1993, ApJ 415
563
\bibitem[Wills et al.(1995)]{wil95} Wills, B., Thompson, K., Han, M. et al 1995, ApJ 437
\bibitem[Wills and Brotherton(1995)]{wbr95} Wills, B., Brotherton, M 1995, ApJL
448 81
\bibitem[Wills and Brotherton(1996)]{wbr96} Wills, B., Brotherton, M 1995, "Jets and QSO Spectra"
in \emph{Jets from Stars and Galactic Nuclei}, Kundt, W. ed
(Springer lecture Note Series)
http://xxx.lanl.gov/abs/astro-ph/9605161
\bibitem[Wills and Browne(1986)]{bro86} Wills, B.J., Browne, I.W.A. 1986 ApJ 302
56
\bibitem[Woo et al.(2010)]{woo10} Woo, J-H., Treu. T., Barth, A. et al.2010 ApJ 716 269
\bibitem[Zheng et al.(1997)]{zhe97} Zheng, W., Kriss, G., Telfer, R. et al. 1997 ApJ 475 469
\end{thebibliography}
\end{document}